\documentclass[aps, prb, twocolumn, superscriptaddress, bibnotes]{revtex4-2}

\pdfoutput=1
\usepackage[english]{babel}
\usepackage[utf8]{inputenc}

\usepackage{lmodern}
\usepackage[T1]{fontenc}

\usepackage{bm}
\usepackage{graphicx}
\usepackage{mathtools}
\usepackage{microtype}
\usepackage{slantsc}
\usepackage{upgreek}
\usepackage{xcolor}

\usepackage[absolute]{textpos}

\usepackage[colorlinks, allcolors=blue]{hyperref}
\usepackage[figure, figure*]{hypcap}

\def\D{\mathrm d}
\def\E{\mathrm e}
\def\I{\mathrm i}

\def\abs#1{|#1|}
\def\bra#1{\langle#1|}
\def\ket#1{|#1\rangle}
\def\sub#1{_{\text{#1}}}

\def\topparens#1#2{%
    \overset{\clap{\smash[b]{\raisebox{-3.5pt}{$%
        \displaystyle%
        \scalebox{0.4}{\bf(\hspace{#1}}%
        \hphantom{#2}%
        \scalebox{0.4}{\bf\hspace{#1})}%
        $}}}}{#2}}

\def\TaS2{1H-TaS\s2}

\let\Pi\varPi
\let\Theta\varTheta
\let\epsilon\varepsilon
\let\phi\varphi
\let\s\textsubscript
\let\tilde\widetilde
\let\vec\bm

\def\cdwthin{Lin2018, Hall2019}
\def\cdwbulk{Tidman1974, Scholz1982, Coleman1988, Wang1990, Wang1991, Nagata1992, Tonjes2001}

\def\dopingcdwhightc{Fradkin2015}
\def\dopingcdwtas2{Sanders2016, Albertini2017, Hall2019, Shao2019}
\def\dopingcdwtmdc{Wilson1974, DiSalvo1975, Chen2015, Yu2015, Shao2016}

\def\cuprates{Bednorz1986, Shen2002, Gunnarsson2008, Fradkin2015, Cavalleri2018}
\def\febased{Kamihara2008, Stewart2011, Huang2017}
\def\hydrides{Ashcroft1968, Gorkov2018, Drozdov2019}

\def\origincdw{Withers1986, Rossnagel2011, Pasquier2019}

\def\agreement{Doran1978, CastroNeto2001, Johannes2008, Ge2012}
\def\bonding{McMillan1977, Haas1978, Inglesfield1980a, Inglesfield1980b, Whangbo1992, SilvaGuillen2016}
\def\matrixelements{Varma1983, Wang1990, Rossnagel2001, Valla2004, Johannes2006, Calandra2009, Weber2011, Soumyanarayanan2013, Arguello2014, Arguello2015, Flicker2015, Xi2015, Zhu2015, Ugeda2016, Nakata2018}
\def\nesting{Wilson1974, Wilson1975, Wexler1976, Wilson1977, Straub1999, Shen2008, Borisenko2009}
\def\vhs{Rice1975, Tonjes2001, Kiss2007}

\def\phrenorm{Varma1977, Inglesfield1980a, Inglesfield1980b, Varma1983, Weber2013, Flicker2015}
\def\nonadiabatic{Maksimov1996, Lazzeri2006, Caudal2007, Pisana2007, Piscanec2007, Saitta2008, Calandra2010}

\def\tise2{Wilson1978, Rossnagel2002, Cercellier2007, vanWezel2010, Monney2011, Chen2018, Kaneko2018, Pasquier2018}

\def\ITP{
    Institut f\"ur Theoretische Physik,
    Universit\"at Bremen,
    Otto-Hahn-Allee 1,
    D-28359 Bremen,
    Germany}

\def\BCCMS{
    Bremen Center for Computational Materials Science,
    Universit\"at Bremen,
    Am Fallturm 1a,
    D-28359 Bremen,
    Germany}

\def\IMM{
    Institute for Molecules and Materials,
    Radboud University,
    Heyendaalseweg 135,
    NL-6525 AJ Nijmegen,
    The Netherlands}

\begin{document}

\title{\emph{Ab initio} phonon self-energies and fluctuation diagnostics of phonon anomalies:\\
    Lattice instabilities from Dirac pseudospin physics in transition metal dichalcogenides}

\author{Jan Berges}
\affiliation\ITP
\affiliation\BCCMS

\author{Erik G. C. P. van Loon}
\affiliation\ITP
\affiliation\BCCMS

\author{Arne Schobert}
\affiliation\ITP
\affiliation\BCCMS

\author{Malte R\"osner}
\affiliation\IMM

\author{Tim O. Wehling}
\affiliation\ITP
\affiliation\BCCMS

\begin{abstract}
We present an \emph{ab initio} approach for the calculation of phonon self-energies and their fluctuation diagnostics, which allows us to identify the electronic processes behind phonon anomalies. Application to the transition-metal-dichalcogenide monolayer \TaS2 reveals that coupling between the longitudinal--acoustic phonons and the electrons from an isolated low-energy metallic band is entirely responsible for phonon anomalies such as the mode softening and associated charge-density waves observed in this material. Our analysis allows us to distinguish between different mode-softening mechanisms including matrix-element effects, Fermi-surface nesting, and Van Hove scenarios. We find that matrix-element effects originating from a peculiar type of Dirac pseudospin textures control the charge-density-wave physics in \TaS2 and similar transition metal dichalcogenides.
\end{abstract}

\maketitle

\begin{textblock*}{\paperwidth}(0mm, \paperheight-18mm)
    \centering
    \small
    Published in \href{https://doi.org/10.1103/PhysRevB.101.155107}{Phys.\@ Rev.\@ B \textbf{101}, 155107 (2020)} \copyright~American Physical Society
\end{textblock*}

\section{Introduction}

\begin{figure*}
    \includegraphics[width=\linewidth]{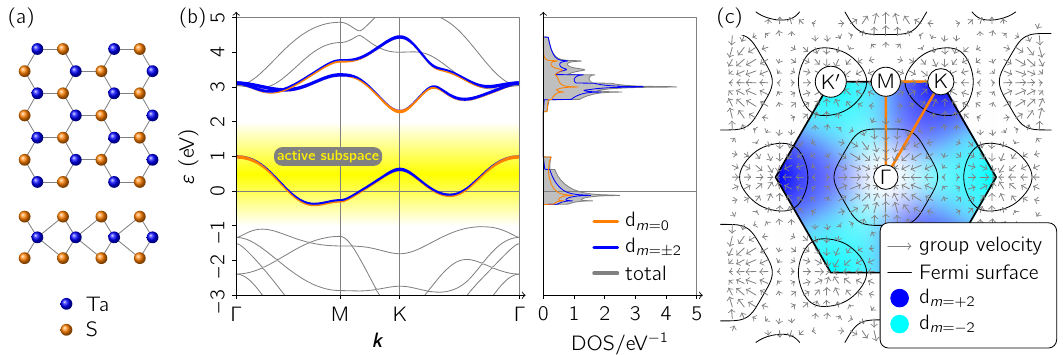}
    \caption{Crystal and electronic structure of \TaS2. (a)~Top and side view of the crystal structure. (b)~Orbitally resolved band structure and density of states. The active subspace with the isolated low-energy metallic band is highlighted in yellow. (c)~Fermi surface, orbital character, and group velocity of the low-energy band.}
    \label{fig:electrons}
\end{figure*}

Different states of electronic quantum matter are often tightly linked to lattice degrees of freedom. Examples include superconductivity, periodic lattice distortions and charge-density waves (CDWs), metal--insulator transitions, and nematic, magnetic, ``stripe\rlap,'' or excitonic order across vastly different material classes ranging from cuprate \cite{\cuprates} and Fe-based high-temperature superconductors \cite{\febased} to hydride compounds \cite{\hydrides}. Disentangling the interplay of lattice and electronic degrees of freedom has remained a formidable challenge in many cases.

Phonon anomalies and mode softening are often an indicator of instabilities of the electronic system. However, the question of whether and which electronic processes are responsible for a given phonon anomaly is the source of many controversies in the literature. Often suggestions for very different mechanisms such as matrix-element effects \cite{Varma1983}, Fermi-surface nesting \cite{Wilson1975}, or Van Hove scenarios \cite{Rice1975} are made for a phonon anomaly in one and the same material \cite{\origincdw}. Unambiguously distinguishing between such mechanisms is complicated and has typically required the combination of experimental probes of lattice and electron dynamics \cite{Cavalleri2004, Hellmann2012} with theoretical modeling \cite{Perfetti2006}.

Here, we present \emph{ab initio} calculations of phonon self-energies, and we introduce the concept of fluctuation diagnostics \cite{Gunnarsson2015, Gunnarsson2016} to the domain of lattice dynamics. This scheme can distinguish between different strong- and weak-coupling effects and combinations thereof in the context of phonon anomalies in a quantitative and material-specific way.

One prototypical class of materials hosting phonon anomalies are the hexagonal polytypes of the layered group-V transition metal dichalcogenides (TMDCs) \cite{Wilson1969}. Bulk and monolayer [Fig.~\ref{fig:electrons}\,(a)] are denoted by 2H- and 1H-$MX_2$, where $M$ stands for Nb or Ta and $X$ for S or Se. Temperature-dependent phonon-mode softening and CDWs are ubiquitous in these materials, but explanations have remained controversial for several decades, and suggestions include strong-coupling arguments based on matrix elements \cite{\matrixelements} or local chemical bonding \cite{\bonding} as well as weak-coupling arguments based on Fermi-surface nesting \cite{\nesting} or Van Hove scenarios \cite{\vhs}. Our phonon-self-energy calculations and fluctuation diagnostics for monolayer \TaS2 reveal that coupling between the longitudinal--acoustic (LA) phonons and the electrons from an isolated low-energy metallic band [Fig.~\ref{fig:electrons}\,(b,\,c)] is entirely responsible for the mode softening and associated CDWs observed in this material. A combination of imperfect Fermiology conditions and matrix-element effects resulting from Dirac pseudospin textures is pinpointed as the cause of the CDW phase diagram of \TaS2 and similar TMDCs.

\section{Bare and screened phonons}
\label{sec:bare_screened}

\begin{subequations}
\label{eq:H}
A general Hamiltonian describing systems of interacting electrons and phonons reads
\begin{equation}
    H = H \sub{el} + H \sub{el--el} + H \sub{ph} + H \sub{el--ph}
\end{equation}
and contains one-body electron terms $H \sub{el}$, the Coulomb interaction $H \sub{el--el}$, pure phonon terms $H \sub{ph}$, and the electron--phonon interaction $H \sub{el--ph}$. The necessity to account simultaneously for the complexity of the single-particle electronic wave functions and the difficulties arising from the interactions present in $H$ render realistic descriptions of solid-state systems notoriously complicated. One way to proceed and to gain insights in practice is via material-realistic low-energy Hamiltonians, where the electronic degrees of freedom accounted for in $H$ are restricted to some low-energy subspace, often also dubbed correlated subspace, target subspace, or active subspace. We will adopt the latter nomenclature. In the presented case of \TaS2, we take as a natural choice for the active subspace the electronic states of the low-energy band highlighted in Fig.~\ref{fig:electrons}\,(b), where spin--orbit coupling is disregarded for simplicity. A discussion of spin--orbit-coupling effects is given in Appendix~\ref{app:SOC}.

Then, all quantities entering $H$ have to be partially renormalized to account for the elimination of the higher-energy degrees of freedom \cite{Aryasetiawan2004, Nomura2015}. More precisely, the phonon energies entering
\begin{equation}
    H \sub{ph} = \sum_{\vec q \nu}
    \omega_{\vec q \nu}^{\phantom \dagger} \,
    [b_{\vec q \nu}^\dagger \,
    b_{\vec q \nu}^{\phantom \dagger}
    + \tfrac 1 2]
\end{equation}
and the electron--phonon couplings in
\begin{equation}
    H \sub{el--ph} = \frac 1 {\sqrt N} \,
    \sum_{\mathclap{\vec q \nu \vec k m n}}
    g_{\vec q \nu \vec k m n}^{\phantom \dagger} \,
    [b_{-\vec q \nu}^\dagger + b_{\vec q \nu}^{\phantom \dagger}] \,
    c_{\vec k + \vec q m}^\dagger \,
    c_{\vec k n}^{\phantom \dagger}
\end{equation}
are partially renormalized from the viewpoint of a full first-principles Hamiltonian and ``bare'' from the viewpoint of the model.
\end{subequations}

Directly experimentally observable are the fully renormalized quantities, which can be obtained by either solving the model Hamiltonian $H$ [Eqs.~\eqref{eq:H}] or by direct treatment of the full system from first principles. For instance, density-functional perturbation theory (DFPT) \cite{Baroni2001} yields the (in practice approximate) fully renormalized phonon dispersions and electron--phonon couplings from first principles.

Partially screened phonon dispersions and electron--phonon couplings can be obtained from the constrained density-functional perturbation theory (cDFPT) \cite{Nomura2015}. Analogously to the constrained random-phase approximation (cRPA) \cite{Aryasetiawan2004} to partially screened Coulomb interactions, cDFPT excludes polarization processes taking place inside the active subspace from the screening of phonon dispersions and electron--phonon couplings. In the following, we distinguish fully from partially screened quantities by a tilde (\raisebox{-1ex}{$\tilde{\phantom x}$}) on top of symbols for the former, and we refer to them as ``screened'' and ``bare'' for brevity.

Selected bare (cDFPT) and screened (DFPT) phonon dispersions are shown in Fig.~\ref{fig:phonons}\,(a). The bare phonon dispersion of \TaS2 (middle), which excludes screening intrinsic to the active subspace, is smooth in reciprocal space indicating correspondingly short-range force constants, and it does not show any Kohn anomalies \cite{Kohn1959} or dynamical lattice instabilities. In contrast, the screened phonon dispersion of \TaS2 (left) shows strong Kohn anomalies, softening, and instabilities in the LA phonon branch in extended regions of the Brillouin zone (BZ). The instability regions include the wave vector $\vec q = 2/3 \, \mathrm M$ associated with the $3 \times 3$ CDW observed in bulk \cite{\cdwbulk} and thin \TaS2 \cite{\cdwthin}. The leading instability in the screened dispersion as signaled by the (in absolute value) largest imaginary phonon energy is indeed close to $\vec q = 2/3 \, \mathrm M$. As the bare phononic system is dynamically stable and has a smooth LA dispersion in contrast to the screened one, renormalization processes taking place inside the low-energy band must be fully responsible for the mode softening and the CDW physics observed in \TaS2.

While the bare phonon dispersion is not directly experimentally observable, screening due to the low-energy band can be suppressed also in experiment, for instance by effective doping. If the low-energy band is completely filled, no intraband screening processes are possible. Such a situation is realized in group-VI TMDCs such as 1H-WS\s2, which is isostructural to the undistorted high-temperature phase of \TaS2 and has one additional electron per primitive cell but otherwise a similar electronic band structure. As seen in Fig.~\ref{fig:phonons}\,(a), the screened phonon dispersion of 1H-WS\s2 (right) is indeed very similar to the bare phonon dispersion of \TaS2 (middle). Thus, studies of isostructural compounds with different filling of the electronic bands present a route toward experimental estimates of bare phonon dispersions.

\section{\emph{Ab initio} phonon self-energies}
\label{sec:self-energies}

\begin{figure*}
    \includegraphics[width=\linewidth]{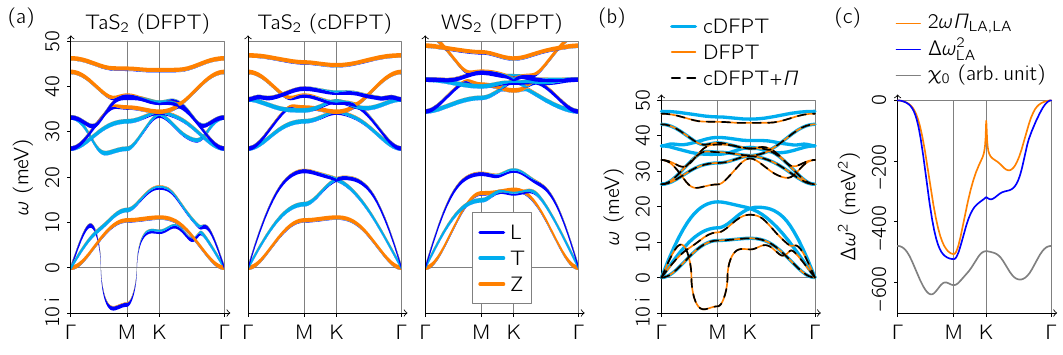}
    \caption{(a)~Phonon dispersions of \TaS2 (DFPT, cDFPT) and 1H-WS\s2 (DFPT). (b)~Bare phonon dispersion of \TaS2 from cDFPT compared to screened ones from DFPT and according to Eqs.~\eqref{eq:scheme} (cDFPT phonon energies renormalized \emph{a posteriori} with the adiabatic phonon self-energy). The two screened dispersions are identical, showing that Eqs.~\eqref{eq:scheme} provide the exact link between cDFPT and DFPT\@. (c)~LA diagonal matrix element of the phonon self-energy, squared-energy change of the (predominantly) LA phonon band, and bare electronic susceptibility of \TaS2.}
    \label{fig:phonons}
\end{figure*}

There are two different ways to calculate screened phonon dispersions: first, with DFPT, and second, by approximately solving the model Hamiltonian $H$. In the latter case, the experimentally observable lattice dynamics is encoded in the screened phonon Green function.

\begin{subequations}
\label{eq:scheme}
In the adiabatic approximation, the changes in phonon normal modes and energies induced by electron--phonon coupling can be obtained from the renormalized dynamical matrix
\begin{equation}
    \label{eq:dyson}
    \tilde \omega^2_{\vec q \mu \nu}
    = \omega^2_{\vec q \nu} \delta_{\mu \nu}^{\phantom 2}
    + 2 \sqrt{\omega_{\vec q \mu} \omega_{\vec q \nu}} \Pi_{\vec q \mu \nu},
\end{equation}
which follows from the bare dynamical matrix $\omega^2_{\vec q \nu} \delta_{\mu \nu}$, here written in its eigenbasis labeled by $\mu$ and $\nu$, and a correction determined by the phonon self-energy
\begin{equation}
    \label{eq:Pi}
    \Pi_{\vec q \mu \nu} = \frac 2 N \sum_{\vec k m n}
    g_{\vec q \mu \vec k m n}^* \frac
        {f(\epsilon_{\vec k + \vec q m}) - f(\epsilon_{\vec k n})}
        {  \epsilon_{\vec k + \vec q m}  -   \epsilon_{\vec k n} }
    \tilde g_{\vec q \nu \vec k m n}^{\phantom *}.
\end{equation}
Here, $\epsilon$ and $f$ are electronic band energies and occupations, $m$ and $n$ label the electronic bands that constitute our active subspace, the factor of $2$ comes from the spin degeneracy, and $N$ is the number of $\vec k$~points summed over. The electron--phonon coupling $g$ appears in both bare and screened form and reads~\footnote{Note that in the present formalism, the screened electron--phonon coupling
\begin{equation*}
    \protect \tilde g_{\vec q \nu \vec k m n}
    = \frac 1 {\sqrt{2 \omega_{\vec q \nu}}}
    \sum_i e_{\vec q i \nu}
    \frac 1 {\sqrt{M_i}}
    \protect \bra{\vec k {+} \vec q m}
        \frac{\tilde{\partial V}}{\partial u_{\vec q i}}
    \protect \ket{\vec k n}
\end{equation*}
depends on the screened potential change $\protect \tilde{\partial V}$ but on the \emph{bare} phonon energies $\omega$ and eigenvectors $e$.}
\begin{equation}
    \label{eq:g}
    \topparens{2pt}{\tilde g}_{\vec q \nu \vec k m n}
    = \frac 1 {\sqrt{2 \omega_{\vec q \nu}}}
    \sum_i e_{\vec q i \nu}
    \frac 1 {\sqrt{M_i}}
    \bra{\vec k {+} \vec q m}
        \smash{\frac{\topparens{-3pt}{\tilde{\partial V}}}{\partial u_{\vec q i}}}
    \ket{\vec k n},
\end{equation}
where the combined index $i$ runs over the three Cartesian displacement directions of each atom, $e$ is an eigenvector of the bare dynamical matrix, $M$ is the atomic mass, and $\partial V$ is the change of the self-consistent Kohn--Sham potential upon an atomic displacement $\partial u$. For detailed information on Eqs.~\eqref{eq:scheme}, we refer to Ref.~\onlinecite{Giustino2017}, in particular Section~V\,A.
\end{subequations}
We note that the coupling $g$ is a complex quantity that contains \emph{a priori} arbitrary phase factors from the electronic eigenstates at $\vec k$ and $\vec k + \vec q$. Therefore, care has to be taken that a consistent gauge is applied when obtaining $g$ and $\tilde g$ from independent cDFPT and DFPT calculations. We address this problem by fixing the gauge in a localized basis of Wannier functions, which as an additional advantage allows for the Fourier interpolation to arbitrary $\vec q$ and $\vec k$~resolutions \cite{Giustino2007}.

We calculated the phonon self-energy [Eq.~\eqref{eq:Pi}] for the case of \TaS2 and renormalized the bare phonon dispersions obtained from cDFPT accordingly [Eq.~\eqref{eq:dyson}]. A comparison of the bare phonon dispersion to the screened ones as obtained from DFPT and from the phonon self-energy is shown in Fig.~\ref{fig:phonons}\,(b). We see that both screened dispersions are the same throughout the BZ path. Indeed, the approximations involved in the DFPT calculation (adiabaticity and a semilocal exchange--correlation functional) can be shown to be equivalent to Eqs.~\eqref{eq:scheme} \cite{Nomura2015}. Exchange--correlation effects beyond RPA only enter through the difference between $g$ and $\tilde g$.

The phonon self-energy is a matrix in the space of atomic displacement coordinates. The renormalization according to Eqs.~\eqref{eq:scheme} accounts for this full matrix structure. To lowest order, corrections to the bare phonon energies from low-energy electronic screening are contained in the diagonal components $\Pi_{\vec q \nu \nu}$ of the phonon self-energy. Fig.~\ref{fig:phonons}\,(c) shows a comparison of the change $\Delta \omega^2 \sub{LA} = \tilde \omega^2 \sub{LA} - \omega^2 \sub{LA}$ of the (predominantly) LA phonon band upon low-energy electronic screening to the corresponding diagonal matrix element $2\omega \Pi_{\text{LA}, \text{LA}}$ in the eigenbasis of the bare phonons. We see that $\Delta \omega^2 \sub{LA}$ and $2 \omega \Pi_{\text{LA}, \text{LA}}$ show a qualitatively similar $\vec q$~dependence, but there are also deviations between the two, particularly close to the $\mathrm K$~point. This is due to changes in the normal-mode eigenvectors upon renormalization. Thus, the diagonal part of the phonon self-energy can serve as a qualitative guide for the understanding of phonon-renormalization phenomena, but quantitative calculations must account for its full matrix structure.

Phonon self-energies, screened, and bare phonon dispersions for \TaS2 have also been calculated in Ref.~\onlinecite{Albertini2017}, albeit with a different procedure. While the screened phonon dispersions in Ref.~\onlinecite{Albertini2017} are similar to those depicted in Fig.~\ref{fig:phonons}\,(a), the bare phonon dispersions in Ref.~\onlinecite{Albertini2017}, which have not been obtained from cDFPT but were estimated from DFPT data, differ from those found here by not being smooth but still displaying a dip at the wave vector associated with the $3 \times 3$ CDW\@. Possible origins of this discrepancy are that the analysis in Ref.~\onlinecite{Albertini2017} has been restricted to the diagonal components of the phonon self-energy and that the calculations involved only screened electron--phonon vertices instead of the required combination of bare and screened vertices as in Eq.~\eqref{eq:Pi}.

Several works, e.g., Refs.~\onlinecite{\phrenorm}, addressed the renormalization of phonons due to chosen subsets of interaction processes, but they had to rely on further, often semiempirical models or assumptions on the shape of the ``bare'' phonon dispersion. In this context, partially screened phonons and electron--phonon couplings from cDFPT as considered in this work are very helpful, since cDFPT delivers an unambiguous bare starting point. Additionally, when analyzing nonadiabaticity \cite{\nonadiabatic}, cDFPT can provide a well-defined adiabatic starting point together with the correct coupling for nonadiabatic correction terms.

\section{Fluctuation diagnostics of phonon self-energies}

\begin{figure}
    \includegraphics[width=\linewidth]{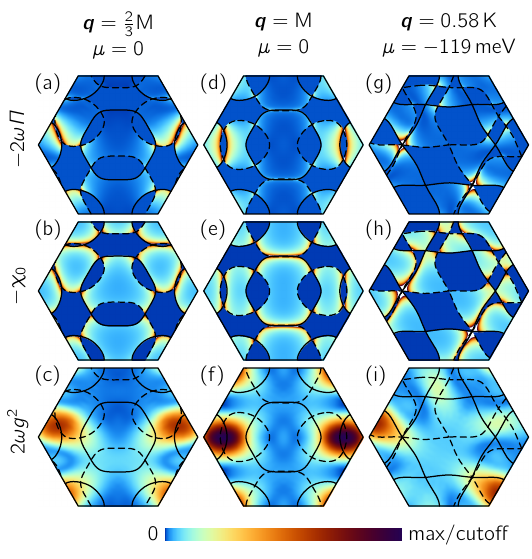}
    \caption{Momentum-resolved fluctuation diagnostics of LA-phonon-mode softening at $\vec q = 2/3 \, \mathrm M$ (left) and $\vec q = \mathrm M$ (middle) in undoped \TaS2 as well as at $\vec q = 0.58 \, \mathrm K$ (right) at Van Hove filling (chemical potential $\mu = -119$\,meV). The $\vec k$-dependent contributions to the phonon self-energy $2 \omega \Pi$, the bare electronic susceptibility $\chi_0$, and the coupling matrix elements $2 \omega g^2$ are shown color-coded. Solid (dashed) lines indicate the Fermi surface (shifted by the respective $\vec q$~vectors).}
    \label{fig:kdep}
\end{figure}

\begin{figure}
    \includegraphics[width=\linewidth]{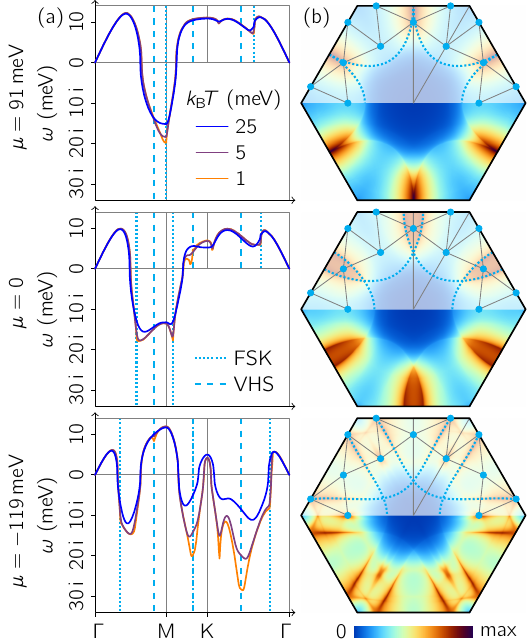}
    \caption{(a)~LA phonon dispersion and (b)~$\vec q$-dependent LA phonon self-energy ($-2 \omega \Pi$) at different chemical potentials~$\mu$. At $\mu = -119$\,meV, the VHS is at the Fermi level; at $\mu = 0$, we have no doping; at $\mu = 91$\,meV, we have touching $\mathrm K$ and $\mathrm K'$~pockets for $\vec q = \mathrm M$. The dispersion is shown for different broadenings $k \sub B T$, the phonon self-energy only for $k \sub B T = 1$\,meV\@. Special $\vec q$~points are marked in cyan, where FSK and VHS stands for touching $\mathrm K$ and $\mathrm K'$~pockets and superimposed saddle points, respectively.}
    \label{fig:qdep}
\end{figure}

\begin{figure*}
    \includegraphics[width=\linewidth]{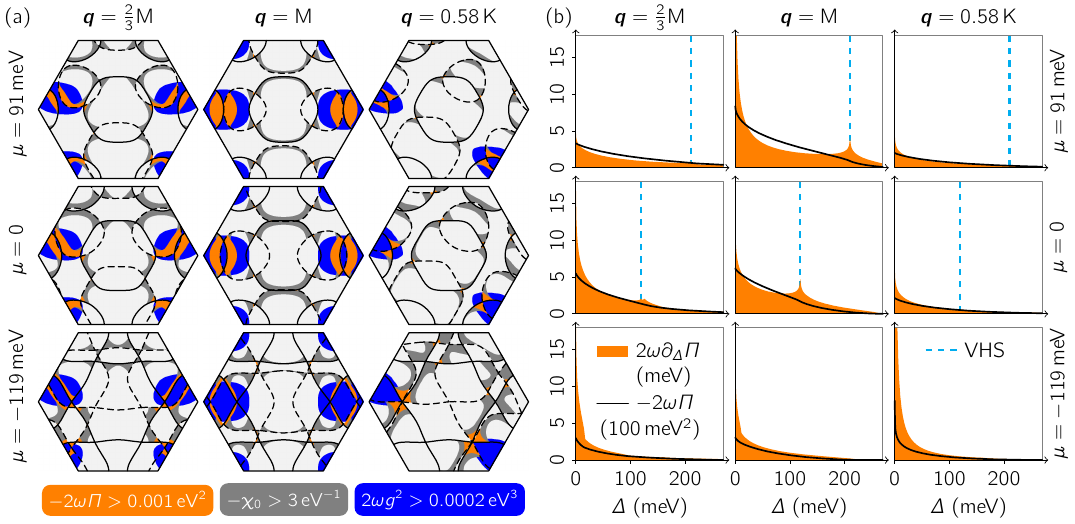}
    \caption{(a)~Momentum- ($\vec k$-) and (b)~energy-resolved fluctuation diagnostics of LA-phonon-mode softening in \TaS2 as a function of charge doping (the electronic chemical potential $\mu$) at $\vec q = 2/3 \, \mathrm M$, $\mathrm M$, and $\mathrm K$. The blue-, gray-, and orange-shaded regions in panel (a) indicate $\vec k$~points with significant contributions to the phonon self-energy, the bare electronic susceptibility, and the electron--phonon coupling as defined in Eqs.~\eqref{eq:fluctuations}.}
    \label{fig:combined}
\end{figure*}

We seek to understand unambiguously how the electrons renormalize the phonon dispersion. The expression for the phonon self-energy makes this possible: Each summand in Eq.~\eqref{eq:Pi} quantifies how much specific electronic states contribute to the phonon self-energy and allows us to identify the mechanism responsible for the phonon renormalization. A similar kind of ``fluctuation diagnostics'' has previously been applied in correlated electron systems to identify antiferromagnetic correlations as the mechanism responsible for the pseudogap in the Hubbard model \cite{Gunnarsson2015}.

\begin{subequations}
\label{eq:fluctuations}
The phonon self-energy as approximated by Eq.~\eqref{eq:Pi} is a BZ, band, and spin sum of electronic fluctuations
\begin{equation}
    \chi^0_{\vec q \vec k m n} = \frac
        {f(\epsilon_{\vec k + \vec q m}) - f(\epsilon_{\vec k n})}
        {  \epsilon_{\vec k + \vec q m}  -   \epsilon_{\vec k n} }
\end{equation}
weighted by the coupling matrix elements, for which we now adopt a symmetrized, Hermitian representation~\footnote{From the fact that the phonon self-energy $\Pi_{\mu \nu}$ is Hermitian, it follows immediately that the result of Eq.~\eqref{eq:Pi} will not change if we replace $g_\mu^* \protect \tilde g_\nu^{\protect \phantom *}$ by $\protect \tilde g_\mu^* g_\nu^{\protect \phantom *}$ or, as a consequence, the right-hand side of Eq.~\eqref{eq:g2}. All anti-Hermitian parts of the summands in Eq.~\eqref{eq:Pi} cancel.},
\begin{equation}
    \label{eq:g2}
    g^2_{\vec q \mu \nu \vec k m n} =
    \frac {
        g_{\vec q \mu \vec k m n}^* \cdot \tilde g_{\vec q \nu \vec k m n}^{\phantom *} +
        \tilde g_{\vec q \mu \vec k m n}^* \cdot g_{\vec q \nu \vec k m n}^{\phantom *}
    } 2.
\end{equation}
By analyzing
\begin{equation}
    \Pi_{\vec q  \mu \nu \vec k m n} =
    g^2_{\vec q \mu \nu \vec k m n}
    \cdot \chi^0_{\vec q \vec k m n}
\end{equation}
as a function of electronic momenta $\vec k$ and $\vec k + \vec q$, we can perform ``fluctuation diagnostics'' and identify which fermionic fluctuations contribute most dominantly to the phonon self-energy. Furthermore, by comparison of $\Pi$ to $\chi_0$ and $g^2$ we can directly distinguish purely electronic band-structure and ``Fermiology'' effects from matrix-element effects.
\end{subequations}

\begin{subequations}
\label{eq:Delta}
One can similarly quantify the contribution of electronic states from a certain energy range to the phonon renormalization. Considering a single band for brevity,
\begin{equation}
    \Pi(\Delta) = \frac 2 A \int \D^2 k \, \Pi_{\vec k} \,
    \Theta(\abs{\epsilon_{\vec k}} - \Delta) \, \Theta(\abs{\epsilon_{\vec k + \vec q}} - \Delta)
\end{equation}
with the BZ area $A$ accounts only for fermionic modes with energies outside of an energy window $[-\Delta, +\Delta]$ around the Fermi level. In the renormalization-group spirit, this corresponds to integrating out all electrons outside of the energy window. The full result of the calculation is recovered by letting $\Delta \to 0$. The derivative
\begin{multline}
    \partial_\Delta \Pi(\Delta) = -\frac 2 A \int \D^2 k \, \Pi_{\vec k} \,
    [ \delta(\abs{\epsilon_{\vec k}} - \Delta) \, \Theta(\abs{\epsilon_{\vec k + \vec q}} - \Delta) \\
    + \Theta(\abs{\epsilon_{\vec k}} - \Delta) \, \delta(\abs{\epsilon_{\vec k + \vec q}} - \Delta) ]
\end{multline}
then quantifies the contributions of electronic states with energies from the shell $\abs{\epsilon_{\vec k}} = \Delta$ to $\Pi$~\footnote{The information contained in $\partial_\Delta \Pi(\Delta)$ is similar to the information contained in $\mathrm{Im} \, \Pi(\omega + \I 0^+)$.}.
\end{subequations}

Matrix-element effects, Fermi-surface nesting, and Van Hove scenarios as possible causes of phonon-mode softening manifest markedly differently in the fluctuation diagnostics. For a generic $d$-dimensional system, in the absence of any Fermi-surface anomalies like nesting, $\Pi(\Delta)$ remains finite as $\Delta \rightarrow 0$. On the other hand, Fermi-surface nesting, as realized in one-dimensional systems for $q \sub c = 2 k \sub F$ \cite{Peierls1955} or also in higher-dimensional systems for parallel sheets of the Fermi surface linked by some nesting vector $\vec{q} \sub c$, leads to diverging $\Pi_{\vec q \sub c}(\Delta) \sim \log \abs \Delta$ and $\partial_\Delta \Pi_{\vec q \sub c}(\Delta) \sim 1 / \Delta$ \cite{Roth1966}. We expect the same kind of divergences in so-called Van Hove scenarios, where the Fermi level is at the energy of Van Hove singularities (VHS) in the electronic density of states, e.g., for wave vectors $\vec q \sub c = \vec q \sub{VHS}$ connecting two saddle points \cite{Rice1975} as found in \TaS2. In turn, nesting and Van Hove scenarios can be clearly distinguished in $\vec k$-space fluctuation diagnostics by dominant contributions to $\Pi$ originating from line segments in the case of nesting and being centered around the Van Hove points in the latter case. The role of the matrix elements is seen by comparing $\chi_0$ to $\Pi$.

\section{Fluctuation~diagnostics of LA-phonon-mode~softening and CDW~formation in \protect\TaS2}

The CDW physics of \TaS2 is associated with softening of LA phonon modes in the undistorted phase toward dynamical lattice instabilities. As shown in Section~\ref{sec:bare_screened}, this softening is entirely due to coupling of the phonons to the electrons from the active subspace in Fig.~\ref{fig:electrons}\,(b). In the following, we identify which processes inside this active subspace contribute most dominantly to the phonon-mode softening for pristine [Section~\ref{sec:pristine}] and doped \TaS2 [Section~\ref{sec:doped}]. Only the LA diagonal elements of $2 \omega g^2$ and $2 \omega \Pi$ in the eigenbasis of the bare phonons are shown. The prefactor of $2 \omega$ cancels with the prefactor in Eq.~\eqref{eq:g}. Most subscripts are omitted for brevity. All computational parameters are listed in Appendix~\ref{app:parameters}.

\subsection{Pristine \protect\TaS2}
\label{sec:pristine}

The wave vector associated with the $3\times 3$ CDW in pristine \TaS2, i.e., at the chemical potential $\mu = 0$, is $\vec q = 2/3 \, \mathrm M$. The fluctuation diagnostics of the corresponding phonon self-energy $2 \omega \Pi$ [Fig.~\ref{fig:kdep}\,(a)] reveals that the dominant contributions to $2 \omega \Pi$ are peaked in distinct regions of $\vec k$~space: The Fermi surface (contour) of undoped \TaS2 consists of three hole pockets encircling $\Gamma$, $\mathrm K$, and $\mathrm K'$, respectively. The strongest contributions to $2 \omega \Pi$ originate from regions where the original pocket around $\mathrm K$ approximately touches the pocket around $\mathrm K'$, shifted by $-\vec q$, and vice versa. There are two such regions of touching $\mathrm K$ and $\mathrm K'$~pockets (intervalley processes) and two regions of touching $\mathrm K$ and shifted $\mathrm K$~pockets or touching $\mathrm K'$ and shifted $\mathrm K'$~pockets (intravalley processes) in the BZ\@. While all four of these regions contribute to the bare electronic susceptibility $\chi_0$ [Fig.~\ref{fig:kdep}\,(b)], only the two regions associated with the intervalley coupling contribute significantly to the phonon self-energy $2 \omega \Pi$ [Fig.~\ref{fig:kdep}\,(a)]. The coupling matrix elements $2 \omega g^2$ [Fig.~\ref{fig:kdep}\,(c)] filter out these two of the four regions of (approximately) touching hole pockets.

One might be tempted to explain the contributions to $2 \omega \Pi$ from the two remaining regions with touching $\mathrm K$ and $\mathrm K'$~pockets in terms of nesting. However, our results rule out such a nesting scenario: As a first indication already seen in Fig.~\ref{fig:phonons}\,(c), $\vec q$~dependencies in $\chi_0$ are much less pronounced than those in $2 \omega \Pi$, which is opposite to the expectation of a logarithmically divergent $\chi_0$ in a nesting scenario. $2 \omega \Pi$ shows a pronounced extremum for $\vec q = \mathrm M$, while $\chi_0$ shows significantly smaller and opposite variations. The $\vec k$-resolved fluctuation diagnostics of $2 \omega \Pi$ at $\vec q = \mathrm M$ [Fig.~\ref{fig:kdep}\,(d)] shows that dominant contributions come again from $\mathrm K$ and $\mathrm K'$~pockets, which are now slightly overlapping rather than approximately touching and clearly not nested at $\vec q = \mathrm M$. Indeed, there is nesting for the hole pocket around $\Gamma$ which contributes to $\chi_0$ [Fig.~\ref{fig:kdep}\,(e)]. However, the resultant approximate divergence is logarithmic, thus already weak on the level of $\chi_0$, and fully masked by matrix-element effects in the phonon self-energy. Consequently, matrix-element effects clearly dominate here.

A basic and widely used model \cite{Varma1979, Flicker2015} of the interaction between electrons and LA phonons assumes that $g_{\vec q \vec k} \sim (\vec v_{\vec k} - \vec v_{\vec k + \vec q}) \cdot \vec q$, which suggests that coupling is most effective if the group velocities $\vec v_{\vec k} = \nabla_{\vec k} \epsilon_{\vec k}$ and $\vec v_{\vec k + \vec q}$ of the coupled electronic states are opposed to each other and parallel to $\vec q$. Interestingly, those regions with strong (suppressed) electron--phonon coupling identified by our analysis [Fig.~\ref{fig:kdep}\,(c,\,f,\,i)] are characterized by group velocities mainly orthogonal (parallel) to the phonon momentum $\vec q$, which is exactly opposite to the expectation from the model. The reason for this deviation is the massive-Dirac-fermion nature of the low-energy-band states and associated pseudospin textures, as will be explained in Section~\ref{sec:Dirac}.

\subsection{Doping dependence and Van Hove scenarios}
\label{sec:doped}

Charge doping is known to affect CDW instabilities by shifting the ordering wave vectors and suppressing or supporting CDW order in many materials from high-$T \sub c$ superconductors \cite{\dopingcdwhightc} to TMDCs \cite{\dopingcdwtmdc} and \TaS2 in particular \cite{\dopingcdwtas2}. We studied the dependence of the LA phonon mode on charge doping in the phonon-self-energy formalism [Eqs.~\eqref{eq:scheme}] by changing the electronic chemical potential $\mu$ in the model Hamiltonian $H$. In doing so, we disregard changes in the screened coupling $\tilde g$. This approximation is justified by the small relative difference between $g$ and $\tilde g$ in the undoped case. A comparison of phonon self-energies and resulting screened phonon dispersions for hole doping ($\mu = -119$\,meV), the charge-neutral system ($\mu = 0$), and electron doping ($\mu = 91$\,meV) is given in Fig.~\ref{fig:qdep}. In line with former theoretical results \cite{Albertini2017} and experiments \cite{Hall2019}, we find that electron doping pushes the wave vector of the leading lattice instability toward the $\mathrm M$~point. Hole doping of $\mu = -119$\,meV, on the other hand, shifts the instabilities further away from $\mathrm M$ and also lets additional ``fragile'' instabilities between $\Gamma$ and $\mathrm K$ emerge, which depend very sensitively on the thermal broadening $k \sub B T$ of the electronic Fermi distribution function. At smallest temperatures $k \sub B T \approx 1$\,meV, the leading instabilities occurring in the phonon dispersions [Fig.~\ref{fig:qdep}\,(a)] coincide with the extrema of the corresponding phonon self-energies [Fig.~\ref{fig:qdep}\,(b)]. The latter are fully determined by the Fermiology conditions of touching $\mathrm K$ and $\mathrm K'$~pockets and (approximately) superimposed Van Hove points over the whole range of doping levels. However, these extrema are by no means isolated points of enhanced/divergent phonon self-energies but embedded in extended $\vec q$-space regions with appreciable mode softening.

Fluctuation diagnostics [Fig.~\ref{fig:combined}] reveals the mechanisms behind the doping dependencies found in Fig.~\ref{fig:qdep}: While the filter determined by the electron--phonon coupling remains the same regardless of the doping level [blue-shaded regions in Fig.~\ref{fig:combined}\,(a)], electron doping shrinks the hole pockets around $\mathrm K$, $\mathrm K'$, and $\Gamma$. Correspondingly, touching or partially overlapping $\mathrm K$ and $\mathrm K'$~pockets are realized for CDW wave vectors larger than $\vec q = 2/3 \, \mathrm M$. For $\mu = 91$\,meV, $\vec q = \mathrm M$ leads to touching $\mathrm K$ and $\mathrm K'$~pockets. Therefore, contributions to $2 \omega \Pi$ are correspondingly enhanced at $\mathrm M$, while the mode softening at $\vec q = 2/3 \, \mathrm M$ is weaker in this electron-doped case. Hence, nearly overlapping hole pockets and corresponding contributions to $\chi_0$ are necessary for effective mode softening. If the instability at $\mathrm M$ in this case was a pure nesting effect, one should find a logarithmic divergence in the energy-resolved fluctuation diagnostics in $\Pi(\Delta)$ and a corresponding $1 / \Delta$ divergence in $\partial_\Delta \Pi(\Delta)$. As one can see from Fig.~\ref{fig:combined}\,(b), we do not find such divergences for the electron-doped case at $\vec q = \mathrm M$ nor for the undoped case at $\vec q = 2/3 \, \mathrm M$ [cf. Appendix~\ref{app:divergence}]. Thus, sufficiently large matrix elements of the electron--phonon coupling and sufficiently large albeit finite bare electronic susceptibilities are of central importance in both cases. This finding is in line with Refs.~\onlinecite{\agreement}.

For an electronic chemical potential of $\mu = -119$\,meV one realizes ``Van Hove doping\rlap,'' i.e., hole doping such that the electronic VHS at $\vec k = 0.58 \, \mathrm K$ and symmetry-equivalent $\vec k$~points are directly at the Fermi level. In this situation, it is possible to realize phonon-mode softening as put forward in the so-called Van Hove scenario \cite{Rice1975}, where low-energy electronic fluctuations from the vicinity of VHS yield diverging contributions to the bare electronic susceptibility and possibly to the phonon self-energy. However, the situation in \TaS2 at Van Hove doping is intricate: There are dynamical lattice instabilities in large parts of the BZ, particularly between $\Gamma$ and $\mathrm K$. For most unstable parts of the $\Gamma$--$\mathrm K$ section, one has imperfect nesting. The notable exception is $\vec q_\text{VHS} = 0.58 \, \mathrm K$, which realizes a Van Hove scenario on top of imperfect nesting. For this $\vec q$~vector there are sizable contributions to $\chi_0$ from the vicinity of the VHS [Fig.~\ref{fig:kdep}\,(h)]. Also the electron--phonon matrix elements are nonzero in the vicinity of the matched VHS [Fig.~\ref{fig:kdep}\,(i)]. Thus, at sufficiently small energies, the VHS-induced logarithmic divergence in $\chi_0$ manifests also in $2 \omega \Pi$ [Fig.~\ref{fig:kdep}\,(g)]. That becomes more obvious in the energy-resolved fluctuation diagnostics [Fig.~\ref{fig:combined}\,(b)]: We find divergences $\partial_\Delta \Pi \sim 1 / \Delta$ and $\Pi \sim \log \Delta$ as expected in the Van Hove scenario [cf.\@ Appendix~\ref{app:divergence}]. However, in absolute numbers, very small energy scales have to be reached for the Van Hove contribution to dominate over more conventional effects (e.g., imperfect nesting and matrix-element effects): That can be seen from the dependence of the screened phonon dispersions on electronic broadening in the VHS-doped case [Fig.~\ref{fig:qdep}\,(a)]. $\vec q_\text{VHS}$ clearly defines the leading instability only for electronic temperatures below $k \sub B T \approx 1$\,meV, while instabilities in large parts of the BZ exist already at $k \sub B T = 25$\,meV.

Taken together, our fluctuation diagnostics confirms that Fermiology alone is insufficient to understand the phonon renormalization. The matrix-element filter provided by the electron--phonon coupling in its interplay with the Fermiology determines the phonon self-energies and mode softening in \TaS2.

\section{Tight-binding and Dirac model of electron--phonon coupling in \protect\TaS2}

To obtain a microscopic understanding of the matrix-element effects, we calculate the electron--phonon coupling for a nearest-neighbor tight-binding (TB) model, following two widely used approaches by Varma \emph{et al.} \cite{Varma1979, Flicker2015}, and we compare it to the \emph{ab initio} results. We find that the momentum dependence of the electron--phonon coupling results from the multiorbital nature of the active subspace and can be understood in terms of pseudospin textures of massive Dirac fermions.

\subsection{Tight binding}
\label{sec:TB}

We consider a nearest-neighbor TB model
\begin{equation}
    \label{eq:TB}
    H_{\vec k}^{\alpha \smash[b] \beta} = \epsilon_0^{\alpha \smash[b] \beta} + \sum_{n = 1}^6
    t_n^{\alpha \smash[b] \beta} \E^{\I \vec a_n \vec k}
\end{equation}
with the bond vectors $\vec a_n$ and orbital indices $\alpha$ and $\beta$. The on-site energy $\epsilon_0$ and the hopping $t_n$ are specified in Appendix~\ref{app:TB}. As seen in Fig.~\ref{fig:TB}\,(a), the resulting band structure fits the reference from density-functional theory (DFT) quite well, particularly the low-energy band.

In Appendix~\ref{app:TB} we briefly review how to derive an approximate expression for the corresponding electron--phonon coupling \cite{Varma1979}. Transforming Eq.~\eqref{eq:gTB_orbital} into the band eigenbasis of $H_{\vec k}$ and $H_{\vec k + \vec q}$ yields
\begin{equation}
    \label{eq:gTB}
    g_{\vec q \nu \vec k m n} \sim \sum_l
    \big(
        A_{\vec q \vec k}^{m l} \vec v_{\vec k}^{l n} -
        \vec v_{\vec k + \vec q}^{m l} A_{\vec q \vec k}^{l n}
    \big)
    \cdot \vec e_{\vec q \nu},
\end{equation}
where $\vec v_{\vec k}^{m n} = \sum_{\alpha \beta} (U_{\vec k}^{\alpha m})^* \, (\nabla_{\vec k}^{\vphantom \alpha} H_{\vec k}^{\alpha \smash[b] \beta}) \, U_{\vec k}^{\smash[b] \beta n}$ is the velocity operator in the band basis and $A_{\vec q \vec k}^{m n} = \sum_\alpha (U_{\vec k + \vec q}^{\alpha m})^* \, U_{\vec k}^{\alpha n}$ is a unitary matrix which describes the overlap of the lattice-periodic part of band state $n$ at $\vec k$ with band state $m$ at $\vec k + \vec q$. Here, $U_{\vec k}$ is the matrix of right eigenvectors of $H_{\vec k}$ which fulfills $\sum_{\alpha \beta} (U_{\vec k}^{\alpha m})^* \, H_{\vec k}^{\alpha \smash[b] \beta} \, U_{\vec k}^{\smash[b] \beta n} = \epsilon_{\vec k n} \delta_{m n}$.

Another widely used approximation [cf.\@ Eqs.~(A.11) and (A.12) of Ref.~\onlinecite{Varma1979} and Eq.~(2) of Ref.~\onlinecite{Flicker2015}] neglects the off-diagonal matrix elements of the velocity operator in the band basis and thus only accounts for the standard electronic group velocities $\vec v_{\vec k}^{n n} = \nabla_{\vec k} \epsilon_{\vec k n}$,
\begin{equation}
    \label{eq:gTB_simple}
    g_{\vec q \nu \vec k m n} \sim
    A_{\vec q \vec k}^{m n}
    \big(
        \vec v_{\vec k}^{n n} -
        \vec v_{\vec k + \vec q}^{m m}
    \big)
    \cdot \vec e_{\vec q \nu}.
\end{equation}

\begin{figure}
    \includegraphics[width=\linewidth]{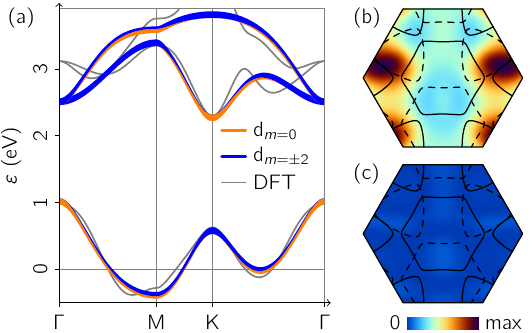}
    \caption{(a)~Electronic band structure from the TB model [Eq.~\eqref{eq:TB}] compared to DFT, (b)~the resulting $\vec k$-resolved coupling $2 \omega g^2$ of LA phonons with the low-energy band for $\vec q = 2/3 \, \mathrm M$ according to Eq.~\eqref{eq:gTB}, and (c)~the corresponding simplified coupling defined in Eq.~\eqref{eq:gTB_simple}.}
    \label{fig:TB}
\end{figure}

In Fig.~\ref{fig:TB}\,(b,\,c), we show the $\vec k$-dependent coupling of the LA phonons with the low-energy band for $\vec q = 2/3 \, \mathrm M$ within the approximations of Eqs.~\eqref{eq:gTB} and \eqref{eq:gTB_simple}. According to (c)DFPT [Fig.~\ref{fig:kdep}\,(c)], there are only two spots in the vicinity of $\mathrm K$ and $\mathrm K'$ where the coupling is large. Eq.~\eqref{eq:gTB} reproduces the structure of the coupling found in (c)DFPT qualitatively [Fig.~\ref{fig:TB}\,(b)]. However, the simplified Eq.~\eqref{eq:gTB_simple} does not capture the relevant physics, as the resulting coupling is much weaker and has its maximum in the hole pocket around $\Gamma$ [Fig.~\ref{fig:TB}\,(c)]. Therefore, intraband variation of orbital characters and the full matrix structure of the velocity operator must be decisive in determining the electron--LA-phonon coupling.

\subsection{Massive Dirac fermions}
\label{sec:Dirac}

\begin{figure*}
    \includegraphics[width=\linewidth]{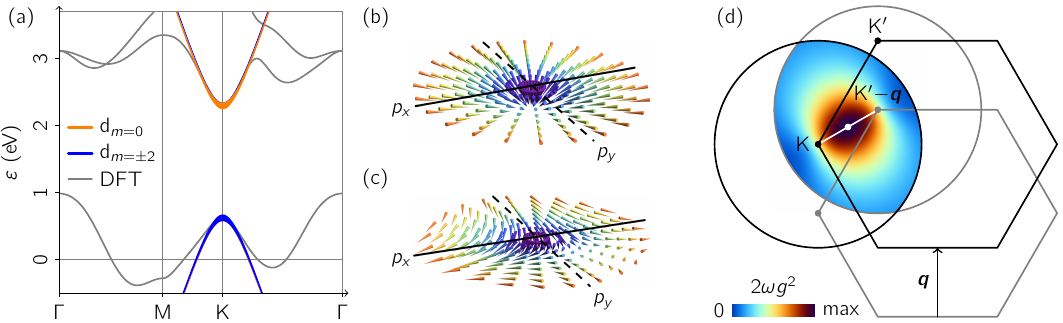}
    \caption{(a)~Electronic band structure from a model of massive Dirac fermions compared to DFT\@. (b,\,c)~Dirac pseudospin texture of the lower band around (b)~$\mathrm K$ and (c)~$\mathrm K'$. (d)~The resulting intervalley electron--phonon coupling strength $2 \omega g^2$ for $\vec q = 2/3 \, \mathrm M$ exhibits a clear maximum halfway between $\mathrm K$ and $\mathrm K' - \vec q$.}
    \label{fig:Dirac}
\end{figure*}

Both the TB model and the first-principles calculations [Section~\ref{sec:pristine}] suggest that the coupling between electrons and LA phonons is strong when the group velocities of the electronic states are (largely) orthogonal to the phonon momentum $\vec q$. This is exactly opposite to the expectation from Eq.~\eqref{eq:gTB_simple}. To understand the origin of this behavior, we resort to an even simpler model and describe the low-energy band of \TaS2 around $\mathrm K$ and $\mathrm K'$ in terms of massive Dirac fermions \cite{Xiao2012}:
\begin{equation}
    \label{eq:Dirac}
    H \sub D = v_0 (\tau p_x \sigma_x + p_y \sigma_y) + \frac \Delta 2 \sigma_z.
\end{equation}
Here, $\tau = \pm 1$ is the valley index, which selects between regions around $\mathrm K$ and $\mathrm K'$ [Fig.~\ref{fig:electrons}\,(c)], $v_0$ is an effective velocity playing the role of the speed of light in the relativistic Dirac equation, and $\Delta$ is the band gap. The Pauli matrices $\sigma_i$ with $i \in \{ x, y, z \}$ act on pseudospinors $\chi_{\tau \vec p}$ from a space of two Ta $d$ orbitals: The upper (lower) component of $\chi_{\tau \vec p}$ describes the $d$ orbitals with orbital angular momentum $m = 0$ ($m = 2 \tau$). The eigenvalues of $H \sub D$ at momentum $\vec p$ relative to $\mathrm K$ or $\mathrm K'$ are $\epsilon_{\vec p} = \pm \sqrt{\Delta^2 / 4 + v_0^2 p^2}$, as shown in Fig.~\ref{fig:Dirac}\,(a).

The velocity operator resulting from the Dirac Hamiltonian [Eq.~\eqref{eq:Dirac}] is
\begin{equation}
    \label{eq:velocity}
    \vec v \sub D
    = \nabla_{\vec p} H \sub D
    = v_0 (\tau \vec e_x \sigma_x + \vec e_y \sigma_y).
\end{equation}
This operator describes the change in both eigenvalues \emph{and} the change in eigenvectors of $H \sub D$. As in relativistic quantum theory, the Dirac velocity operator is independent of $\vec k = \mathrm K^{(\prime)} + \vec p$ inside either valley. Therefore, the intravalley electron--phonon coupling according to Eq.~\eqref{eq:gTB} vanishes. Thus, the massive-Dirac-fermion nature of the quasiparticles and the resultant pseudospin--momentum coupling causes the smallness of the matrix elements of the electron--phonon coupling associated with intravalley scattering, as seen in Figs.~\ref{fig:kdep}\,(c) and \ref{fig:TB}\,(b).

Regarding the intervalley coupling, we first note that the operator $\vec v \sub D$ has contributions perpendicular and parallel to the equal-energy contours of $H \sub D$~\footnote{It is instructive to rewrite the valley Hamiltonians and velocity operators in polar coordinates:
\begin{align*}
    H_{p \phi} &= v_0 p \, \vec e_p \cdot \vec \sigma_\tau + \frac \Delta 2 \sigma_z, \\
    \vec v_{p \phi} &= v_0 \left[
        (\vec e_p \cdot \vec \sigma_\tau) \vec e_p +
        (\vec e_\phi \cdot \vec \sigma_\tau) \vec e_\phi
    \right],
\end{align*}
where $\vec \sigma_\tau = (\tau \sigma_x, \sigma_y)$, $p$ ($\phi$) is the radial (angular) momentum coordinate, and $\vec e_p$ ($\vec e_\phi$) are the corresponding unit vectors in the radial (angular) direction. The equal-energy contours are circles with constant $p$. We see that $\vec v_{p \phi}$ has contributions perpendicular ($\vec e_p$ direction) and parallel ($\vec e_\phi$ direction) to the equal-energy contours.}, resulting from the pseudospin--momentum coupling intrinsic to the Dirac equation [Eq.~\eqref{eq:Dirac}]. Hence, the Dirac velocity operator $\vec v \sub D$ is very different from the na\"ive expectation for the group velocity $\vec v_{\vec k} = \nabla_{\vec k} \epsilon_{\vec k} = v_0^2 \, \vec k / \epsilon_{\vec k}$, which only describes the change in energy eigenvalues and always points perpendicular to the equal-energy contours.

The rotation of pseudospins in the lower band around $\mathrm K$ and $\mathrm K'$ indeed gives rise to a velocity component \emph{parallel} to the equal-energy contours. This allows, generically, for intervalley coupling between electrons and LA phonons at arbitrary angles between equal-energy contours and phonon momentum $\vec q$, in contrast to the simplified model of Eq.~\eqref{eq:gTB_simple}. For the specific analysis of intervalley scattering in TMDCs, one has to be careful since in the model of Eq.~\eqref{eq:Dirac} the orbital character associated with one of the pseudospin components changes from $m = +2$ to $-2$ when going from $\mathrm K$ to $\mathrm K'$.

Let $\chi_{\mathrm K}$ and $\chi_{\mathrm K'}$ be pseudospinors belonging to lower-band states at $\vec k = \mathrm K + \vec p$ and $\vec k + \vec q = \mathrm K' + \vec p'$ located in the $\mathrm K$ and $\mathrm K'$~valleys, respectively. That is, $\chi_{\mathrm K}$ ($\chi_{\mathrm K'}$) is the negative-eigenvalue eigenvector of $H \sub D$ for $\tau = +1$ ($-1$) and for momentum $\vec p$ ($\vec p'$). Then, the electron--phonon coupling according to Eq.~\eqref{eq:gTB} can be expressed using projection operators $P_0$ on the $m = 0$ orbitals:
\begin{equation}
    g_{\vec q \vec k} \sim
    \chi_{\mathrm K'}^\dagger
    ( \vec v_{\mathrm K}^{\phantom \dagger} P_0
    - P_0 \vec v_{\mathrm K'}^{\phantom \dagger})
    \chi_{\mathrm K}^{\phantom \dagger}
    \cdot \vec e_{\vec q \nu}.
\end{equation}
In the same matrix representation as used in Eq.~\eqref{eq:Dirac}, we have $P_0 = (1 + \sigma_z) / 2$, where $1$ is the $2 \times 2$ identity matrix. With Eq.~\eqref{eq:velocity} we then find
\begin{equation}
    \label{eq:g_Dirac}
    g_{\vec q \vec k} \sim
    \chi_{\mathrm K'}^\dagger \sigma_x \chi_{\mathrm K}^{\phantom \dagger}
    \, (\vec e_x + \I \vec e_y) \cdot \vec e_{\vec q \nu}.
\end{equation}
The direction of the phonon eigenvector $\vec e_{\vec q \nu}$ enters merely as a phase factor. For the absolute value of the LA coupling, the angle between the equal-energy contours and $\vec q$ is not decisive. Instead, the coupling strength is determined by $\smash{\chi_{\mathrm K'}^\dagger \sigma_x \chi_{\mathrm K}^{\phantom \dagger}}$ and thus results from the pseudospin textures of the lower-band states around $\mathrm K$ and $\mathrm K'$ shown in Fig.~\ref{fig:Dirac}\,(b,\,c) alone. Contributions to $g$ according to Eq.~\eqref{eq:g_Dirac} arise from opposite-sign pseudospin projections $\smash{\chi^\dagger \sigma_i \chi}$ of $\chi_{\mathrm K}$ and $\chi_{\mathrm K'}$ in the $y$ or $z$ direction and equal-sign pseudospin projection in the $x$ direction. The $z$ projection of all lower-band pseudospins is ``down\rlap,'' which does not lead to contributions to $g$. The winding of the in-plane pseudospin component is opposite in the $\mathrm K$ and $\mathrm K'$~valleys [Fig.~\ref{fig:Dirac}\,(b,\,c)]. The condition of maximum coupling $\abs g$ is fulfilled if $\chi_{\mathrm K}$ and $\chi_{\mathrm K'}$ belong to time-reversed crystal momenta, $\vec p = -\vec p'$, preferably with large absolute value. This is realized for $\vec k$ being halfway between $\mathrm K$ and the shifted $\mathrm K'$~point [Fig.~\ref{fig:Dirac}\,(d)], i.e., exactly in the region where we find enhanced electron--phonon coupling in (c)DFPT [Fig.~\ref{fig:kdep}\,(c)] and also in the TB model [Fig.~\ref{fig:TB}\,(b)]. Hence, the momentum-space selection rule for the electron--phonon coupling, which we identified using fluctuation diagnostics, originates from the massive-Dirac-fermion nature of the low-energy-band states in \TaS2.

\section{Conclusions}

We have presented an \emph{ab initio}--based scheme for the calculation of phonon self-energies from material-realistic low-energy models. This approach allows us to generalize the idea of fluctuation diagnostics \cite{Gunnarsson2015} to electron--phonon-coupled systems and to identify the contributions of the different electronic fluctuation channels to the renormalization of phonons in complex materials.

Application of this scheme to the model CDW compound of \TaS2 showed that polarization processes taking place in an isolated low-energy metallic band are entirely responsible for the mode softening and CDW physics in this material. While Fermi-surface effects originating from this band can affect the CDW ordering wave vector, we give direct proof that lattice instabilities in \TaS2 are largely controlled by electron--phonon-coupling matrix-element effects. The origin of these matrix-element effects is shown to be the massive-Dirac-fermion nature, the resultant pseudospin textures, and associated anomalous intervalley velocity matrix elements of electronic quasiparticles in the low-energy band of \TaS2. Thus, the fluctuation diagnostics provides a purely \emph{ab initio} way to settle the long-standing debate on the nature of CDW physics in group-V TMDCs and to reveal its physical origin.

The scheme for the calculation of phonon self-energies and for performing fluctuation diagnostics outlined here should be generally insightful to disentangle the interplay of electronic and lattice degrees of freedom in complex materials. Promising areas of future application range from materials like TiSe\s2, where excitonic physics intertwines with lattice instabilities \cite{\tise2}, to strongly correlated electron systems with coupled lattice, spin, and superconducting phenomena as take place in Fe-based superconductors \cite{\febased} or in proximity to stripe phases in cuprate high-temperature superconductors \cite{\cuprates}. Finally, generalizations beyond the adiabatic limit should be conveniently possible and allow, e.g., for the description of phonons in electronic flat-band systems such as twisted bilayer graphene.

\begin{acknowledgments}
Financial support by the Deutsche Forschungsgemeinschaft (DFG) through GRK 2247, the European Graphene Flagship, and computational resources of the North-German Supercomputing Alliance (HLRN) are gratefully acknowledged. Moreover, we owe special thanks to \mbox{Ryotaro Arita} and \mbox{Yusuke Nomura} for providing us with the cDFPT source code described in Ref.~\citenum{Nomura2015}.
\end{acknowledgments}

\appendix

\section{Divergence in the Van Hove scenario}
\label{app:divergence}

\begin{figure}
    \includegraphics[width=\linewidth]{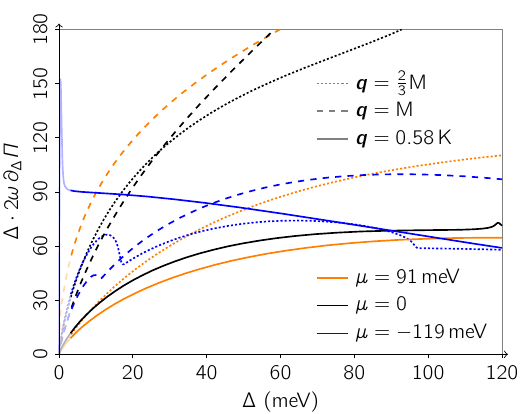}
    \caption{Product of energy-window size $\Delta$ and derivative of phonon self-energy $\partial_\Delta \Pi$ for all situations shown in Fig.~\ref{fig:combined}\,(b). Different line styles (dotted, dashed, solid) refer to different $\vec q$~points while different line colors (orange, black, blue) refer to different chemical potentials. For a diverging $\partial_\Delta \Pi \sim 1 / \Delta$, the displayed quantity is a nonzero constant in the limit of small $\Delta$. As expected, this is only the case in the Van Hove scenario (solid, blue line). The artefacts in the grayed-out region where $\Delta < 3$\,meV are due to smearing-related numerical inaccuracies.}
    \label{fig:divergence}
\end{figure}

In Section~\ref{sec:doped} we state that we find the expected divergences $\Pi \sim \log \Delta$ and $\partial_\Delta \Pi \sim 1 / \Delta$ of the phonon self-energy and its derivative for small energy-window sizes $\Delta$. While from Fig.~\ref{fig:combined}\,(b) the exact type of divergence does not become evident, the representation of the data shown Fig.~\ref{fig:divergence} is more suitable for this purpose.

\section{Details of the TB model}
\label{app:TB}

In the tight-binding model discussed in Section~\ref{sec:TB}, we only consider transitions between $d$ orbitals localized at neighboring Ta atoms. The hopping to the $n$th neighbor at the relative position $\vec a_n = r_n (\cos \phi_n, \sin \phi_n)$ is
\begin{equation}
    \label{eq:hopping}
    t_n = \Big[ \frac {r_n} a \Big]^\lambda
    R_{\phi_n} \cdot M^n \cdot t_0 \cdot M^n \cdot R_{-\phi_n},
\end{equation}
where $r_n = a$ and $\phi_n = n \pi / 3$ are bond length and angle, $\lambda = -5$ quantifies the distance dependence of $d$--$d$ bonds \cite{Harrison2004, Rossnagel2011}, $t_0$ is the hopping in the $x$ direction, $M$ is the reflection $y \mapsto -y$, and $R$ is a rotation about the $z$ axis.

In the basis of complex harmonics $d_{m = 0} = d_{z^2}$ and $d_{m = \pm 2} = (d_{x^2 - y^2} \pm \I d_{x y}) / \sqrt 2$, we have $M = \mathrm{diag}(1, \sigma_x)$ with the Pauli matrix $\sigma_x$ and $R_\phi = \mathrm{diag}(1, \E^{2 \I \phi}, \E^{-2 \I \phi})$. The on-site energy and zeroth hopping read
\begin{align}
    \epsilon_0 &= \mathrm{diag}(\xi = 1.85, \eta = 2.30, \eta)\,\text{eV},
    \\
    t_0 &=
    \begin{bmatrix*}[l]
        \alpha = -0.14 & \gamma = 0.34 - 0.27\,\I & \gamma^* \\
        \gamma         & \beta  = 0.03 + 0.31\,\I & \delta = -0.29 \\
        \gamma^*       & \delta                   & \beta^*
    \end{bmatrix*}\text{eV}. \notag
\end{align}
The number of independent parameters has been reduced using the point symmetries of the crystal. Their given values have been obtained by fitting to data from DFT.

To analyze the electron--phonon coupling, we rescale the hopping matrix elements in the distorted structures according to Eq.~\eqref{eq:hopping}. With the neighbor index $n$ of all quantities understood,
\begin{align}
    \label{eq:hopping_derivative}
    \frac{\partial t}{\partial \vec a} &=
    \frac{\partial t}{\partial r}
    \frac{\partial r}{\partial \vec a} +
    \frac{\partial t}{\partial \phi}
    \frac{\partial \phi}{\partial \vec a}
    \\
    &= \frac {\lambda t} r \frac {\vec a} r
    + \left[
        \frac{\partial R_{\phi}}{\partial \phi} \cdot R_{-\phi} \cdot t +
        t \cdot R_{\phi} \cdot \frac{\partial R_{-\phi}}{\partial \phi}
    \right]
    \frac{R_{\frac \pi 2} \vec a}{r^2}.
    \notag
\end{align}
Starting from the Hamiltonian in Eq.~\eqref{eq:TB}, it is straightforward to derive the formula for the electron--phonon coupling in the orbital basis [cf.\@ Eq.~(2.29) of Ref.~\onlinecite{Varma1979}],
\begin{equation}
    g_{\vec q \nu \vec k}
    = \frac 1 {\sqrt{2 \omega_{\vec q \nu} M}} \sum_n
    \frac
      {\partial t_n}
      {\partial \vec a_n}
    \big(
        \E^{\I \vec a_n \vec k} -
        \E^{\I \vec a_n (\vec k + \vec q)}
    \big)
    \cdot \vec e_{\vec q \nu},
\end{equation}
where $M$ is the atomic mass. Neglecting the $\phi$~derivatives in Eq.~\eqref{eq:hopping_derivative} and noting that both $\nabla_{\vec k} \E^{\I \vec a \vec k} = \I \vec a \E^{\I \vec a \vec k}$ and $\nabla_{\vec a} t = \lambda t \vec a / a^2$ are parallel to $\vec a$, we can rewrite the coupling as [cf.\@ Eq.~(A.8) of Ref.~\onlinecite{Varma1979}]
\begin{equation}
   \label{eq:gTB_orbital}
   g_{\vec q \nu \vec k} = \frac{-\I \lambda}{a^2 \sqrt{2 \omega_{\vec q \nu} M}}
    \big( \vec v_{\vec k} - \vec v_{\vec k + \vec q} \big)
    \cdot \vec e_{\vec q \nu},
\end{equation}
where $\vec v_{\vec k} = \nabla_{\vec k} H_{\vec k}$ is the velocity operator in the orbital basis.

\section{Computational parameters}
\label{app:parameters}

All DFT and DFPT calculations are carried out using \textsc{Quantum EPSRESSO} \cite{Giannozzi2009, Giannozzi2017}. The modification that is required for cDFPT is described in detail in Ref.~\onlinecite{Nomura2015}. For the transformation of the electron energies and electron--phonon couplings to the Wannier basis, we use \textsc{Wannier90} \cite{Mostofi2014} and the \textsc{EPW} code \cite{Giustino2007, Ponce2016}.

We apply the generalized gradient approximation (GGA) by Perdew, Burke, and Ernzerhof (PBE) \cite{Perdew1996, Perdew1997} and use norm-conserving Hartwigsen--Goedecker--Hutter (HGH) \cite{Goedecker1996, Hartwigsen1998} pseudopotentials at a plane-wave cutoff of 100\,Ry. Uniform meshes (including $\Gamma$) of $12 \times 12$ $\vec q$ and $36 \times 36$ $\vec k$~points are combined with a Fermi--Dirac smearing of 5\,mRy. For a fixed cell height of 15\,\AA, minimizing forces and in-plane pressure to below 1\,$\upmu$Ry/Bohr and 0.1\,kbar yields a lattice constant of 3.39\,\AA.

On the model level, for the DFPT comparison shown in Fig.~\ref{fig:phonons}\,(b,\,c), we use the same meshes and smearings as stated above. For the $\vec q$-dependent results shown in Fig.~\ref{fig:qdep}, we use $360 \times 360$ $\vec q$ and $\vec k$~points together with a Fermi--Dirac smearing of 1\,meV, if not stated otherwise. For the fluctuation diagnostics for selected $\vec q$~points shown in Figs.~\ref{fig:kdep}, \ref{fig:combined}, \ref{fig:divergence}, and \ref{fig:SOC}, we use $5040 \times 5040$ $\vec k$~points together with a Fermi--Dirac smearing of 0.07\,meV and a Gaussian smearing of 0.7\,meV for the $\Theta$ and $\delta$ functions in Eqs.~\eqref{eq:Delta}. The Fermi level is not recalculated for each mesh and smearing but kept fixed at the \emph{ab initio} value throughout this work.

\section{Effect of spin--orbit coupling}
\label{app:SOC}

\begin{figure}
    \includegraphics[width=\linewidth]{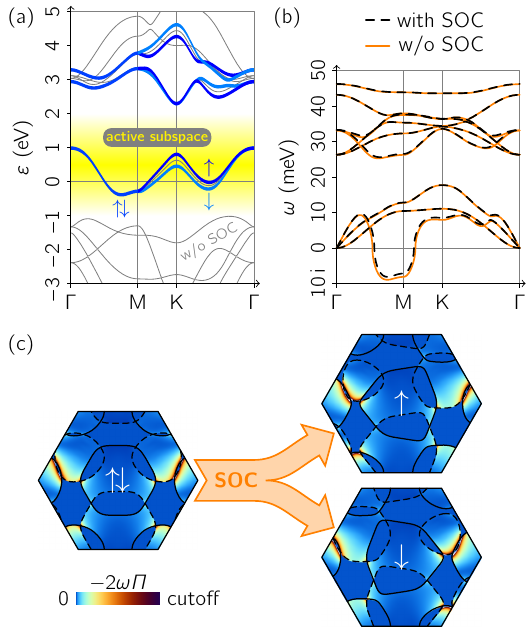}
    \caption{Effect of SOC on band structures and fluctuations. (a)~Spin-resolved electronic band structure. (b)~Screened phonon dispersion with and without SOC\@. (c)~Spin- and momentum-resolved LA phonon self-energy for $\vec q = 2/3 \, \mathrm M$ in the undoped case with (right) and without SOC (left). Solid (dashed) lines indicate the Fermi surface (shifted by $\vec q$) for the respective spin direction(s). To match color scales, the data on the right are multiplied by a factor of 2.}
    \label{fig:SOC}
\end{figure}

We have so far disregarded the spin splitting of the electronic bands due to spin--orbit coupling (SOC). In the following, we will briefly outline how SOC affects our findings.

We include SOC by modifying the electronic Hamiltonian, where each spin direction is treated separately \cite{Liu2013}. In the basis of real harmonics $d_{z^2}$, $d_{x^2 - y^2}$, and $d_{x y}$,
\begin{equation}
    H^{\uparrow, \downarrow} \sub{el} \rightarrow H^{\uparrow, \downarrow} \sub{el} \pm
    \begin{bmatrix*}[r]
        0 & 0 & 0 \\
        0 & 0 & -\I \\
        0 & \phantom+\I & 0
    \end{bmatrix*}
    \lambda.
\end{equation}
We choose a SOC strength $\lambda = 174$\,meV to reproduce the splitting of the experimental and theoretical bands in Ref.~\onlinecite{Sanders2016}.

Fig.~\ref{fig:SOC}\,(a) shows the resulting modified electron dispersion [cf.\@ Fig.~\ref{fig:electrons}\,(b)]. Along $\Gamma$--$\mathrm M$, the $d$ bands remain spin-degenerate. Beyond this line, the bands split with a maximum splitting of $2 \lambda$ of the low-energy band at $\mathrm K$.

The screened phonon dispersions with and without SOC according to Eqs.~\eqref{eq:scheme} are compared in Fig.~\ref{fig:SOC}\,(b) [cf.\@ Fig.~\ref{fig:phonons}\,(b)]. We note that there is no prefactor of~2 in Eq.~\eqref{eq:Pi} when the spin is explicitly included in the band summations. As expected, SOC does not cause qualitative changes in the phonon dispersion.

This can be explained by the $\vec k$-space fluctuation diagnostics. In Fig.~\ref{fig:SOC}\,(c), the LA phonon self-energy for $\vec q = 2/3 \, \mathrm M$ in the undoped case is displayed together with the Fermi surface, both with and without SOC [cf.\@ Fig.~\ref{fig:kdep}\,(a)]. The argument of touching $\mathrm K$ and $\mathrm K'$~pockets is hardly affected by SOC: For each spin direction, one pocket shrinks while the other expands, leaving their overlap region and thus the resulting phonon self-energy essentially unchanged.

\nocite{*}

\bibliography{ms}

\begin{thebibliography}{110}%
\makeatletter
\providecommand \@ifxundefined [1]{%
 \@ifx{#1\undefined}
}%
\providecommand \@ifnum [1]{%
 \ifnum #1\expandafter \@firstoftwo
 \else \expandafter \@secondoftwo
 \fi
}%
\providecommand \@ifx [1]{%
 \ifx #1\expandafter \@firstoftwo
 \else \expandafter \@secondoftwo
 \fi
}%
\providecommand \natexlab [1]{#1}%
\providecommand \enquote  [1]{``#1''}%
\providecommand \bibnamefont  [1]{#1}%
\providecommand \bibfnamefont [1]{#1}%
\providecommand \citenamefont [1]{#1}%
\providecommand \href@noop [0]{\@secondoftwo}%
\providecommand \href [0]{\begingroup \@sanitize@url \@href}%
\providecommand \@href[1]{\@@startlink{#1}\@@href}%
\providecommand \@@href[1]{\endgroup#1\@@endlink}%
\providecommand \@sanitize@url [0]{\catcode `\\12\catcode `\$12\catcode
  `\&12\catcode `\#12\catcode `\^12\catcode `\_12\catcode `\%12\relax}%
\providecommand \@@startlink[1]{}%
\providecommand \@@endlink[0]{}%
\providecommand \url  [0]{\begingroup\@sanitize@url \@url }%
\providecommand \@url [1]{\endgroup\@href {#1}{\urlprefix }}%
\providecommand \urlprefix  [0]{URL }%
\providecommand \Eprint [0]{\href }%
\providecommand \doibase [0]{https://doi.org/}%
\providecommand \selectlanguage [0]{\@gobble}%
\providecommand \bibinfo  [0]{\@secondoftwo}%
\providecommand \bibfield  [0]{\@secondoftwo}%
\providecommand \translation [1]{[#1]}%
\providecommand \BibitemOpen [0]{}%
\providecommand \bibitemStop [0]{}%
\providecommand \bibitemNoStop [0]{.\EOS\space}%
\providecommand \EOS [0]{\spacefactor3000\relax}%
\providecommand \BibitemShut  [1]{\csname bibitem#1\endcsname}%
\let\auto@bib@innerbib\@empty
%</preamble>
\bibitem [{\citenamefont {Bednorz}\ and\ \citenamefont
  {M\"uller}(1986)}]{Bednorz1986}%
  \BibitemOpen
  \bibfield  {author} {\bibinfo {author} {\bibfnamefont {J.~G.}\ \bibnamefont
  {Bednorz}}\ and\ \bibinfo {author} {\bibfnamefont {K.~A.}\ \bibnamefont
  {M\"uller}},\ }\bibfield  {title} {\bibinfo {title} {\emph {Possible
  high-{$T_c$} superconductivity in the {Ba}--{La}--{Cu}--{O} system}},\ }\href
  {https://doi.org/10.1007/BF01303701} {\bibfield  {journal} {\bibinfo
  {journal} {Z. Phys. B}\ }\textbf {\bibinfo {volume} {64}},\ \bibinfo {pages}
  {189} (\bibinfo {year} {1986})}\BibitemShut {NoStop}%
\bibitem [{\citenamefont {Shen}\ \emph {et~al.}(2002)\citenamefont {Shen},
  \citenamefont {Lanzara}, \citenamefont {Ishihara},\ and\ \citenamefont
  {Nagaosa}}]{Shen2002}%
  \BibitemOpen
  \bibfield  {author} {\bibinfo {author} {\bibfnamefont {Z.-X.}\ \bibnamefont
  {Shen}}, \bibinfo {author} {\bibfnamefont {A.}~\bibnamefont {Lanzara}},
  \bibinfo {author} {\bibfnamefont {S.}~\bibnamefont {Ishihara}},\ and\
  \bibinfo {author} {\bibfnamefont {N.}~\bibnamefont {Nagaosa}},\ }\bibfield
  {title} {\bibinfo {title} {\emph {Role of the electron-phonon interaction in
  the strongly correlated cuprate superconductors}},\ }\href
  {https://doi.org/10.1080/13642810208220725} {\bibfield  {journal} {\bibinfo
  {journal} {Philos. Mag. B}\ }\textbf {\bibinfo {volume} {82}},\ \bibinfo
  {pages} {1349} (\bibinfo {year} {2002})}\BibitemShut {NoStop}%
\bibitem [{\citenamefont {Gunnarsson}\ and\ \citenamefont
  {R\"osch}(2008)}]{Gunnarsson2008}%
  \BibitemOpen
  \bibfield  {author} {\bibinfo {author} {\bibfnamefont {O.}~\bibnamefont
  {Gunnarsson}}\ and\ \bibinfo {author} {\bibfnamefont {O.}~\bibnamefont
  {R\"osch}},\ }\bibfield  {title} {\bibinfo {title} {\emph {Interplay between
  electron--phonon and {Coulomb} interactions in cuprates}},\ }\href
  {https://doi.org/10.1088/0953-8984/20/04/043201} {\bibfield  {journal}
  {\bibinfo  {journal} {J. Phys. Condens. Matter}\ }\textbf {\bibinfo {volume}
  {20}},\ \bibinfo {pages} {043201} (\bibinfo {year} {2008})},\ \Eprint
  {https://arxiv.org/abs/0708.1407} {arXiv:0708.1407}\BibitemShut {NoStop}%
\bibitem [{\citenamefont {Fradkin}\ \emph {et~al.}(2015)\citenamefont
  {Fradkin}, \citenamefont {Kivelson},\ and\ \citenamefont
  {Tranquada}}]{Fradkin2015}%
  \BibitemOpen
  \bibfield  {author} {\bibinfo {author} {\bibfnamefont {E.}~\bibnamefont
  {Fradkin}}, \bibinfo {author} {\bibfnamefont {S.~A.}\ \bibnamefont
  {Kivelson}},\ and\ \bibinfo {author} {\bibfnamefont {J.~M.}\ \bibnamefont
  {Tranquada}},\ }\bibfield  {title} {\bibinfo {title} {\emph {Colloquium:
  Theory of intertwined orders in high temperature superconductors}},\ }\href
  {https://doi.org/10.1103/RevModPhys.87.457} {\bibfield  {journal} {\bibinfo
  {journal} {Rev. Mod. Phys.}\ }\textbf {\bibinfo {volume} {87}},\ \bibinfo
  {pages} {457} (\bibinfo {year} {2015})},\ \Eprint
  {https://arxiv.org/abs/1407.4480} {arXiv:1407.4480}\BibitemShut {NoStop}%
\bibitem [{\citenamefont {Cavalleri}(2018)}]{Cavalleri2018}%
  \BibitemOpen
  \bibfield  {author} {\bibinfo {author} {\bibfnamefont {A.}~\bibnamefont
  {Cavalleri}},\ }\bibfield  {title} {\bibinfo {title} {\emph {Photo-induced
  superconductivity}},\ }\href {https://doi.org/10.1080/00107514.2017.1406623}
  {\bibfield  {journal} {\bibinfo  {journal} {Contemp. Phys.}\ }\textbf
  {\bibinfo {volume} {59}},\ \bibinfo {pages} {31} (\bibinfo {year}
  {2018})}\BibitemShut {NoStop}%
\bibitem [{\citenamefont {Kamihara}\ \emph {et~al.}(2008)\citenamefont
  {Kamihara}, \citenamefont {Watanabe}, \citenamefont {Hirano},\ and\
  \citenamefont {Hosono}}]{Kamihara2008}%
  \BibitemOpen
  \bibfield  {author} {\bibinfo {author} {\bibfnamefont {Y.}~\bibnamefont
  {Kamihara}}, \bibinfo {author} {\bibfnamefont {T.}~\bibnamefont {Watanabe}},
  \bibinfo {author} {\bibfnamefont {M.}~\bibnamefont {Hirano}},\ and\ \bibinfo
  {author} {\bibfnamefont {H.}~\bibnamefont {Hosono}},\ }\bibfield  {title}
  {\bibinfo {title} {\emph {Iron-based layered superconductor
  {La}[{O\s{1\ensuremath-x}F\s{x}}]{FeAs} (x~=~0.05--0.12) with
  {$T_c$}~=~26\,{K}}},\ }\href {https://doi.org/10.1021/ja800073m} {\bibfield
  {journal} {\bibinfo  {journal} {J. Am. Chem. Soc.}\ }\textbf {\bibinfo
  {volume} {130}},\ \bibinfo {pages} {3296} (\bibinfo {year}
  {2008})}\BibitemShut {NoStop}%
\bibitem [{\citenamefont {Stewart}(2011)}]{Stewart2011}%
  \BibitemOpen
  \bibfield  {author} {\bibinfo {author} {\bibfnamefont {G.~R.}\ \bibnamefont
  {Stewart}},\ }\bibfield  {title} {\bibinfo {title} {\emph {Superconductivity
  in iron compounds}},\ }\href {https://doi.org/10.1103/RevModPhys.83.1589}
  {\bibfield  {journal} {\bibinfo  {journal} {Rev. Mod. Phys.}\ }\textbf
  {\bibinfo {volume} {83}},\ \bibinfo {pages} {1589} (\bibinfo {year}
  {2011})},\ \Eprint {https://arxiv.org/abs/1106.1618}
  {arXiv:1106.1618}\BibitemShut {NoStop}%
\bibitem [{\citenamefont {Huang}\ and\ \citenamefont
  {Hoffman}(2017)}]{Huang2017}%
  \BibitemOpen
  \bibfield  {author} {\bibinfo {author} {\bibfnamefont {D.}~\bibnamefont
  {Huang}}\ and\ \bibinfo {author} {\bibfnamefont {J.~E.}\ \bibnamefont
  {Hoffman}},\ }\bibfield  {title} {\bibinfo {title} {\emph {Monolayer {FeSe}
  on {SrTiO\s3}}},\ }\href
  {https://doi.org/10.1146/annurev-conmatphys-031016-025242} {\bibfield
  {journal} {\bibinfo  {journal} {Annu. Rev. Condens. Matter Phys.}\ }\textbf
  {\bibinfo {volume} {8}},\ \bibinfo {pages} {311} (\bibinfo {year} {2017})},\
  \Eprint {https://arxiv.org/abs/1703.09306} {arXiv:1703.09306}\BibitemShut
  {NoStop}%
\bibitem [{\citenamefont {Ashcroft}(1968)}]{Ashcroft1968}%
  \BibitemOpen
  \bibfield  {author} {\bibinfo {author} {\bibfnamefont {N.~W.}\ \bibnamefont
  {Ashcroft}},\ }\bibfield  {title} {\bibinfo {title} {\emph {Metallic
  hydrogen: A high-temperature superconductor?}},\ }\href
  {https://doi.org/10.1103/PhysRevLett.21.1748} {\bibfield  {journal} {\bibinfo
   {journal} {Phys. Rev. Lett.}\ }\textbf {\bibinfo {volume} {21}},\ \bibinfo
  {pages} {1748} (\bibinfo {year} {1968})}\BibitemShut {NoStop}%
\bibitem [{\citenamefont {Gor'kov}\ and\ \citenamefont
  {Kresin}(2018)}]{Gorkov2018}%
  \BibitemOpen
  \bibfield  {author} {\bibinfo {author} {\bibfnamefont {L.~P.}\ \bibnamefont
  {Gor'kov}}\ and\ \bibinfo {author} {\bibfnamefont {V.~Z.}\ \bibnamefont
  {Kresin}},\ }\bibfield  {title} {\bibinfo {title} {\emph {Colloquium: High
  pressure and road to room temperature superconductivity}},\ }\href
  {https://doi.org/10.1103/RevModPhys.90.011001} {\bibfield  {journal}
  {\bibinfo  {journal} {Rev. Mod. Phys.}\ }\textbf {\bibinfo {volume} {90}},\
  \bibinfo {pages} {011001} (\bibinfo {year} {2018})},\ \Eprint
  {https://arxiv.org/abs/1802.02296} {arXiv:1802.02296}\BibitemShut {NoStop}%
\bibitem [{\citenamefont {Drozdov}\ \emph {et~al.}(2019)\citenamefont
  {Drozdov}, \citenamefont {Kong}, \citenamefont {Minkov}, \citenamefont
  {Besedin}, \citenamefont {Kuzovnikov}, \citenamefont {Mozaffari},
  \citenamefont {Balicas}, \citenamefont {Balakirev}, \citenamefont {Graf},
  \citenamefont {Prakapenka}, \citenamefont {Greenberg}, \citenamefont
  {Knyazev}, \citenamefont {Tkacz},\ and\ \citenamefont
  {Eremets}}]{Drozdov2019}%
  \BibitemOpen
  \bibfield  {author} {\bibinfo {author} {\bibfnamefont {A.~P.}\ \bibnamefont
  {Drozdov}}, \bibinfo {author} {\bibfnamefont {P.~P.}\ \bibnamefont {Kong}},
  \bibinfo {author} {\bibfnamefont {V.~S.}\ \bibnamefont {Minkov}}, \bibinfo
  {author} {\bibfnamefont {S.~P.}\ \bibnamefont {Besedin}}, \bibinfo {author}
  {\bibfnamefont {M.~A.}\ \bibnamefont {Kuzovnikov}}, \bibinfo {author}
  {\bibfnamefont {S.}~\bibnamefont {Mozaffari}}, \bibinfo {author}
  {\bibfnamefont {L.}~\bibnamefont {Balicas}}, \bibinfo {author} {\bibfnamefont
  {F.~F.}\ \bibnamefont {Balakirev}}, \bibinfo {author} {\bibfnamefont {D.~E.}\
  \bibnamefont {Graf}}, \bibinfo {author} {\bibfnamefont {V.~B.}\ \bibnamefont
  {Prakapenka}}, \bibinfo {author} {\bibfnamefont {E.}~\bibnamefont
  {Greenberg}}, \bibinfo {author} {\bibfnamefont {D.~A.}\ \bibnamefont
  {Knyazev}}, \bibinfo {author} {\bibfnamefont {M.}~\bibnamefont {Tkacz}},\
  and\ \bibinfo {author} {\bibfnamefont {M.~I.}\ \bibnamefont {Eremets}},\
  }\bibfield  {title} {\bibinfo {title} {\emph {Superconductivity at 250\,{K}
  in lanthanum hydride under high pressures}},\ }\href
  {https://doi.org/10.1038/s41586-019-1201-8} {\bibfield  {journal} {\bibinfo
  {journal} {Nature}\ }\textbf {\bibinfo {volume} {569}},\ \bibinfo {pages}
  {528} (\bibinfo {year} {2019})},\ \Eprint {https://arxiv.org/abs/1812.01561}
  {arXiv:1812.01561}\BibitemShut {NoStop}%
\bibitem [{\citenamefont {Varma}\ and\ \citenamefont
  {Simons}(1983)}]{Varma1983}%
  \BibitemOpen
  \bibfield  {author} {\bibinfo {author} {\bibfnamefont {C.~M.}\ \bibnamefont
  {Varma}}\ and\ \bibinfo {author} {\bibfnamefont {A.~L.}\ \bibnamefont
  {Simons}},\ }\bibfield  {title} {\bibinfo {title} {\emph {Strong-coupling
  theory of charge-density-wave transitions}},\ }\href
  {https://doi.org/10.1103/PhysRevLett.51.138} {\bibfield  {journal} {\bibinfo
  {journal} {Phys. Rev. Lett.}\ }\textbf {\bibinfo {volume} {51}},\ \bibinfo
  {pages} {138} (\bibinfo {year} {1983})}\BibitemShut {NoStop}%
\bibitem [{\citenamefont {Wilson}\ \emph {et~al.}(1975)\citenamefont {Wilson},
  \citenamefont {Di~Salvo},\ and\ \citenamefont {Mahajan}}]{Wilson1975}%
  \BibitemOpen
  \bibfield  {author} {\bibinfo {author} {\bibfnamefont {J.}~\bibnamefont
  {Wilson}}, \bibinfo {author} {\bibfnamefont {F.}~\bibnamefont {Di~Salvo}},\
  and\ \bibinfo {author} {\bibfnamefont {S.}~\bibnamefont {Mahajan}},\
  }\bibfield  {title} {\bibinfo {title} {\emph {Charge-density waves and
  superlattices in the metallic layered transition metal dichalcogenides}},\
  }\href {https://doi.org/10.1080/00018737500101391} {\bibfield  {journal}
  {\bibinfo  {journal} {Adv. Phys.}\ }\textbf {\bibinfo {volume} {24}},\
  \bibinfo {pages} {117} (\bibinfo {year} {1975})}\BibitemShut {NoStop}%
\bibitem [{\citenamefont {Rice}\ and\ \citenamefont {Scott}(1975)}]{Rice1975}%
  \BibitemOpen
  \bibfield  {author} {\bibinfo {author} {\bibfnamefont {T.~M.}\ \bibnamefont
  {Rice}}\ and\ \bibinfo {author} {\bibfnamefont {G.~K.}\ \bibnamefont
  {Scott}},\ }\bibfield  {title} {\bibinfo {title} {\emph {New mechanism for a
  charge-density-wave instability}},\ }\href
  {https://doi.org/10.1103/PhysRevLett.35.120} {\bibfield  {journal} {\bibinfo
  {journal} {Phys. Rev. Lett.}\ }\textbf {\bibinfo {volume} {35}},\ \bibinfo
  {pages} {120} (\bibinfo {year} {1975})}\BibitemShut {NoStop}%
\bibitem [{\citenamefont {Withers}\ and\ \citenamefont
  {Wilson}(1986)}]{Withers1986}%
  \BibitemOpen
  \bibfield  {author} {\bibinfo {author} {\bibfnamefont {R.~L.}\ \bibnamefont
  {Withers}}\ and\ \bibinfo {author} {\bibfnamefont {J.~A.}\ \bibnamefont
  {Wilson}},\ }\bibfield  {title} {\bibinfo {title} {\emph {An examination of
  the formation and characteristics of charge-density waves in inorganic
  materials with special reference to the two- and one-dimensional
  transition-metal chalcogenides}},\ }\href
  {https://doi.org/10.1088/0022-3719/19/25/005} {\bibfield  {journal} {\bibinfo
   {journal} {J. Phys. C: Solid State Phys.}\ }\textbf {\bibinfo {volume}
  {19}},\ \bibinfo {pages} {4809} (\bibinfo {year} {1986})}\BibitemShut
  {NoStop}%
\bibitem [{\citenamefont {Rossnagel}(2011)}]{Rossnagel2011}%
  \BibitemOpen
  \bibfield  {author} {\bibinfo {author} {\bibfnamefont {K.}~\bibnamefont
  {Rossnagel}},\ }\bibfield  {title} {\bibinfo {title} {\emph {On the origin of
  charge-density waves in select layered transition-metal dichalcogenides}},\
  }\href {https://doi.org/10.1088/0953-8984/23/21/213001} {\bibfield  {journal}
  {\bibinfo  {journal} {J. Phys. Condens. Matter}\ }\textbf {\bibinfo {volume}
  {23}},\ \bibinfo {pages} {213001} (\bibinfo {year} {2011})}\BibitemShut
  {NoStop}%
\bibitem [{\citenamefont {Pasquier}\ and\ \citenamefont
  {Yazyev}(2019)}]{Pasquier2019}%
  \BibitemOpen
  \bibfield  {author} {\bibinfo {author} {\bibfnamefont {D.}~\bibnamefont
  {Pasquier}}\ and\ \bibinfo {author} {\bibfnamefont {O.~V.}\ \bibnamefont
  {Yazyev}},\ }\bibfield  {title} {\bibinfo {title} {\emph {Unified picture of
  lattice instabilities in metallic transition metal dichalcogenides}},\ }\href
  {https://doi.org/10.1103/PhysRevB.100.201103} {\bibfield  {journal} {\bibinfo
   {journal} {Phys. Rev. B}\ }\textbf {\bibinfo {volume} {100}},\ \bibinfo
  {pages} {201103(R)} (\bibinfo {year} {2019})},\ \Eprint
  {https://arxiv.org/abs/1901.10588} {arXiv:1901.10588}\BibitemShut {NoStop}%
\bibitem [{\citenamefont {Cavalleri}\ \emph {et~al.}(2004)\citenamefont
  {Cavalleri}, \citenamefont {Dekorsy}, \citenamefont {Chong}, \citenamefont
  {Kieffer},\ and\ \citenamefont {Schoenlein}}]{Cavalleri2004}%
  \BibitemOpen
  \bibfield  {author} {\bibinfo {author} {\bibfnamefont {A.}~\bibnamefont
  {Cavalleri}}, \bibinfo {author} {\bibfnamefont {T.}~\bibnamefont {Dekorsy}},
  \bibinfo {author} {\bibfnamefont {H.~H.~W.}\ \bibnamefont {Chong}}, \bibinfo
  {author} {\bibfnamefont {J.~C.}\ \bibnamefont {Kieffer}},\ and\ \bibinfo
  {author} {\bibfnamefont {R.~W.}\ \bibnamefont {Schoenlein}},\ }\bibfield
  {title} {\bibinfo {title} {\emph {Evidence for a structurally-driven
  insulator-to-metal transition in {VO\s2}: A view from the ultrafast
  timescale}},\ }\href {https://doi.org/10.1103/PhysRevB.70.161102} {\bibfield
  {journal} {\bibinfo  {journal} {Phys. Rev. B}\ }\textbf {\bibinfo {volume}
  {70}},\ \bibinfo {pages} {161102(R)} (\bibinfo {year} {2004})},\ \Eprint
  {https://arxiv.org/abs/cond-mat/0403214} {arXiv:cond-mat/0403214}\BibitemShut
  {NoStop}%
\bibitem [{\citenamefont {Hellmann}\ \emph {et~al.}(2012)\citenamefont
  {Hellmann}, \citenamefont {Rohwer}, \citenamefont {Kall\"ane}, \citenamefont
  {Hanff}, \citenamefont {Sohrt}, \citenamefont {Stange}, \citenamefont {Carr},
  \citenamefont {Murnane}, \citenamefont {Kapteyn}, \citenamefont {Kipp},
  \citenamefont {Bauer},\ and\ \citenamefont {Rossnagel}}]{Hellmann2012}%
  \BibitemOpen
  \bibfield  {author} {\bibinfo {author} {\bibfnamefont {S.}~\bibnamefont
  {Hellmann}}, \bibinfo {author} {\bibfnamefont {T.}~\bibnamefont {Rohwer}},
  \bibinfo {author} {\bibfnamefont {M.}~\bibnamefont {Kall\"ane}}, \bibinfo
  {author} {\bibfnamefont {K.}~\bibnamefont {Hanff}}, \bibinfo {author}
  {\bibfnamefont {C.}~\bibnamefont {Sohrt}}, \bibinfo {author} {\bibfnamefont
  {A.}~\bibnamefont {Stange}}, \bibinfo {author} {\bibfnamefont
  {A.}~\bibnamefont {Carr}}, \bibinfo {author} {\bibfnamefont {M.~M.}\
  \bibnamefont {Murnane}}, \bibinfo {author} {\bibfnamefont {H.~C.}\
  \bibnamefont {Kapteyn}}, \bibinfo {author} {\bibfnamefont {L.}~\bibnamefont
  {Kipp}}, \bibinfo {author} {\bibfnamefont {M.}~\bibnamefont {Bauer}},\ and\
  \bibinfo {author} {\bibfnamefont {K.}~\bibnamefont {Rossnagel}},\ }\bibfield
  {title} {\bibinfo {title} {\emph {Time-domain classification of
  charge-density-wave insulators}},\ }\href
  {https://doi.org/10.1038/ncomms2078} {\bibfield  {journal} {\bibinfo
  {journal} {Nat. Commun.}\ }\textbf {\bibinfo {volume} {3}},\ \bibinfo {pages}
  {1} (\bibinfo {year} {2012})}\BibitemShut {NoStop}%
\bibitem [{\citenamefont {Perfetti}\ \emph {et~al.}(2006)\citenamefont
  {Perfetti}, \citenamefont {Loukakos}, \citenamefont {Lisowski}, \citenamefont
  {Bovensiepen}, \citenamefont {Berger}, \citenamefont {Biermann},
  \citenamefont {Cornaglia}, \citenamefont {Georges},\ and\ \citenamefont
  {Wolf}}]{Perfetti2006}%
  \BibitemOpen
  \bibfield  {author} {\bibinfo {author} {\bibfnamefont {L.}~\bibnamefont
  {Perfetti}}, \bibinfo {author} {\bibfnamefont {P.~A.}\ \bibnamefont
  {Loukakos}}, \bibinfo {author} {\bibfnamefont {M.}~\bibnamefont {Lisowski}},
  \bibinfo {author} {\bibfnamefont {U.}~\bibnamefont {Bovensiepen}}, \bibinfo
  {author} {\bibfnamefont {H.}~\bibnamefont {Berger}}, \bibinfo {author}
  {\bibfnamefont {S.}~\bibnamefont {Biermann}}, \bibinfo {author}
  {\bibfnamefont {P.~S.}\ \bibnamefont {Cornaglia}}, \bibinfo {author}
  {\bibfnamefont {A.}~\bibnamefont {Georges}},\ and\ \bibinfo {author}
  {\bibfnamefont {M.}~\bibnamefont {Wolf}},\ }\bibfield  {title} {\bibinfo
  {title} {\emph {Time evolution of the electronic structure of {1T}-{TaS\s2}
  through the insulator-metal transition}},\ }\href
  {https://doi.org/10.1103/PhysRevLett.97.067402} {\bibfield  {journal}
  {\bibinfo  {journal} {Phys. Rev. Lett.}\ }\textbf {\bibinfo {volume} {97}},\
  \bibinfo {pages} {067402} (\bibinfo {year} {2006})}\BibitemShut {NoStop}%
\bibitem [{\citenamefont {Gunnarsson}\ \emph {et~al.}(2015)\citenamefont
  {Gunnarsson}, \citenamefont {Sch\"afer}, \citenamefont {LeBlanc},
  \citenamefont {Gull}, \citenamefont {Merino}, \citenamefont {Sangiovanni},
  \citenamefont {Rohringer},\ and\ \citenamefont {Toschi}}]{Gunnarsson2015}%
  \BibitemOpen
  \bibfield  {author} {\bibinfo {author} {\bibfnamefont {O.}~\bibnamefont
  {Gunnarsson}}, \bibinfo {author} {\bibfnamefont {T.}~\bibnamefont
  {Sch\"afer}}, \bibinfo {author} {\bibfnamefont {J.~P.~F.}\ \bibnamefont
  {LeBlanc}}, \bibinfo {author} {\bibfnamefont {E.}~\bibnamefont {Gull}},
  \bibinfo {author} {\bibfnamefont {J.}~\bibnamefont {Merino}}, \bibinfo
  {author} {\bibfnamefont {G.}~\bibnamefont {Sangiovanni}}, \bibinfo {author}
  {\bibfnamefont {G.}~\bibnamefont {Rohringer}},\ and\ \bibinfo {author}
  {\bibfnamefont {A.}~\bibnamefont {Toschi}},\ }\bibfield  {title} {\bibinfo
  {title} {\emph {Fluctuation diagnostics of the electron self-energy: Origin
  of the pseudogap physics}},\ }\href
  {https://doi.org/10.1103/PhysRevLett.114.236402} {\bibfield  {journal}
  {\bibinfo  {journal} {Phys. Rev. Lett.}\ }\textbf {\bibinfo {volume} {114}},\
  \bibinfo {pages} {236402} (\bibinfo {year} {2015})},\ \Eprint
  {https://arxiv.org/abs/1411.6947} {arXiv:1411.6947}\BibitemShut {NoStop}%
\bibitem [{\citenamefont {Gunnarsson}\ \emph {et~al.}(2016)\citenamefont
  {Gunnarsson}, \citenamefont {Sch\"afer}, \citenamefont {LeBlanc},
  \citenamefont {Merino}, \citenamefont {Sangiovanni}, \citenamefont
  {Rohringer},\ and\ \citenamefont {Toschi}}]{Gunnarsson2016}%
  \BibitemOpen
  \bibfield  {author} {\bibinfo {author} {\bibfnamefont {O.}~\bibnamefont
  {Gunnarsson}}, \bibinfo {author} {\bibfnamefont {T.}~\bibnamefont
  {Sch\"afer}}, \bibinfo {author} {\bibfnamefont {J.~P.~F.}\ \bibnamefont
  {LeBlanc}}, \bibinfo {author} {\bibfnamefont {J.}~\bibnamefont {Merino}},
  \bibinfo {author} {\bibfnamefont {G.}~\bibnamefont {Sangiovanni}}, \bibinfo
  {author} {\bibfnamefont {G.}~\bibnamefont {Rohringer}},\ and\ \bibinfo
  {author} {\bibfnamefont {A.}~\bibnamefont {Toschi}},\ }\bibfield  {title}
  {\bibinfo {title} {\emph {Parquet decomposition calculations of the
  electronic self-energy}},\ }\href
  {https://doi.org/10.1103/PhysRevB.93.245102} {\bibfield  {journal} {\bibinfo
  {journal} {Phys. Rev. B}\ }\textbf {\bibinfo {volume} {93}},\ \bibinfo
  {pages} {245102} (\bibinfo {year} {2016})},\ \Eprint
  {https://arxiv.org/abs/1604.01614} {arXiv:1604.01614}\BibitemShut {NoStop}%
\bibitem [{\citenamefont {Wilson}\ and\ \citenamefont
  {Yoffe}(1969)}]{Wilson1969}%
  \BibitemOpen
  \bibfield  {author} {\bibinfo {author} {\bibfnamefont {J.}~\bibnamefont
  {Wilson}}\ and\ \bibinfo {author} {\bibfnamefont {A.}~\bibnamefont {Yoffe}},\
  }\bibfield  {title} {\bibinfo {title} {\emph {The transition metal
  dichalcogenides: Discussion and interpretation of the observed optical,
  electrical and structural properties}},\ }\href
  {https://doi.org/10.1080/00018736900101307} {\bibfield  {journal} {\bibinfo
  {journal} {Adv. Phys.}\ }\textbf {\bibinfo {volume} {18}},\ \bibinfo {pages}
  {193} (\bibinfo {year} {1969})}\BibitemShut {NoStop}%
\bibitem [{\citenamefont {Wang}\ \emph {et~al.}(1990)\citenamefont {Wang},
  \citenamefont {Giambattista}, \citenamefont {Slough}, \citenamefont
  {Coleman},\ and\ \citenamefont {Subramanian}}]{Wang1990}%
  \BibitemOpen
  \bibfield  {author} {\bibinfo {author} {\bibfnamefont {C.}~\bibnamefont
  {Wang}}, \bibinfo {author} {\bibfnamefont {B.}~\bibnamefont {Giambattista}},
  \bibinfo {author} {\bibfnamefont {C.~G.}\ \bibnamefont {Slough}}, \bibinfo
  {author} {\bibfnamefont {R.~V.}\ \bibnamefont {Coleman}},\ and\ \bibinfo
  {author} {\bibfnamefont {M.~A.}\ \bibnamefont {Subramanian}},\ }\bibfield
  {title} {\bibinfo {title} {\emph {Energy gaps measured by scanning tunneling
  microscopy}},\ }\href {https://doi.org/10.1103/PhysRevB.42.8890} {\bibfield
  {journal} {\bibinfo  {journal} {Phys. Rev. B}\ }\textbf {\bibinfo {volume}
  {42}},\ \bibinfo {pages} {8890} (\bibinfo {year} {1990})}\BibitemShut
  {NoStop}%
\bibitem [{\citenamefont {Rossnagel}\ \emph {et~al.}(2001)\citenamefont
  {Rossnagel}, \citenamefont {Seifarth}, \citenamefont {Kipp}, \citenamefont
  {Skibowski}, \citenamefont {Vo{\ss}}, \citenamefont {Kr\"uger}, \citenamefont
  {Mazur},\ and\ \citenamefont {Pollmann}}]{Rossnagel2001}%
  \BibitemOpen
  \bibfield  {author} {\bibinfo {author} {\bibfnamefont {K.}~\bibnamefont
  {Rossnagel}}, \bibinfo {author} {\bibfnamefont {O.}~\bibnamefont {Seifarth}},
  \bibinfo {author} {\bibfnamefont {L.}~\bibnamefont {Kipp}}, \bibinfo {author}
  {\bibfnamefont {M.}~\bibnamefont {Skibowski}}, \bibinfo {author}
  {\bibfnamefont {D.}~\bibnamefont {Vo{\ss}}}, \bibinfo {author} {\bibfnamefont
  {P.}~\bibnamefont {Kr\"uger}}, \bibinfo {author} {\bibfnamefont
  {A.}~\bibnamefont {Mazur}},\ and\ \bibinfo {author} {\bibfnamefont
  {J.}~\bibnamefont {Pollmann}},\ }\bibfield  {title} {\bibinfo {title} {\emph
  {Fermi surface of {2H}-{NbSe\s2} and its implications on the
  charge-density-wave mechanism}},\ }\href
  {https://doi.org/10.1103/PhysRevB.64.235119} {\bibfield  {journal} {\bibinfo
  {journal} {Phys. Rev. B}\ }\textbf {\bibinfo {volume} {64}},\ \bibinfo
  {pages} {235119} (\bibinfo {year} {2001})}\BibitemShut {NoStop}%
\bibitem [{\citenamefont {Valla}\ \emph {et~al.}(2004)\citenamefont {Valla},
  \citenamefont {Fedorov}, \citenamefont {Johnson}, \citenamefont {Glans},
  \citenamefont {McGuinness}, \citenamefont {Smith}, \citenamefont {Andrei},\
  and\ \citenamefont {Berger}}]{Valla2004}%
  \BibitemOpen
  \bibfield  {author} {\bibinfo {author} {\bibfnamefont {T.}~\bibnamefont
  {Valla}}, \bibinfo {author} {\bibfnamefont {A.~V.}\ \bibnamefont {Fedorov}},
  \bibinfo {author} {\bibfnamefont {P.~D.}\ \bibnamefont {Johnson}}, \bibinfo
  {author} {\bibfnamefont {P.-A.}\ \bibnamefont {Glans}}, \bibinfo {author}
  {\bibfnamefont {C.}~\bibnamefont {McGuinness}}, \bibinfo {author}
  {\bibfnamefont {K.~E.}\ \bibnamefont {Smith}}, \bibinfo {author}
  {\bibfnamefont {E.~Y.}\ \bibnamefont {Andrei}},\ and\ \bibinfo {author}
  {\bibfnamefont {H.}~\bibnamefont {Berger}},\ }\bibfield  {title} {\bibinfo
  {title} {\emph {Quasiparticle spectra, charge-density waves,
  superconductivity, and electron-phonon coupling in {2H}-{NbSe\s2}}},\ }\href
  {https://doi.org/10.1103/PhysRevLett.92.086401} {\bibfield  {journal}
  {\bibinfo  {journal} {Phys. Rev. Lett.}\ }\textbf {\bibinfo {volume} {92}},\
  \bibinfo {pages} {086401} (\bibinfo {year} {2004})},\ \Eprint
  {https://arxiv.org/abs/cond-mat/0308278} {arXiv:cond-mat/0308278}\BibitemShut
  {NoStop}%
\bibitem [{\citenamefont {Johannes}\ \emph {et~al.}(2006)\citenamefont
  {Johannes}, \citenamefont {Mazin},\ and\ \citenamefont
  {Howells}}]{Johannes2006}%
  \BibitemOpen
  \bibfield  {author} {\bibinfo {author} {\bibfnamefont {M.~D.}\ \bibnamefont
  {Johannes}}, \bibinfo {author} {\bibfnamefont {I.~I.}\ \bibnamefont
  {Mazin}},\ and\ \bibinfo {author} {\bibfnamefont {C.~A.}\ \bibnamefont
  {Howells}},\ }\bibfield  {title} {\bibinfo {title} {\emph {Fermi-surface
  nesting and the origin of the charge-density wave in {NbSe\s2}}},\ }\href
  {https://doi.org/10.1103/PhysRevB.73.205102} {\bibfield  {journal} {\bibinfo
  {journal} {Phys. Rev. B}\ }\textbf {\bibinfo {volume} {73}},\ \bibinfo
  {pages} {205102} (\bibinfo {year} {2006})},\ \Eprint
  {https://arxiv.org/abs/cond-mat/0510390} {arXiv:cond-mat/0510390}\BibitemShut
  {NoStop}%
\bibitem [{\citenamefont {Calandra}\ \emph {et~al.}(2009)\citenamefont
  {Calandra}, \citenamefont {Mazin},\ and\ \citenamefont
  {Mauri}}]{Calandra2009}%
  \BibitemOpen
  \bibfield  {author} {\bibinfo {author} {\bibfnamefont {M.}~\bibnamefont
  {Calandra}}, \bibinfo {author} {\bibfnamefont {I.~I.}\ \bibnamefont
  {Mazin}},\ and\ \bibinfo {author} {\bibfnamefont {F.}~\bibnamefont {Mauri}},\
  }\bibfield  {title} {\bibinfo {title} {\emph {Effect of dimensionality on the
  charge-density wave in few-layer {2H}-{NbSe\s2}}},\ }\href
  {https://doi.org/10.1103/PhysRevB.80.241108} {\bibfield  {journal} {\bibinfo
  {journal} {Phys. Rev. B}\ }\textbf {\bibinfo {volume} {80}},\ \bibinfo
  {pages} {241108(R)} (\bibinfo {year} {2009})},\ \Eprint
  {https://arxiv.org/abs/0910.0956} {arXiv:0910.0956}\BibitemShut {NoStop}%
\bibitem [{\citenamefont {Weber}\ \emph {et~al.}(2011)\citenamefont {Weber},
  \citenamefont {Rosenkranz}, \citenamefont {Castellan}, \citenamefont
  {Osborn}, \citenamefont {Hott}, \citenamefont {Heid}, \citenamefont {Bohnen},
  \citenamefont {Egami}, \citenamefont {Said},\ and\ \citenamefont
  {Reznik}}]{Weber2011}%
  \BibitemOpen
  \bibfield  {author} {\bibinfo {author} {\bibfnamefont {F.}~\bibnamefont
  {Weber}}, \bibinfo {author} {\bibfnamefont {S.}~\bibnamefont {Rosenkranz}},
  \bibinfo {author} {\bibfnamefont {J.-P.}\ \bibnamefont {Castellan}}, \bibinfo
  {author} {\bibfnamefont {R.}~\bibnamefont {Osborn}}, \bibinfo {author}
  {\bibfnamefont {R.}~\bibnamefont {Hott}}, \bibinfo {author} {\bibfnamefont
  {R.}~\bibnamefont {Heid}}, \bibinfo {author} {\bibfnamefont {K.-P.}\
  \bibnamefont {Bohnen}}, \bibinfo {author} {\bibfnamefont {T.}~\bibnamefont
  {Egami}}, \bibinfo {author} {\bibfnamefont {A.~H.}\ \bibnamefont {Said}},\
  and\ \bibinfo {author} {\bibfnamefont {D.}~\bibnamefont {Reznik}},\
  }\bibfield  {title} {\bibinfo {title} {\emph {Extended phonon collapse and
  the origin of the charge-density wave in {2H}-{NbSe\s2}}},\ }\href
  {https://doi.org/10.1103/PhysRevLett.107.107403} {\bibfield  {journal}
  {\bibinfo  {journal} {Phys. Rev. Lett.}\ }\textbf {\bibinfo {volume} {107}},\
  \bibinfo {pages} {107403} (\bibinfo {year} {2011})},\ \Eprint
  {https://arxiv.org/abs/1103.5755} {arXiv:1103.5755}\BibitemShut {NoStop}%
\bibitem [{\citenamefont {Soumyanarayanan}\ \emph {et~al.}(2013)\citenamefont
  {Soumyanarayanan}, \citenamefont {Yee}, \citenamefont {He}, \citenamefont
  {van Wezel}, \citenamefont {Rahn}, \citenamefont {Rossnagel}, \citenamefont
  {Hudson}, \citenamefont {Norman},\ and\ \citenamefont
  {Hoffman}}]{Soumyanarayanan2013}%
  \BibitemOpen
  \bibfield  {author} {\bibinfo {author} {\bibfnamefont {A.}~\bibnamefont
  {Soumyanarayanan}}, \bibinfo {author} {\bibfnamefont {M.~M.}\ \bibnamefont
  {Yee}}, \bibinfo {author} {\bibfnamefont {Y.}~\bibnamefont {He}}, \bibinfo
  {author} {\bibfnamefont {J.}~\bibnamefont {van Wezel}}, \bibinfo {author}
  {\bibfnamefont {D.~J.}\ \bibnamefont {Rahn}}, \bibinfo {author}
  {\bibfnamefont {K.}~\bibnamefont {Rossnagel}}, \bibinfo {author}
  {\bibfnamefont {E.~W.}\ \bibnamefont {Hudson}}, \bibinfo {author}
  {\bibfnamefont {M.~R.}\ \bibnamefont {Norman}},\ and\ \bibinfo {author}
  {\bibfnamefont {J.~E.}\ \bibnamefont {Hoffman}},\ }\bibfield  {title}
  {\bibinfo {title} {\emph {Quantum phase transition from triangular to stripe
  charge order in {NbSe\s2}}},\ }\href
  {https://doi.org/10.1073/pnas.1211387110} {\bibfield  {journal} {\bibinfo
  {journal} {Proc. Natl. Acad. Sci. USA}\ }\textbf {\bibinfo {volume} {110}},\
  \bibinfo {pages} {1623} (\bibinfo {year} {2013})},\ \Eprint
  {https://arxiv.org/abs/1212.4087} {arXiv:1212.4087}\BibitemShut {NoStop}%
\bibitem [{\citenamefont {Arguello}\ \emph {et~al.}(2014)\citenamefont
  {Arguello}, \citenamefont {Chockalingam}, \citenamefont {Rosenthal},
  \citenamefont {Zhao}, \citenamefont {Guti\'errez}, \citenamefont {Kang},
  \citenamefont {Chung}, \citenamefont {Fernandes}, \citenamefont {Jia},
  \citenamefont {Millis}, \citenamefont {Cava},\ and\ \citenamefont
  {Pasupathy}}]{Arguello2014}%
  \BibitemOpen
  \bibfield  {author} {\bibinfo {author} {\bibfnamefont {C.~J.}\ \bibnamefont
  {Arguello}}, \bibinfo {author} {\bibfnamefont {S.~P.}\ \bibnamefont
  {Chockalingam}}, \bibinfo {author} {\bibfnamefont {E.~P.}\ \bibnamefont
  {Rosenthal}}, \bibinfo {author} {\bibfnamefont {L.}~\bibnamefont {Zhao}},
  \bibinfo {author} {\bibfnamefont {C.}~\bibnamefont {Guti\'errez}}, \bibinfo
  {author} {\bibfnamefont {J.~H.}\ \bibnamefont {Kang}}, \bibinfo {author}
  {\bibfnamefont {W.~C.}\ \bibnamefont {Chung}}, \bibinfo {author}
  {\bibfnamefont {R.~M.}\ \bibnamefont {Fernandes}}, \bibinfo {author}
  {\bibfnamefont {S.}~\bibnamefont {Jia}}, \bibinfo {author} {\bibfnamefont
  {A.~J.}\ \bibnamefont {Millis}}, \bibinfo {author} {\bibfnamefont {R.~J.}\
  \bibnamefont {Cava}},\ and\ \bibinfo {author} {\bibfnamefont {A.~N.}\
  \bibnamefont {Pasupathy}},\ }\bibfield  {title} {\bibinfo {title} {\emph
  {Visualizing the charge density wave transition in {2H}-{NbSe\s2} in real
  space}},\ }\href {https://doi.org/10.1103/PhysRevB.89.235115} {\bibfield
  {journal} {\bibinfo  {journal} {Phys. Rev. B}\ }\textbf {\bibinfo {volume}
  {89}},\ \bibinfo {pages} {235115} (\bibinfo {year} {2014})}\BibitemShut
  {NoStop}%
\bibitem [{\citenamefont {Arguello}\ \emph {et~al.}(2015)\citenamefont
  {Arguello}, \citenamefont {Rosenthal}, \citenamefont {Andrade}, \citenamefont
  {Jin}, \citenamefont {Yeh}, \citenamefont {Zaki}, \citenamefont {Jia},
  \citenamefont {Cava}, \citenamefont {Fernandes}, \citenamefont {Millis},
  \citenamefont {Valla}, \citenamefont {Osgood},\ and\ \citenamefont
  {Pasupathy}}]{Arguello2015}%
  \BibitemOpen
  \bibfield  {author} {\bibinfo {author} {\bibfnamefont {C.~J.}\ \bibnamefont
  {Arguello}}, \bibinfo {author} {\bibfnamefont {E.~P.}\ \bibnamefont
  {Rosenthal}}, \bibinfo {author} {\bibfnamefont {E.~F.}\ \bibnamefont
  {Andrade}}, \bibinfo {author} {\bibfnamefont {W.}~\bibnamefont {Jin}},
  \bibinfo {author} {\bibfnamefont {P.~C.}\ \bibnamefont {Yeh}}, \bibinfo
  {author} {\bibfnamefont {N.}~\bibnamefont {Zaki}}, \bibinfo {author}
  {\bibfnamefont {S.}~\bibnamefont {Jia}}, \bibinfo {author} {\bibfnamefont
  {R.~J.}\ \bibnamefont {Cava}}, \bibinfo {author} {\bibfnamefont {R.~M.}\
  \bibnamefont {Fernandes}}, \bibinfo {author} {\bibfnamefont {A.~J.}\
  \bibnamefont {Millis}}, \bibinfo {author} {\bibfnamefont {T.}~\bibnamefont
  {Valla}}, \bibinfo {author} {\bibfnamefont {R.~M.}\ \bibnamefont {Osgood}},\
  and\ \bibinfo {author} {\bibfnamefont {A.~N.}\ \bibnamefont {Pasupathy}},\
  }\bibfield  {title} {\bibinfo {title} {\emph {Quasiparticle interference,
  quasiparticle interactions, and the origin of the charge density wave in
  {2H}-{NbSe\s2}}},\ }\href {https://doi.org/10.1103/PhysRevLett.114.037001}
  {\bibfield  {journal} {\bibinfo  {journal} {Phys. Rev. Lett.}\ }\textbf
  {\bibinfo {volume} {114}},\ \bibinfo {pages} {037001} (\bibinfo {year}
  {2015})},\ \Eprint {https://arxiv.org/abs/1408.4432}
  {arXiv:1408.4432}\BibitemShut {NoStop}%
\bibitem [{\citenamefont {Flicker}\ and\ \citenamefont {van
  Wezel}(2015)}]{Flicker2015}%
  \BibitemOpen
  \bibfield  {author} {\bibinfo {author} {\bibfnamefont {F.}~\bibnamefont
  {Flicker}}\ and\ \bibinfo {author} {\bibfnamefont {J.}~\bibnamefont {van
  Wezel}},\ }\bibfield  {title} {\bibinfo {title} {\emph {Charge order from
  orbital-dependent coupling evidenced by {NbSe\s2}}},\ }\href
  {https://doi.org/10.1038/ncomms8034} {\bibfield  {journal} {\bibinfo
  {journal} {Nat. Commun.}\ }\textbf {\bibinfo {volume} {6}},\ \bibinfo {pages}
  {7034} (\bibinfo {year} {2015})},\ \Eprint {https://arxiv.org/abs/1502.06816}
  {arXiv:1502.06816}\BibitemShut {NoStop}%
\bibitem [{\citenamefont {Xi}\ \emph {et~al.}(2015)\citenamefont {Xi},
  \citenamefont {Zhao}, \citenamefont {Wang}, \citenamefont {Berger},
  \citenamefont {Forr\'o}, \citenamefont {Shan},\ and\ \citenamefont
  {Mak}}]{Xi2015}%
  \BibitemOpen
  \bibfield  {author} {\bibinfo {author} {\bibfnamefont {X.}~\bibnamefont
  {Xi}}, \bibinfo {author} {\bibfnamefont {L.}~\bibnamefont {Zhao}}, \bibinfo
  {author} {\bibfnamefont {Z.}~\bibnamefont {Wang}}, \bibinfo {author}
  {\bibfnamefont {H.}~\bibnamefont {Berger}}, \bibinfo {author} {\bibfnamefont
  {L.}~\bibnamefont {Forr\'o}}, \bibinfo {author} {\bibfnamefont
  {J.}~\bibnamefont {Shan}},\ and\ \bibinfo {author} {\bibfnamefont {K.~F.}\
  \bibnamefont {Mak}},\ }\bibfield  {title} {\bibinfo {title} {\emph {Strongly
  enhanced charge-density-wave order in monolayer {NbSe\s2}}},\ }\href
  {https://doi.org/10.1038/nnano.2015.143} {\bibfield  {journal} {\bibinfo
  {journal} {Nat. Nanotechnol.}\ }\textbf {\bibinfo {volume} {10}},\ \bibinfo
  {pages} {765} (\bibinfo {year} {2015})},\ \Eprint
  {https://arxiv.org/abs/1507.05595} {arXiv:1507.05595}\BibitemShut {NoStop}%
\bibitem [{\citenamefont {Zhu}\ \emph {et~al.}(2015)\citenamefont {Zhu},
  \citenamefont {Cao}, \citenamefont {Zhang}, \citenamefont {Plummer},\ and\
  \citenamefont {Guo}}]{Zhu2015}%
  \BibitemOpen
  \bibfield  {author} {\bibinfo {author} {\bibfnamefont {X.}~\bibnamefont
  {Zhu}}, \bibinfo {author} {\bibfnamefont {Y.}~\bibnamefont {Cao}}, \bibinfo
  {author} {\bibfnamefont {J.}~\bibnamefont {Zhang}}, \bibinfo {author}
  {\bibfnamefont {E.~W.}\ \bibnamefont {Plummer}},\ and\ \bibinfo {author}
  {\bibfnamefont {J.}~\bibnamefont {Guo}},\ }\bibfield  {title} {\bibinfo
  {title} {\emph {Classification of charge density waves based on their
  nature}},\ }\href {https://doi.org/10.1073/pnas.1424791112} {\bibfield
  {journal} {\bibinfo  {journal} {Proc. Natl. Acad. Sci. USA}\ }\textbf
  {\bibinfo {volume} {112}},\ \bibinfo {pages} {2367} (\bibinfo {year}
  {2015})},\ \Eprint {https://arxiv.org/abs/1503.01858}
  {arXiv:1503.01858}\BibitemShut {NoStop}%
\bibitem [{\citenamefont {Ugeda}\ \emph {et~al.}(2016)\citenamefont {Ugeda},
  \citenamefont {Bradley}, \citenamefont {Zhang}, \citenamefont {Onishi},
  \citenamefont {Chen}, \citenamefont {Ruan}, \citenamefont
  {Ojeda-Aristizabal}, \citenamefont {Ryu}, \citenamefont {Edmonds},
  \citenamefont {Tsai}, \citenamefont {Riss}, \citenamefont {Mo}, \citenamefont
  {Lee}, \citenamefont {Zettl}, \citenamefont {Hussain}, \citenamefont {Shen},\
  and\ \citenamefont {Crommie}}]{Ugeda2016}%
  \BibitemOpen
  \bibfield  {author} {\bibinfo {author} {\bibfnamefont {M.~M.}\ \bibnamefont
  {Ugeda}}, \bibinfo {author} {\bibfnamefont {A.~J.}\ \bibnamefont {Bradley}},
  \bibinfo {author} {\bibfnamefont {Y.}~\bibnamefont {Zhang}}, \bibinfo
  {author} {\bibfnamefont {S.}~\bibnamefont {Onishi}}, \bibinfo {author}
  {\bibfnamefont {Y.}~\bibnamefont {Chen}}, \bibinfo {author} {\bibfnamefont
  {W.}~\bibnamefont {Ruan}}, \bibinfo {author} {\bibfnamefont {C.}~\bibnamefont
  {Ojeda-Aristizabal}}, \bibinfo {author} {\bibfnamefont {H.}~\bibnamefont
  {Ryu}}, \bibinfo {author} {\bibfnamefont {M.~T.}\ \bibnamefont {Edmonds}},
  \bibinfo {author} {\bibfnamefont {H.-Z.}\ \bibnamefont {Tsai}}, \bibinfo
  {author} {\bibfnamefont {A.}~\bibnamefont {Riss}}, \bibinfo {author}
  {\bibfnamefont {S.-K.}\ \bibnamefont {Mo}}, \bibinfo {author} {\bibfnamefont
  {D.}~\bibnamefont {Lee}}, \bibinfo {author} {\bibfnamefont {A.}~\bibnamefont
  {Zettl}}, \bibinfo {author} {\bibfnamefont {Z.}~\bibnamefont {Hussain}},
  \bibinfo {author} {\bibfnamefont {Z.-X.}\ \bibnamefont {Shen}},\ and\
  \bibinfo {author} {\bibfnamefont {M.~F.}\ \bibnamefont {Crommie}},\
  }\bibfield  {title} {\bibinfo {title} {\emph {Characterization of collective
  ground states in single-layer {NbSe\s2}}},\ }\href
  {https://doi.org/10.1038/nphys3527} {\bibfield  {journal} {\bibinfo
  {journal} {Nat. Phys.}\ }\textbf {\bibinfo {volume} {12}},\ \bibinfo {pages}
  {92} (\bibinfo {year} {2016})},\ \Eprint {https://arxiv.org/abs/1506.08460}
  {arXiv:1506.08460}\BibitemShut {NoStop}%
\bibitem [{\citenamefont {Nakata}\ \emph {et~al.}(2018)\citenamefont {Nakata},
  \citenamefont {Sugawara}, \citenamefont {Ichinokura}, \citenamefont {Okada},
  \citenamefont {Hitosugi}, \citenamefont {Koretsune}, \citenamefont {Ueno},
  \citenamefont {Hasegawa}, \citenamefont {Takahashi},\ and\ \citenamefont
  {Sato}}]{Nakata2018}%
  \BibitemOpen
  \bibfield  {author} {\bibinfo {author} {\bibfnamefont {Y.}~\bibnamefont
  {Nakata}}, \bibinfo {author} {\bibfnamefont {K.}~\bibnamefont {Sugawara}},
  \bibinfo {author} {\bibfnamefont {S.}~\bibnamefont {Ichinokura}}, \bibinfo
  {author} {\bibfnamefont {Y.}~\bibnamefont {Okada}}, \bibinfo {author}
  {\bibfnamefont {T.}~\bibnamefont {Hitosugi}}, \bibinfo {author}
  {\bibfnamefont {T.}~\bibnamefont {Koretsune}}, \bibinfo {author}
  {\bibfnamefont {K.}~\bibnamefont {Ueno}}, \bibinfo {author} {\bibfnamefont
  {S.}~\bibnamefont {Hasegawa}}, \bibinfo {author} {\bibfnamefont
  {T.}~\bibnamefont {Takahashi}},\ and\ \bibinfo {author} {\bibfnamefont
  {T.}~\bibnamefont {Sato}},\ }\bibfield  {title} {\bibinfo {title} {\emph
  {Anisotropic band splitting in monolayer {NbSe\s2}: Implications for
  superconductivity and charge density wave}},\ }\href
  {https://doi.org/10.1038/s41699-018-0057-3} {\bibfield  {journal} {\bibinfo
  {journal} {npj 2D Mater. Appl.}\ }\textbf {\bibinfo {volume} {2}},\ \bibinfo
  {pages} {12} (\bibinfo {year} {2018})}\BibitemShut {NoStop}%
\bibitem [{\citenamefont {McMillan}(1977)}]{McMillan1977}%
  \BibitemOpen
  \bibfield  {author} {\bibinfo {author} {\bibfnamefont {W.~L.}\ \bibnamefont
  {McMillan}},\ }\bibfield  {title} {\bibinfo {title} {\emph {Microscopic model
  of charge-density waves in {2H}-{TaSe\s2}}},\ }\href
  {https://doi.org/10.1103/PhysRevB.16.643} {\bibfield  {journal} {\bibinfo
  {journal} {Phys. Rev. B}\ }\textbf {\bibinfo {volume} {16}},\ \bibinfo
  {pages} {643} (\bibinfo {year} {1977})}\BibitemShut {NoStop}%
\bibitem [{\citenamefont {Haas}(1978)}]{Haas1978}%
  \BibitemOpen
  \bibfield  {author} {\bibinfo {author} {\bibfnamefont {C.}~\bibnamefont
  {Haas}},\ }\bibfield  {title} {\bibinfo {title} {\emph {Chemical bond model
  of lattice distortions in hexagonal layers}},\ }\href
  {https://doi.org/10.1016/0038-1098(78)90725-1} {\bibfield  {journal}
  {\bibinfo  {journal} {Solid State Commun.}\ }\textbf {\bibinfo {volume}
  {26}},\ \bibinfo {pages} {709} (\bibinfo {year} {1978})}\BibitemShut
  {NoStop}%
\bibitem [{\citenamefont {Inglesfield}(1980{\natexlab{a}})}]{Inglesfield1980a}%
  \BibitemOpen
  \bibfield  {author} {\bibinfo {author} {\bibfnamefont {J.~E.}\ \bibnamefont
  {Inglesfield}},\ }\bibfield  {title} {\bibinfo {title} {\emph {Bonding and
  phase transitions in transition metal dichalcogenide layer compounds}},\
  }\href {https://doi.org/10.1088/0022-3719/13/1/007} {\bibfield  {journal}
  {\bibinfo  {journal} {J. Phys. C: Solid State Phys.}\ }\textbf {\bibinfo
  {volume} {13}},\ \bibinfo {pages} {17} (\bibinfo {year}
  {1980}{\natexlab{a}})}\BibitemShut {NoStop}%
\bibitem [{\citenamefont {Inglesfield}(1980{\natexlab{b}})}]{Inglesfield1980b}%
  \BibitemOpen
  \bibfield  {author} {\bibinfo {author} {\bibfnamefont {J.~E.}\ \bibnamefont
  {Inglesfield}},\ }\bibfield  {title} {\bibinfo {title} {\emph {Bonding and
  charge density wave phase transitions}},\ }\href
  {https://doi.org/10.1016/0378-4363(80)90238-7} {\bibfield  {journal}
  {\bibinfo  {journal} {Physica B+C}\ }\textbf {\bibinfo {volume} {99}},\
  \bibinfo {pages} {238} (\bibinfo {year} {1980}{\natexlab{b}})}\BibitemShut
  {NoStop}%
\bibitem [{\citenamefont {Whangbo}\ and\ \citenamefont
  {Canadell}(1992)}]{Whangbo1992}%
  \BibitemOpen
  \bibfield  {author} {\bibinfo {author} {\bibfnamefont {M.~H.}\ \bibnamefont
  {Whangbo}}\ and\ \bibinfo {author} {\bibfnamefont {E.}~\bibnamefont
  {Canadell}},\ }\bibfield  {title} {\bibinfo {title} {\emph {Analogies between
  the concepts of molecular chemistry and solid-state physics concerning
  structural instabilities. {Electronic} origin of the structural modulations
  in layered transition metal dichalcogenides}},\ }\href
  {https://doi.org/10.1021/ja00050a044} {\bibfield  {journal} {\bibinfo
  {journal} {J. Am. Chem. Soc.}\ }\textbf {\bibinfo {volume} {114}},\ \bibinfo
  {pages} {9587} (\bibinfo {year} {1992})}\BibitemShut {NoStop}%
\bibitem [{\citenamefont {Silva-Guill\'en}\ \emph {et~al.}(2016)\citenamefont
  {Silva-Guill\'en}, \citenamefont {Ordej\'on}, \citenamefont {Guinea},\ and\
  \citenamefont {Canadell}}]{SilvaGuillen2016}%
  \BibitemOpen
  \bibfield  {author} {\bibinfo {author} {\bibfnamefont {J.~A.}\ \bibnamefont
  {Silva-Guill\'en}}, \bibinfo {author} {\bibfnamefont {P.}~\bibnamefont
  {Ordej\'on}}, \bibinfo {author} {\bibfnamefont {F.}~\bibnamefont {Guinea}},\
  and\ \bibinfo {author} {\bibfnamefont {E.}~\bibnamefont {Canadell}},\
  }\bibfield  {title} {\bibinfo {title} {\emph {Electronic structure of
  {2H}-{NbSe\s2} single-layers in the {CDW} state}},\ }\href
  {https://doi.org/10.1088/2053-1583/3/3/035028} {\bibfield  {journal}
  {\bibinfo  {journal} {2D Mater.}\ }\textbf {\bibinfo {volume} {3}},\ \bibinfo
  {pages} {035028} (\bibinfo {year} {2016})},\ \Eprint
  {https://arxiv.org/abs/1603.09072} {arXiv:1603.09072}\BibitemShut {NoStop}%
\bibitem [{\citenamefont {Wilson}\ \emph {et~al.}(1974)\citenamefont {Wilson},
  \citenamefont {Di~Salvo},\ and\ \citenamefont {Mahajan}}]{Wilson1974}%
  \BibitemOpen
  \bibfield  {author} {\bibinfo {author} {\bibfnamefont {J.~A.}\ \bibnamefont
  {Wilson}}, \bibinfo {author} {\bibfnamefont {F.~J.}\ \bibnamefont
  {Di~Salvo}},\ and\ \bibinfo {author} {\bibfnamefont {S.}~\bibnamefont
  {Mahajan}},\ }\bibfield  {title} {\bibinfo {title} {\emph {Charge-density
  waves in metallic, layered, transition-metal dichalcogenides}},\ }\href
  {https://doi.org/10.1103/PhysRevLett.32.882} {\bibfield  {journal} {\bibinfo
  {journal} {Phys. Rev. Lett.}\ }\textbf {\bibinfo {volume} {32}},\ \bibinfo
  {pages} {882} (\bibinfo {year} {1974})}\BibitemShut {NoStop}%
\bibitem [{\citenamefont {Wexler}\ and\ \citenamefont
  {Woolley}(1976)}]{Wexler1976}%
  \BibitemOpen
  \bibfield  {author} {\bibinfo {author} {\bibfnamefont {G.}~\bibnamefont
  {Wexler}}\ and\ \bibinfo {author} {\bibfnamefont {A.~M.}\ \bibnamefont
  {Woolley}},\ }\bibfield  {title} {\bibinfo {title} {\emph {Fermi surfaces and
  band structures of the {2H} metallic transition-metal dichalcogenides}},\
  }\href {https://doi.org/10.1088/0022-3719/9/7/010} {\bibfield  {journal}
  {\bibinfo  {journal} {J. Phys. C: Solid State Phys.}\ }\textbf {\bibinfo
  {volume} {9}},\ \bibinfo {pages} {1185} (\bibinfo {year} {1976})}\BibitemShut
  {NoStop}%
\bibitem [{\citenamefont {Wilson}(1977)}]{Wilson1977}%
  \BibitemOpen
  \bibfield  {author} {\bibinfo {author} {\bibfnamefont {J.~A.}\ \bibnamefont
  {Wilson}},\ }\bibfield  {title} {\bibinfo {title} {\emph {Charge-density
  waves in the {2H}-{TaSe\s2} family: Action on the {Fermi} surface}},\ }\href
  {https://doi.org/10.1103/PhysRevB.15.5748} {\bibfield  {journal} {\bibinfo
  {journal} {Phys. Rev. B}\ }\textbf {\bibinfo {volume} {15}},\ \bibinfo
  {pages} {5748} (\bibinfo {year} {1977})}\BibitemShut {NoStop}%
\bibitem [{\citenamefont {Straub}\ \emph {et~al.}(1999)\citenamefont {Straub},
  \citenamefont {Finteis}, \citenamefont {Claessen}, \citenamefont {Steiner},
  \citenamefont {H\"ufner}, \citenamefont {Blaha}, \citenamefont {Oglesby},\
  and\ \citenamefont {Bucher}}]{Straub1999}%
  \BibitemOpen
  \bibfield  {author} {\bibinfo {author} {\bibfnamefont {T.}~\bibnamefont
  {Straub}}, \bibinfo {author} {\bibfnamefont {T.}~\bibnamefont {Finteis}},
  \bibinfo {author} {\bibfnamefont {R.}~\bibnamefont {Claessen}}, \bibinfo
  {author} {\bibfnamefont {P.}~\bibnamefont {Steiner}}, \bibinfo {author}
  {\bibfnamefont {S.}~\bibnamefont {H\"ufner}}, \bibinfo {author}
  {\bibfnamefont {P.}~\bibnamefont {Blaha}}, \bibinfo {author} {\bibfnamefont
  {C.~S.}\ \bibnamefont {Oglesby}},\ and\ \bibinfo {author} {\bibfnamefont
  {E.}~\bibnamefont {Bucher}},\ }\bibfield  {title} {\bibinfo {title} {\emph
  {Charge-density-wave mechanism in {2H}-{NbSe\s2}: Photoemission results}},\
  }\href {https://doi.org/10.1103/PhysRevLett.82.4504} {\bibfield  {journal}
  {\bibinfo  {journal} {Phys. Rev. Lett.}\ }\textbf {\bibinfo {volume} {82}},\
  \bibinfo {pages} {4504} (\bibinfo {year} {1999})}\BibitemShut {NoStop}%
\bibitem [{\citenamefont {Shen}\ \emph {et~al.}(2008)\citenamefont {Shen},
  \citenamefont {Zhang}, \citenamefont {Yang}, \citenamefont {Wei},
  \citenamefont {Ou}, \citenamefont {Dong}, \citenamefont {Xie}, \citenamefont
  {He}, \citenamefont {Zhao}, \citenamefont {Zhou}, \citenamefont {Arita},
  \citenamefont {Shimada}, \citenamefont {Namatame}, \citenamefont {Taniguchi},
  \citenamefont {Shi},\ and\ \citenamefont {Feng}}]{Shen2008}%
  \BibitemOpen
  \bibfield  {author} {\bibinfo {author} {\bibfnamefont {D.~W.}\ \bibnamefont
  {Shen}}, \bibinfo {author} {\bibfnamefont {Y.}~\bibnamefont {Zhang}},
  \bibinfo {author} {\bibfnamefont {L.~X.}\ \bibnamefont {Yang}}, \bibinfo
  {author} {\bibfnamefont {J.}~\bibnamefont {Wei}}, \bibinfo {author}
  {\bibfnamefont {H.~W.}\ \bibnamefont {Ou}}, \bibinfo {author} {\bibfnamefont
  {J.~K.}\ \bibnamefont {Dong}}, \bibinfo {author} {\bibfnamefont {B.~P.}\
  \bibnamefont {Xie}}, \bibinfo {author} {\bibfnamefont {C.}~\bibnamefont
  {He}}, \bibinfo {author} {\bibfnamefont {J.~F.}\ \bibnamefont {Zhao}},
  \bibinfo {author} {\bibfnamefont {B.}~\bibnamefont {Zhou}}, \bibinfo {author}
  {\bibfnamefont {M.}~\bibnamefont {Arita}}, \bibinfo {author} {\bibfnamefont
  {K.}~\bibnamefont {Shimada}}, \bibinfo {author} {\bibfnamefont
  {H.}~\bibnamefont {Namatame}}, \bibinfo {author} {\bibfnamefont
  {M.}~\bibnamefont {Taniguchi}}, \bibinfo {author} {\bibfnamefont
  {J.}~\bibnamefont {Shi}},\ and\ \bibinfo {author} {\bibfnamefont {D.~L.}\
  \bibnamefont {Feng}},\ }\bibfield  {title} {\bibinfo {title} {\emph {Primary
  role of the barely occupied states in the charge density wave formation of
  {NbSe\s2}}},\ }\href {https://doi.org/10.1103/PhysRevLett.101.226406}
  {\bibfield  {journal} {\bibinfo  {journal} {Phys. Rev. Lett.}\ }\textbf
  {\bibinfo {volume} {101}},\ \bibinfo {pages} {226406} (\bibinfo {year}
  {2008})},\ \Eprint {https://arxiv.org/abs/0806.1344}
  {arXiv:0806.1344}\BibitemShut {NoStop}%
\bibitem [{\citenamefont {Borisenko}\ \emph {et~al.}(2009)\citenamefont
  {Borisenko}, \citenamefont {Kordyuk}, \citenamefont {Zabolotnyy},
  \citenamefont {Inosov}, \citenamefont {Evtushinsky}, \citenamefont
  {B\"uchner}, \citenamefont {Yaresko}, \citenamefont {Varykhalov},
  \citenamefont {Follath}, \citenamefont {Eberhardt}, \citenamefont {Patthey},\
  and\ \citenamefont {Berger}}]{Borisenko2009}%
  \BibitemOpen
  \bibfield  {author} {\bibinfo {author} {\bibfnamefont {S.~V.}\ \bibnamefont
  {Borisenko}}, \bibinfo {author} {\bibfnamefont {A.~A.}\ \bibnamefont
  {Kordyuk}}, \bibinfo {author} {\bibfnamefont {V.~B.}\ \bibnamefont
  {Zabolotnyy}}, \bibinfo {author} {\bibfnamefont {D.~S.}\ \bibnamefont
  {Inosov}}, \bibinfo {author} {\bibfnamefont {D.}~\bibnamefont {Evtushinsky}},
  \bibinfo {author} {\bibfnamefont {B.}~\bibnamefont {B\"uchner}}, \bibinfo
  {author} {\bibfnamefont {A.~N.}\ \bibnamefont {Yaresko}}, \bibinfo {author}
  {\bibfnamefont {A.}~\bibnamefont {Varykhalov}}, \bibinfo {author}
  {\bibfnamefont {R.}~\bibnamefont {Follath}}, \bibinfo {author} {\bibfnamefont
  {W.}~\bibnamefont {Eberhardt}}, \bibinfo {author} {\bibfnamefont
  {L.}~\bibnamefont {Patthey}},\ and\ \bibinfo {author} {\bibfnamefont
  {H.}~\bibnamefont {Berger}},\ }\bibfield  {title} {\bibinfo {title} {\emph
  {Two energy gaps and {Fermi}-surface ``arcs'' in {NbSe\s2}}},\ }\href
  {https://doi.org/10.1103/PhysRevLett.102.166402} {\bibfield  {journal}
  {\bibinfo  {journal} {Phys. Rev. Lett.}\ }\textbf {\bibinfo {volume} {102}},\
  \bibinfo {pages} {166402} (\bibinfo {year} {2009})},\ \Eprint
  {https://arxiv.org/abs/0904.3277} {arXiv:0904.3277}\BibitemShut {NoStop}%
\bibitem [{\citenamefont {Tonjes}\ \emph {et~al.}(2001)\citenamefont {Tonjes},
  \citenamefont {Greanya}, \citenamefont {Liu}, \citenamefont {Olson},\ and\
  \citenamefont {Molini\'e}}]{Tonjes2001}%
  \BibitemOpen
  \bibfield  {author} {\bibinfo {author} {\bibfnamefont {W.~C.}\ \bibnamefont
  {Tonjes}}, \bibinfo {author} {\bibfnamefont {V.~A.}\ \bibnamefont {Greanya}},
  \bibinfo {author} {\bibfnamefont {R.}~\bibnamefont {Liu}}, \bibinfo {author}
  {\bibfnamefont {C.~G.}\ \bibnamefont {Olson}},\ and\ \bibinfo {author}
  {\bibfnamefont {P.}~\bibnamefont {Molini\'e}},\ }\bibfield  {title} {\bibinfo
  {title} {\emph {Charge-density-wave mechanism in the {2H}-{NbSe\s2} family:
  Angle-resolved photoemission studies}},\ }\href
  {https://doi.org/10.1103/PhysRevB.63.235101} {\bibfield  {journal} {\bibinfo
  {journal} {Phys. Rev. B}\ }\textbf {\bibinfo {volume} {63}},\ \bibinfo
  {pages} {235101} (\bibinfo {year} {2001})}\BibitemShut {NoStop}%
\bibitem [{\citenamefont {Kiss}\ \emph {et~al.}(2007)\citenamefont {Kiss},
  \citenamefont {Yokoya}, \citenamefont {Chainani}, \citenamefont {Shin},
  \citenamefont {Hanaguri}, \citenamefont {Nohara},\ and\ \citenamefont
  {Takagi}}]{Kiss2007}%
  \BibitemOpen
  \bibfield  {author} {\bibinfo {author} {\bibfnamefont {T.}~\bibnamefont
  {Kiss}}, \bibinfo {author} {\bibfnamefont {T.}~\bibnamefont {Yokoya}},
  \bibinfo {author} {\bibfnamefont {A.}~\bibnamefont {Chainani}}, \bibinfo
  {author} {\bibfnamefont {S.}~\bibnamefont {Shin}}, \bibinfo {author}
  {\bibfnamefont {T.}~\bibnamefont {Hanaguri}}, \bibinfo {author}
  {\bibfnamefont {M.}~\bibnamefont {Nohara}},\ and\ \bibinfo {author}
  {\bibfnamefont {H.}~\bibnamefont {Takagi}},\ }\bibfield  {title} {\bibinfo
  {title} {\emph {Charge-order-maximized momentum-dependent
  superconductivity}},\ }\href {https://doi.org/10.1038/nphys699} {\bibfield
  {journal} {\bibinfo  {journal} {Nat. Phys.}\ }\textbf {\bibinfo {volume}
  {3}},\ \bibinfo {pages} {720} (\bibinfo {year} {2007})}\BibitemShut {NoStop}%
\bibitem [{\citenamefont {Aryasetiawan}\ \emph {et~al.}(2004)\citenamefont
  {Aryasetiawan}, \citenamefont {Imada}, \citenamefont {Georges}, \citenamefont
  {Kotliar}, \citenamefont {Biermann},\ and\ \citenamefont
  {Lichtenstein}}]{Aryasetiawan2004}%
  \BibitemOpen
  \bibfield  {author} {\bibinfo {author} {\bibfnamefont {F.}~\bibnamefont
  {Aryasetiawan}}, \bibinfo {author} {\bibfnamefont {M.}~\bibnamefont {Imada}},
  \bibinfo {author} {\bibfnamefont {A.}~\bibnamefont {Georges}}, \bibinfo
  {author} {\bibfnamefont {G.}~\bibnamefont {Kotliar}}, \bibinfo {author}
  {\bibfnamefont {S.}~\bibnamefont {Biermann}},\ and\ \bibinfo {author}
  {\bibfnamefont {A.~I.}\ \bibnamefont {Lichtenstein}},\ }\bibfield  {title}
  {\bibinfo {title} {\emph {Frequency-dependent local interactions and
  low-energy effective models from electronic structure calculations}},\ }\href
  {https://doi.org/10.1103/PhysRevB.70.195104} {\bibfield  {journal} {\bibinfo
  {journal} {Phys. Rev. B}\ }\textbf {\bibinfo {volume} {70}},\ \bibinfo
  {pages} {195104} (\bibinfo {year} {2004})},\ \Eprint
  {https://arxiv.org/abs/cond-mat/0401620} {arXiv:cond-mat/0401620}\BibitemShut
  {NoStop}%
\bibitem [{\citenamefont {Nomura}\ and\ \citenamefont
  {Arita}(2015)}]{Nomura2015}%
  \BibitemOpen
  \bibfield  {author} {\bibinfo {author} {\bibfnamefont {Y.}~\bibnamefont
  {Nomura}}\ and\ \bibinfo {author} {\bibfnamefont {R.}~\bibnamefont {Arita}},\
  }\bibfield  {title} {\bibinfo {title} {\emph {Ab initio downfolding for
  electron-phonon-coupled systems: Constrained density-functional perturbation
  theory}},\ }\href {https://doi.org/10.1103/PhysRevB.92.245108} {\bibfield
  {journal} {\bibinfo  {journal} {Phys. Rev. B}\ }\textbf {\bibinfo {volume}
  {92}},\ \bibinfo {pages} {245108} (\bibinfo {year} {2015})},\ \Eprint
  {https://arxiv.org/abs/1509.01138} {arXiv:1509.01138}\BibitemShut {NoStop}%
\bibitem [{\citenamefont {Baroni}\ \emph {et~al.}(2001)\citenamefont {Baroni},
  \citenamefont {de~Gironcoli}, \citenamefont {Dal~Corso},\ and\ \citenamefont
  {Giannozzi}}]{Baroni2001}%
  \BibitemOpen
  \bibfield  {author} {\bibinfo {author} {\bibfnamefont {S.}~\bibnamefont
  {Baroni}}, \bibinfo {author} {\bibfnamefont {S.}~\bibnamefont
  {de~Gironcoli}}, \bibinfo {author} {\bibfnamefont {A.}~\bibnamefont
  {Dal~Corso}},\ and\ \bibinfo {author} {\bibfnamefont {P.}~\bibnamefont
  {Giannozzi}},\ }\bibfield  {title} {\bibinfo {title} {\emph {Phonons and
  related crystal properties from density-functional perturbation theory}},\
  }\href {https://doi.org/10.1103/RevModPhys.73.515} {\bibfield  {journal}
  {\bibinfo  {journal} {Rev. Mod. Phys.}\ }\textbf {\bibinfo {volume} {73}},\
  \bibinfo {pages} {515} (\bibinfo {year} {2001})},\ \Eprint
  {https://arxiv.org/abs/cond-mat/0012092} {arXiv:cond-mat/0012092}\BibitemShut
  {NoStop}%
\bibitem [{\citenamefont {Kohn}(1959)}]{Kohn1959}%
  \BibitemOpen
  \bibfield  {author} {\bibinfo {author} {\bibfnamefont {W.}~\bibnamefont
  {Kohn}},\ }\bibfield  {title} {\bibinfo {title} {\emph {Image of the {Fermi}
  surface in the vibration spectrum of a metal}},\ }\href
  {https://doi.org/10.1103/PhysRevLett.2.393} {\bibfield  {journal} {\bibinfo
  {journal} {Phys. Rev. Lett.}\ }\textbf {\bibinfo {volume} {2}},\ \bibinfo
  {pages} {393} (\bibinfo {year} {1959})}\BibitemShut {NoStop}%
\bibitem [{\citenamefont {Tidman}\ \emph {et~al.}(1974)\citenamefont {Tidman},
  \citenamefont {Singh}, \citenamefont {Curzon},\ and\ \citenamefont
  {Frindt}}]{Tidman1974}%
  \BibitemOpen
  \bibfield  {author} {\bibinfo {author} {\bibfnamefont {J.~P.}\ \bibnamefont
  {Tidman}}, \bibinfo {author} {\bibfnamefont {O.}~\bibnamefont {Singh}},
  \bibinfo {author} {\bibfnamefont {A.~E.}\ \bibnamefont {Curzon}},\ and\
  \bibinfo {author} {\bibfnamefont {R.~F.}\ \bibnamefont {Frindt}},\ }\bibfield
   {title} {\bibinfo {title} {\emph {The phase transition in {2H}-{TaS\s2} at
  75\,{K}}},\ }\href {https://doi.org/10.1080/14786437408207274} {\bibfield
  {journal} {\bibinfo  {journal} {Philos. Mag.}\ }\textbf {\bibinfo {volume}
  {30}},\ \bibinfo {pages} {1191} (\bibinfo {year} {1974})}\BibitemShut
  {NoStop}%
\bibitem [{\citenamefont {Scholz}\ \emph {et~al.}(1982)\citenamefont {Scholz},
  \citenamefont {Singh}, \citenamefont {Frindt},\ and\ \citenamefont
  {Curzon}}]{Scholz1982}%
  \BibitemOpen
  \bibfield  {author} {\bibinfo {author} {\bibfnamefont {G.~A.}\ \bibnamefont
  {Scholz}}, \bibinfo {author} {\bibfnamefont {O.}~\bibnamefont {Singh}},
  \bibinfo {author} {\bibfnamefont {R.~F.}\ \bibnamefont {Frindt}},\ and\
  \bibinfo {author} {\bibfnamefont {A.~E.}\ \bibnamefont {Curzon}},\ }\bibfield
   {title} {\bibinfo {title} {\emph {Charge density wave commensurability in
  {2H}-{TaS\s2} and {Ag\s{x}TaS\s2}}},\ }\href
  {https://doi.org/10.1016/0038-1098(82)90030-8} {\bibfield  {journal}
  {\bibinfo  {journal} {Solid State Commun.}\ }\textbf {\bibinfo {volume}
  {44}},\ \bibinfo {pages} {1455} (\bibinfo {year} {1982})}\BibitemShut
  {NoStop}%
\bibitem [{\citenamefont {Coleman}\ \emph {et~al.}(1988)\citenamefont
  {Coleman}, \citenamefont {Giambattista}, \citenamefont {Hansma},
  \citenamefont {Johnson}, \citenamefont {McNairy},\ and\ \citenamefont
  {Slough}}]{Coleman1988}%
  \BibitemOpen
  \bibfield  {author} {\bibinfo {author} {\bibfnamefont {R.~V.}\ \bibnamefont
  {Coleman}}, \bibinfo {author} {\bibfnamefont {B.}~\bibnamefont
  {Giambattista}}, \bibinfo {author} {\bibfnamefont {P.~K.}\ \bibnamefont
  {Hansma}}, \bibinfo {author} {\bibfnamefont {A.}~\bibnamefont {Johnson}},
  \bibinfo {author} {\bibfnamefont {W.~W.}\ \bibnamefont {McNairy}},\ and\
  \bibinfo {author} {\bibfnamefont {C.~G.}\ \bibnamefont {Slough}},\ }\bibfield
   {title} {\bibinfo {title} {\emph {Scanning tunnelling microscopy of
  charge-density waves in transition metal chalcogenides}},\ }\href
  {https://doi.org/10.1080/00018738800101439} {\bibfield  {journal} {\bibinfo
  {journal} {Adv. Phys.}\ }\textbf {\bibinfo {volume} {37}},\ \bibinfo {pages}
  {559} (\bibinfo {year} {1988})}\BibitemShut {NoStop}%
\bibitem [{\citenamefont {Wang}\ \emph {et~al.}(1991)\citenamefont {Wang},
  \citenamefont {Slough},\ and\ \citenamefont {Coleman}}]{Wang1991}%
  \BibitemOpen
  \bibfield  {author} {\bibinfo {author} {\bibfnamefont {C.}~\bibnamefont
  {Wang}}, \bibinfo {author} {\bibfnamefont {C.~G.}\ \bibnamefont {Slough}},\
  and\ \bibinfo {author} {\bibfnamefont {R.~V.}\ \bibnamefont {Coleman}},\
  }\bibfield  {title} {\bibinfo {title} {\emph {Spectroscopy of dichalcogenides
  and trichalcogenides using scanning tunneling microscopy}},\ }\href
  {https://doi.org/10.1116/1.585257} {\bibfield  {journal} {\bibinfo  {journal}
  {J. Vac. Sci. Technol. B: Microelectron. Process. Phenom.}\ }\textbf
  {\bibinfo {volume} {9}},\ \bibinfo {pages} {1048} (\bibinfo {year}
  {1991})}\BibitemShut {NoStop}%
\bibitem [{\citenamefont {Nagata}\ \emph {et~al.}(1992)\citenamefont {Nagata},
  \citenamefont {Aochi}, \citenamefont {Abe}, \citenamefont {Ebisu},
  \citenamefont {Hagino}, \citenamefont {Seki},\ and\ \citenamefont
  {Tsutsumi}}]{Nagata1992}%
  \BibitemOpen
  \bibfield  {author} {\bibinfo {author} {\bibfnamefont {S.}~\bibnamefont
  {Nagata}}, \bibinfo {author} {\bibfnamefont {T.}~\bibnamefont {Aochi}},
  \bibinfo {author} {\bibfnamefont {T.}~\bibnamefont {Abe}}, \bibinfo {author}
  {\bibfnamefont {S.}~\bibnamefont {Ebisu}}, \bibinfo {author} {\bibfnamefont
  {T.}~\bibnamefont {Hagino}}, \bibinfo {author} {\bibfnamefont
  {Y.}~\bibnamefont {Seki}},\ and\ \bibinfo {author} {\bibfnamefont
  {K.}~\bibnamefont {Tsutsumi}},\ }\bibfield  {title} {\bibinfo {title} {\emph
  {Superconductivity in the layered compound {2H}-{TaS\s2}}},\ }\href
  {https://doi.org/10.1016/0022-3697(92)90242-6} {\bibfield  {journal}
  {\bibinfo  {journal} {J. Phys. Chem. Solids}\ }\textbf {\bibinfo {volume}
  {53}},\ \bibinfo {pages} {1259} (\bibinfo {year} {1992})}\BibitemShut
  {NoStop}%
\bibitem [{\citenamefont {Lin}\ \emph {et~al.}(2018)\citenamefont {Lin},
  \citenamefont {Huang}, \citenamefont {Zhao}, \citenamefont {Lian},
  \citenamefont {Duan}, \citenamefont {Chen},\ and\ \citenamefont
  {Ji}}]{Lin2018}%
  \BibitemOpen
  \bibfield  {author} {\bibinfo {author} {\bibfnamefont {H.}~\bibnamefont
  {Lin}}, \bibinfo {author} {\bibfnamefont {W.}~\bibnamefont {Huang}}, \bibinfo
  {author} {\bibfnamefont {K.}~\bibnamefont {Zhao}}, \bibinfo {author}
  {\bibfnamefont {C.}~\bibnamefont {Lian}}, \bibinfo {author} {\bibfnamefont
  {W.}~\bibnamefont {Duan}}, \bibinfo {author} {\bibfnamefont {X.}~\bibnamefont
  {Chen}},\ and\ \bibinfo {author} {\bibfnamefont {S.-H.}\ \bibnamefont {Ji}},\
  }\bibfield  {title} {\bibinfo {title} {\emph {Growth of atomically thick
  transition metal sulfide films on graphene/{6H}-{SiC}(0001) by molecular beam
  epitaxy}},\ }\href {https://doi.org/10.1007/s12274-018-2054-4} {\bibfield
  {journal} {\bibinfo  {journal} {Nano Res.}\ }\textbf {\bibinfo {volume}
  {11}},\ \bibinfo {pages} {4722} (\bibinfo {year} {2018})}\BibitemShut
  {NoStop}%
\bibitem [{\citenamefont {Hall}\ \emph {et~al.}(2019)\citenamefont {Hall},
  \citenamefont {Ehlen}, \citenamefont {Berges}, \citenamefont {van Loon},
  \citenamefont {van Efferen}, \citenamefont {Murray}, \citenamefont
  {R\"osner}, \citenamefont {Li}, \citenamefont {Senkovskiy}, \citenamefont
  {Hell}, \citenamefont {Rolf}, \citenamefont {Heider}, \citenamefont
  {Asensio}, \citenamefont {Avila}, \citenamefont {Plucinski}, \citenamefont
  {Wehling}, \citenamefont {Gr\"uneis},\ and\ \citenamefont
  {Michely}}]{Hall2019}%
  \BibitemOpen
  \bibfield  {author} {\bibinfo {author} {\bibfnamefont {J.}~\bibnamefont
  {Hall}}, \bibinfo {author} {\bibfnamefont {N.}~\bibnamefont {Ehlen}},
  \bibinfo {author} {\bibfnamefont {J.}~\bibnamefont {Berges}}, \bibinfo
  {author} {\bibfnamefont {E.}~\bibnamefont {van Loon}}, \bibinfo {author}
  {\bibfnamefont {C.}~\bibnamefont {van Efferen}}, \bibinfo {author}
  {\bibfnamefont {C.}~\bibnamefont {Murray}}, \bibinfo {author} {\bibfnamefont
  {M.}~\bibnamefont {R\"osner}}, \bibinfo {author} {\bibfnamefont
  {J.}~\bibnamefont {Li}}, \bibinfo {author} {\bibfnamefont {B.~V.}\
  \bibnamefont {Senkovskiy}}, \bibinfo {author} {\bibfnamefont
  {M.}~\bibnamefont {Hell}}, \bibinfo {author} {\bibfnamefont {M.}~\bibnamefont
  {Rolf}}, \bibinfo {author} {\bibfnamefont {T.}~\bibnamefont {Heider}},
  \bibinfo {author} {\bibfnamefont {M.~C.}\ \bibnamefont {Asensio}}, \bibinfo
  {author} {\bibfnamefont {J.}~\bibnamefont {Avila}}, \bibinfo {author}
  {\bibfnamefont {L.}~\bibnamefont {Plucinski}}, \bibinfo {author}
  {\bibfnamefont {T.}~\bibnamefont {Wehling}}, \bibinfo {author} {\bibfnamefont
  {A.}~\bibnamefont {Gr\"uneis}},\ and\ \bibinfo {author} {\bibfnamefont
  {T.}~\bibnamefont {Michely}},\ }\bibfield  {title} {\bibinfo {title} {\emph
  {Environmental control of charge density wave order in monolayer
  {2H}-{TaS\s2}}},\ }\href {https://doi.org/10.1021/acsnano.9b03419} {\bibfield
   {journal} {\bibinfo  {journal} {ACS Nano}\ }\textbf {\bibinfo {volume}
  {13}},\ \bibinfo {pages} {10210} (\bibinfo {year} {2019})}\BibitemShut
  {NoStop}%
\bibitem [{Note1()}]{Note1}%
  \BibitemOpen
  \bibinfo {note} {Note that in the present formalism, the screened
  electron--phonon coupling \begin {equation*} \protect \tilde g_{\protect \bm
  {q} \nu \protect \bm {k} m n} = \protect \frac 1 {\protect \sqrt {2 \omega
  _{\protect \bm {q} \nu }}} \DOTSB \sum@ \slimits@ _i e_{\protect \bm {q} i
  \nu } \protect \frac 1 {\protect \sqrt {M_i}} \protect \bra {\protect \bm {k}
  {+} \protect \bm {q} m} \protect \frac {\protect \widetilde {\partial
  V}}{\partial u_{\protect \bm {q} i}} \protect \ket {\protect \bm {k} n} \end
  {equation*} depends on the screened potential change $\protect \tilde
  {\partial V}$ but on the \protect \emph {bare} phonon energies $\omega $ and
  eigenvectors $e$.}\BibitemShut {Stop}%
\bibitem [{\citenamefont {Giustino}(2017)}]{Giustino2017}%
  \BibitemOpen
  \bibfield  {author} {\bibinfo {author} {\bibfnamefont {F.}~\bibnamefont
  {Giustino}},\ }\bibfield  {title} {\bibinfo {title} {\emph {Electron-phonon
  interactions from first principles}},\ }\href
  {https://doi.org/10.1103/RevModPhys.89.015003} {\bibfield  {journal}
  {\bibinfo  {journal} {Rev. Mod. Phys.}\ }\textbf {\bibinfo {volume} {89}},\
  \bibinfo {pages} {015003} (\bibinfo {year} {2017})},\ \Eprint
  {https://arxiv.org/abs/1603.06965} {arXiv:1603.06965}\BibitemShut {NoStop}%
\bibitem [{\citenamefont {Giustino}\ \emph {et~al.}(2007)\citenamefont
  {Giustino}, \citenamefont {Cohen},\ and\ \citenamefont
  {Louie}}]{Giustino2007}%
  \BibitemOpen
  \bibfield  {author} {\bibinfo {author} {\bibfnamefont {F.}~\bibnamefont
  {Giustino}}, \bibinfo {author} {\bibfnamefont {M.~L.}\ \bibnamefont
  {Cohen}},\ and\ \bibinfo {author} {\bibfnamefont {S.~G.}\ \bibnamefont
  {Louie}},\ }\bibfield  {title} {\bibinfo {title} {\emph {Electron-phonon
  interaction using {Wannier} functions}},\ }\href
  {https://doi.org/10.1103/PhysRevB.76.165108} {\bibfield  {journal} {\bibinfo
  {journal} {Phys. Rev. B}\ }\textbf {\bibinfo {volume} {76}},\ \bibinfo
  {pages} {165108} (\bibinfo {year} {2007})}\BibitemShut {NoStop}%
\bibitem [{\citenamefont {Albertini}\ \emph {et~al.}(2017)\citenamefont
  {Albertini}, \citenamefont {Liu},\ and\ \citenamefont
  {Calandra}}]{Albertini2017}%
  \BibitemOpen
  \bibfield  {author} {\bibinfo {author} {\bibfnamefont {O.~R.}\ \bibnamefont
  {Albertini}}, \bibinfo {author} {\bibfnamefont {A.~Y.}\ \bibnamefont {Liu}},\
  and\ \bibinfo {author} {\bibfnamefont {M.}~\bibnamefont {Calandra}},\
  }\bibfield  {title} {\bibinfo {title} {\emph {Effect of electron doping on
  lattice instabilities in single-layer {1H}-{TaS\s2}}},\ }\href
  {https://doi.org/10.1103/PhysRevB.95.235121} {\bibfield  {journal} {\bibinfo
  {journal} {Phys. Rev. B}\ }\textbf {\bibinfo {volume} {95}},\ \bibinfo
  {pages} {235121} (\bibinfo {year} {2017})},\ \Eprint
  {https://arxiv.org/abs/1702.08588} {arXiv:1702.08588}\BibitemShut {NoStop}%
\bibitem [{\citenamefont {Varma}\ and\ \citenamefont
  {Weber}(1977)}]{Varma1977}%
  \BibitemOpen
  \bibfield  {author} {\bibinfo {author} {\bibfnamefont {C.~M.}\ \bibnamefont
  {Varma}}\ and\ \bibinfo {author} {\bibfnamefont {W.}~\bibnamefont {Weber}},\
  }\bibfield  {title} {\bibinfo {title} {\emph {Phonon dispersion in transition
  metals}},\ }\href {https://doi.org/10.1103/PhysRevLett.39.1094} {\bibfield
  {journal} {\bibinfo  {journal} {Phys. Rev. Lett.}\ }\textbf {\bibinfo
  {volume} {39}},\ \bibinfo {pages} {1094} (\bibinfo {year}
  {1977})}\BibitemShut {NoStop}%
\bibitem [{\citenamefont {Weber}\ \emph {et~al.}(2013)\citenamefont {Weber},
  \citenamefont {Hott}, \citenamefont {Heid}, \citenamefont {Bohnen},
  \citenamefont {Rosenkranz}, \citenamefont {Castellan}, \citenamefont
  {Osborn}, \citenamefont {Said}, \citenamefont {Leu},\ and\ \citenamefont
  {Reznik}}]{Weber2013}%
  \BibitemOpen
  \bibfield  {author} {\bibinfo {author} {\bibfnamefont {F.}~\bibnamefont
  {Weber}}, \bibinfo {author} {\bibfnamefont {R.}~\bibnamefont {Hott}},
  \bibinfo {author} {\bibfnamefont {R.}~\bibnamefont {Heid}}, \bibinfo {author}
  {\bibfnamefont {K.-P.}\ \bibnamefont {Bohnen}}, \bibinfo {author}
  {\bibfnamefont {S.}~\bibnamefont {Rosenkranz}}, \bibinfo {author}
  {\bibfnamefont {J.-P.}\ \bibnamefont {Castellan}}, \bibinfo {author}
  {\bibfnamefont {R.}~\bibnamefont {Osborn}}, \bibinfo {author} {\bibfnamefont
  {A.~H.}\ \bibnamefont {Said}}, \bibinfo {author} {\bibfnamefont {B.~M.}\
  \bibnamefont {Leu}},\ and\ \bibinfo {author} {\bibfnamefont {D.}~\bibnamefont
  {Reznik}},\ }\bibfield  {title} {\bibinfo {title} {\emph {Optical phonons and
  the soft mode in {2H}-{NbSe\s2}}},\ }\href
  {https://doi.org/10.1103/PhysRevB.87.245111} {\bibfield  {journal} {\bibinfo
  {journal} {Phys. Rev. B}\ }\textbf {\bibinfo {volume} {87}},\ \bibinfo
  {pages} {245111} (\bibinfo {year} {2013})},\ \Eprint
  {https://arxiv.org/abs/1303.6503} {arXiv:1303.6503}\BibitemShut {NoStop}%
\bibitem [{\citenamefont {Maksimov}\ and\ \citenamefont
  {Shulga}(1996)}]{Maksimov1996}%
  \BibitemOpen
  \bibfield  {author} {\bibinfo {author} {\bibfnamefont {E.~G.}\ \bibnamefont
  {Maksimov}}\ and\ \bibinfo {author} {\bibfnamefont {S.~V.}\ \bibnamefont
  {Shulga}},\ }\bibfield  {title} {\bibinfo {title} {\emph {Nonadiabatic
  effects in optical phonon self-energy}},\ }\href
  {https://doi.org/10.1016/0038-1098(95)00745-8} {\bibfield  {journal}
  {\bibinfo  {journal} {Solid State Commun.}\ }\textbf {\bibinfo {volume}
  {97}},\ \bibinfo {pages} {553} (\bibinfo {year} {1996})}\BibitemShut
  {NoStop}%
\bibitem [{\citenamefont {Lazzeri}\ and\ \citenamefont
  {Mauri}(2006)}]{Lazzeri2006}%
  \BibitemOpen
  \bibfield  {author} {\bibinfo {author} {\bibfnamefont {M.}~\bibnamefont
  {Lazzeri}}\ and\ \bibinfo {author} {\bibfnamefont {F.}~\bibnamefont
  {Mauri}},\ }\bibfield  {title} {\bibinfo {title} {\emph {Nonadiabatic {Kohn}
  anomaly in a doped graphene monolayer}},\ }\href
  {https://doi.org/10.1103/PhysRevLett.97.266407} {\bibfield  {journal}
  {\bibinfo  {journal} {Phys. Rev. Lett.}\ }\textbf {\bibinfo {volume} {97}},\
  \bibinfo {pages} {266407} (\bibinfo {year} {2006})},\ \Eprint
  {https://arxiv.org/abs/cond-mat/0611708} {arXiv:cond-mat/0611708}\BibitemShut
  {NoStop}%
\bibitem [{\citenamefont {Caudal}\ \emph {et~al.}(2007)\citenamefont {Caudal},
  \citenamefont {Saitta}, \citenamefont {Lazzeri},\ and\ \citenamefont
  {Mauri}}]{Caudal2007}%
  \BibitemOpen
  \bibfield  {author} {\bibinfo {author} {\bibfnamefont {N.}~\bibnamefont
  {Caudal}}, \bibinfo {author} {\bibfnamefont {A.~M.}\ \bibnamefont {Saitta}},
  \bibinfo {author} {\bibfnamefont {M.}~\bibnamefont {Lazzeri}},\ and\ \bibinfo
  {author} {\bibfnamefont {F.}~\bibnamefont {Mauri}},\ }\bibfield  {title}
  {\bibinfo {title} {\emph {Kohn anomalies and nonadiabaticity in doped carbon
  nanotubes}},\ }\href {https://doi.org/10.1103/PhysRevB.75.115423} {\bibfield
  {journal} {\bibinfo  {journal} {Phys. Rev. B}\ }\textbf {\bibinfo {volume}
  {75}},\ \bibinfo {pages} {115423} (\bibinfo {year} {2007})}\BibitemShut
  {NoStop}%
\bibitem [{\citenamefont {Pisana}\ \emph {et~al.}(2007)\citenamefont {Pisana},
  \citenamefont {Lazzeri}, \citenamefont {Casiraghi}, \citenamefont
  {Novoselov}, \citenamefont {Geim}, \citenamefont {Ferrari},\ and\
  \citenamefont {Mauri}}]{Pisana2007}%
  \BibitemOpen
  \bibfield  {author} {\bibinfo {author} {\bibfnamefont {S.}~\bibnamefont
  {Pisana}}, \bibinfo {author} {\bibfnamefont {M.}~\bibnamefont {Lazzeri}},
  \bibinfo {author} {\bibfnamefont {C.}~\bibnamefont {Casiraghi}}, \bibinfo
  {author} {\bibfnamefont {K.~S.}\ \bibnamefont {Novoselov}}, \bibinfo {author}
  {\bibfnamefont {A.~K.}\ \bibnamefont {Geim}}, \bibinfo {author}
  {\bibfnamefont {A.~C.}\ \bibnamefont {Ferrari}},\ and\ \bibinfo {author}
  {\bibfnamefont {F.}~\bibnamefont {Mauri}},\ }\bibfield  {title} {\bibinfo
  {title} {\emph {Breakdown of the adiabatic {Born}--{Oppenheimer}
  approximation in graphene}},\ }\href {https://doi.org/10.1038/nmat1846}
  {\bibfield  {journal} {\bibinfo  {journal} {Nat. Mater.}\ }\textbf {\bibinfo
  {volume} {6}},\ \bibinfo {pages} {198} (\bibinfo {year} {2007})},\ \Eprint
  {https://arxiv.org/abs/cond-mat/0611714} {arXiv:cond-mat/0611714}\BibitemShut
  {NoStop}%
\bibitem [{\citenamefont {Piscanec}\ \emph {et~al.}(2007)\citenamefont
  {Piscanec}, \citenamefont {Lazzeri}, \citenamefont {Robertson}, \citenamefont
  {Ferrari},\ and\ \citenamefont {Mauri}}]{Piscanec2007}%
  \BibitemOpen
  \bibfield  {author} {\bibinfo {author} {\bibfnamefont {S.}~\bibnamefont
  {Piscanec}}, \bibinfo {author} {\bibfnamefont {M.}~\bibnamefont {Lazzeri}},
  \bibinfo {author} {\bibfnamefont {J.}~\bibnamefont {Robertson}}, \bibinfo
  {author} {\bibfnamefont {A.~C.}\ \bibnamefont {Ferrari}},\ and\ \bibinfo
  {author} {\bibfnamefont {F.}~\bibnamefont {Mauri}},\ }\bibfield  {title}
  {\bibinfo {title} {\emph {Optical phonons in carbon nanotubes: Kohn
  anomalies, {Peierls} distortions, and dynamic effects}},\ }\href
  {https://doi.org/10.1103/PhysRevB.75.035427} {\bibfield  {journal} {\bibinfo
  {journal} {Phys. Rev. B}\ }\textbf {\bibinfo {volume} {75}},\ \bibinfo
  {pages} {035427} (\bibinfo {year} {2007})},\ \Eprint
  {https://arxiv.org/abs/cond-mat/0611693} {arXiv:cond-mat/0611693}\BibitemShut
  {NoStop}%
\bibitem [{\citenamefont {Saitta}\ \emph {et~al.}(2008)\citenamefont {Saitta},
  \citenamefont {Lazzeri}, \citenamefont {Calandra},\ and\ \citenamefont
  {Mauri}}]{Saitta2008}%
  \BibitemOpen
  \bibfield  {author} {\bibinfo {author} {\bibfnamefont {A.~M.}\ \bibnamefont
  {Saitta}}, \bibinfo {author} {\bibfnamefont {M.}~\bibnamefont {Lazzeri}},
  \bibinfo {author} {\bibfnamefont {M.}~\bibnamefont {Calandra}},\ and\
  \bibinfo {author} {\bibfnamefont {F.}~\bibnamefont {Mauri}},\ }\bibfield
  {title} {\bibinfo {title} {\emph {Giant nonadiabatic effects in layer metals:
  Raman spectra of intercalated graphite explained}},\ }\href
  {https://doi.org/10.1103/PhysRevLett.100.226401} {\bibfield  {journal}
  {\bibinfo  {journal} {Phys. Rev. Lett.}\ }\textbf {\bibinfo {volume} {100}},\
  \bibinfo {pages} {226401} (\bibinfo {year} {2008})},\ \Eprint
  {https://arxiv.org/abs/0802.4426} {arXiv:0802.4426}\BibitemShut {NoStop}%
\bibitem [{\citenamefont {Calandra}\ \emph {et~al.}(2010)\citenamefont
  {Calandra}, \citenamefont {Profeta},\ and\ \citenamefont
  {Mauri}}]{Calandra2010}%
  \BibitemOpen
  \bibfield  {author} {\bibinfo {author} {\bibfnamefont {M.}~\bibnamefont
  {Calandra}}, \bibinfo {author} {\bibfnamefont {G.}~\bibnamefont {Profeta}},\
  and\ \bibinfo {author} {\bibfnamefont {F.}~\bibnamefont {Mauri}},\ }\bibfield
   {title} {\bibinfo {title} {\emph {Adiabatic and nonadiabatic phonon
  dispersion in a {Wannier} function approach}},\ }\href
  {https://doi.org/10.1103/PhysRevB.82.165111} {\bibfield  {journal} {\bibinfo
  {journal} {Phys. Rev. B}\ }\textbf {\bibinfo {volume} {82}},\ \bibinfo
  {pages} {165111} (\bibinfo {year} {2010})},\ \Eprint
  {https://arxiv.org/abs/1007.2098} {arXiv:1007.2098}\BibitemShut {NoStop}%
\bibitem [{Note2()}]{Note2}%
  \BibitemOpen
  \bibinfo {note} {From the fact that the phonon self-energy $\Pi _{\mu \nu }$
  is Hermitian, it follows immediately that the result of Eq.~\protect \textup
  {\hbox {\mathsurround \z@ \protect \normalfont (\ignorespaces \ref
  {eq:Pi}\unskip \@@italiccorr )}} will not change if we replace $g_\mu ^*
  \protect \tilde g_\nu ^{\protect \phantom *}$ by $\protect \tilde g_\mu ^*
  g_\nu ^{\protect \phantom *}$ or, as a consequence, the right-hand side of
  Eq.~\protect \textup {\hbox {\mathsurround \z@ \protect \normalfont
  (\ignorespaces \ref {eq:g2}\unskip \@@italiccorr )}}. All anti-Hermitian
  parts of the summands in Eq.~\protect \textup {\hbox {\mathsurround \z@
  \protect \normalfont (\ignorespaces \ref {eq:Pi}\unskip \@@italiccorr )}}
  cancel.}\BibitemShut {Stop}%
\bibitem [{Note3()}]{Note3}%
  \BibitemOpen
  \bibinfo {note} {The information contained in $\partial _\Delta \Pi (\Delta
  )$ is similar to the information contained in $\protect \mathrm {Im} \protect
  \tmspace +\thinmuskip {.1667em} \Pi (\omega + \protect \mathrm
  i0^+)$.}\BibitemShut {Stop}%
\bibitem [{\citenamefont {Peierls}(1955)}]{Peierls1955}%
  \BibitemOpen
  \bibfield  {author} {\bibinfo {author} {\bibfnamefont {R.~E.}\ \bibnamefont
  {Peierls}},\ }\href
  {https://doi.org/10.1093/acprof:oso/9780198507819.001.0001} {\emph {\bibinfo
  {title} {Quantum Theory of Solids}}}\ (\bibinfo  {publisher} {Oxford
  University Press},\ \bibinfo {address} {London},\ \bibinfo {year}
  {1955})\BibitemShut {NoStop}%
\bibitem [{\citenamefont {Roth}\ \emph {et~al.}(1966)\citenamefont {Roth},
  \citenamefont {Zeiger},\ and\ \citenamefont {Kaplan}}]{Roth1966}%
  \BibitemOpen
  \bibfield  {author} {\bibinfo {author} {\bibfnamefont {L.~M.}\ \bibnamefont
  {Roth}}, \bibinfo {author} {\bibfnamefont {H.~J.}\ \bibnamefont {Zeiger}},\
  and\ \bibinfo {author} {\bibfnamefont {T.~A.}\ \bibnamefont {Kaplan}},\
  }\bibfield  {title} {\bibinfo {title} {\emph {Generalization of the
  {Ruderman}-{Kittel}-{Kasuya}-{Yosida} interaction for nonspherical {Fermi}
  surfaces}},\ }\href {https://doi.org/10.1103/PhysRev.149.519} {\bibfield
  {journal} {\bibinfo  {journal} {Phys. Rev.}\ }\textbf {\bibinfo {volume}
  {149}},\ \bibinfo {pages} {519} (\bibinfo {year} {1966})}\BibitemShut
  {NoStop}%
\bibitem [{\citenamefont {Varma}\ \emph {et~al.}(1979)\citenamefont {Varma},
  \citenamefont {Blount}, \citenamefont {Vashishta},\ and\ \citenamefont
  {Weber}}]{Varma1979}%
  \BibitemOpen
  \bibfield  {author} {\bibinfo {author} {\bibfnamefont {C.~M.}\ \bibnamefont
  {Varma}}, \bibinfo {author} {\bibfnamefont {E.~I.}\ \bibnamefont {Blount}},
  \bibinfo {author} {\bibfnamefont {P.}~\bibnamefont {Vashishta}},\ and\
  \bibinfo {author} {\bibfnamefont {W.}~\bibnamefont {Weber}},\ }\bibfield
  {title} {\bibinfo {title} {\emph {Electron-phonon interactions in transition
  metals}},\ }\href {https://doi.org/10.1103/PhysRevB.19.6130} {\bibfield
  {journal} {\bibinfo  {journal} {Phys. Rev. B}\ }\textbf {\bibinfo {volume}
  {19}},\ \bibinfo {pages} {6130} (\bibinfo {year} {1979})}\BibitemShut
  {NoStop}%
\bibitem [{\citenamefont {Di~Salvo}\ \emph {et~al.}(1975)\citenamefont
  {Di~Salvo}, \citenamefont {Wilson}, \citenamefont {Bagley},\ and\
  \citenamefont {Waszczak}}]{DiSalvo1975}%
  \BibitemOpen
  \bibfield  {author} {\bibinfo {author} {\bibfnamefont {F.~J.}\ \bibnamefont
  {Di~Salvo}}, \bibinfo {author} {\bibfnamefont {J.~A.}\ \bibnamefont
  {Wilson}}, \bibinfo {author} {\bibfnamefont {B.~G.}\ \bibnamefont {Bagley}},\
  and\ \bibinfo {author} {\bibfnamefont {J.~V.}\ \bibnamefont {Waszczak}},\
  }\bibfield  {title} {\bibinfo {title} {\emph {Effects of doping on
  charge-density waves in layer compounds}},\ }\href
  {https://doi.org/10.1103/PhysRevB.12.2220} {\bibfield  {journal} {\bibinfo
  {journal} {Phys. Rev. B}\ }\textbf {\bibinfo {volume} {12}},\ \bibinfo
  {pages} {2220} (\bibinfo {year} {1975})}\BibitemShut {NoStop}%
\bibitem [{\citenamefont {Chen}\ \emph {et~al.}(2015)\citenamefont {Chen},
  \citenamefont {Miller}, \citenamefont {Nugroho}, \citenamefont {de~la
  Pe\~na}, \citenamefont {Joe}, \citenamefont {Kogar}, \citenamefont {Brock},
  \citenamefont {Geck}, \citenamefont {MacDougall}, \citenamefont {Cooper},
  \citenamefont {Fradkin}, \citenamefont {Van~Harlingen},\ and\ \citenamefont
  {Abbamonte}}]{Chen2015}%
  \BibitemOpen
  \bibfield  {author} {\bibinfo {author} {\bibfnamefont {X.~M.}\ \bibnamefont
  {Chen}}, \bibinfo {author} {\bibfnamefont {A.~J.}\ \bibnamefont {Miller}},
  \bibinfo {author} {\bibfnamefont {C.}~\bibnamefont {Nugroho}}, \bibinfo
  {author} {\bibfnamefont {G.~A.}\ \bibnamefont {de~la Pe\~na}}, \bibinfo
  {author} {\bibfnamefont {Y.~I.}\ \bibnamefont {Joe}}, \bibinfo {author}
  {\bibfnamefont {A.}~\bibnamefont {Kogar}}, \bibinfo {author} {\bibfnamefont
  {J.~D.}\ \bibnamefont {Brock}}, \bibinfo {author} {\bibfnamefont
  {J.}~\bibnamefont {Geck}}, \bibinfo {author} {\bibfnamefont {G.~J.}\
  \bibnamefont {MacDougall}}, \bibinfo {author} {\bibfnamefont {S.~L.}\
  \bibnamefont {Cooper}}, \bibinfo {author} {\bibfnamefont {E.}~\bibnamefont
  {Fradkin}}, \bibinfo {author} {\bibfnamefont {D.~J.}\ \bibnamefont
  {Van~Harlingen}},\ and\ \bibinfo {author} {\bibfnamefont {P.}~\bibnamefont
  {Abbamonte}},\ }\bibfield  {title} {\bibinfo {title} {\emph {Influence of
  {Ti} doping on the incommensurate charge density wave in {1T}-{TaS\s2}}},\
  }\href {https://doi.org/10.1103/PhysRevB.91.245113} {\bibfield  {journal}
  {\bibinfo  {journal} {Phys. Rev. B}\ }\textbf {\bibinfo {volume} {91}},\
  \bibinfo {pages} {245113} (\bibinfo {year} {2015})}\BibitemShut {NoStop}%
\bibitem [{\citenamefont {Yu}\ \emph {et~al.}(2015)\citenamefont {Yu},
  \citenamefont {Yang}, \citenamefont {Lu}, \citenamefont {Yan}, \citenamefont
  {Cho}, \citenamefont {Ma}, \citenamefont {Niu}, \citenamefont {Kim},
  \citenamefont {Son}, \citenamefont {Feng}, \citenamefont {Li}, \citenamefont
  {Cheong}, \citenamefont {Chen},\ and\ \citenamefont {Zhang}}]{Yu2015}%
  \BibitemOpen
  \bibfield  {author} {\bibinfo {author} {\bibfnamefont {Y.}~\bibnamefont
  {Yu}}, \bibinfo {author} {\bibfnamefont {F.}~\bibnamefont {Yang}}, \bibinfo
  {author} {\bibfnamefont {X.~F.}\ \bibnamefont {Lu}}, \bibinfo {author}
  {\bibfnamefont {Y.~J.}\ \bibnamefont {Yan}}, \bibinfo {author} {\bibfnamefont
  {Y.-H.}\ \bibnamefont {Cho}}, \bibinfo {author} {\bibfnamefont
  {L.}~\bibnamefont {Ma}}, \bibinfo {author} {\bibfnamefont {X.}~\bibnamefont
  {Niu}}, \bibinfo {author} {\bibfnamefont {S.}~\bibnamefont {Kim}}, \bibinfo
  {author} {\bibfnamefont {Y.-W.}\ \bibnamefont {Son}}, \bibinfo {author}
  {\bibfnamefont {D.}~\bibnamefont {Feng}}, \bibinfo {author} {\bibfnamefont
  {S.}~\bibnamefont {Li}}, \bibinfo {author} {\bibfnamefont {S.-W.}\
  \bibnamefont {Cheong}}, \bibinfo {author} {\bibfnamefont {X.~H.}\
  \bibnamefont {Chen}},\ and\ \bibinfo {author} {\bibfnamefont
  {Y.}~\bibnamefont {Zhang}},\ }\bibfield  {title} {\bibinfo {title} {\emph
  {Gate-tunable phase transitions in thin flakes of {1T}-{TaS\s2}}},\ }\href
  {https://doi.org/10.1038/nnano.2014.323} {\bibfield  {journal} {\bibinfo
  {journal} {Nat. Nanotechnol.}\ }\textbf {\bibinfo {volume} {10}},\ \bibinfo
  {pages} {270} (\bibinfo {year} {2015})},\ \Eprint
  {https://arxiv.org/abs/1407.3480} {arXiv:1407.3480}\BibitemShut {NoStop}%
\bibitem [{\citenamefont {Shao}\ \emph {et~al.}(2016)\citenamefont {Shao},
  \citenamefont {Xiao}, \citenamefont {Lu}, \citenamefont {Lv}, \citenamefont
  {Li}, \citenamefont {Zhu},\ and\ \citenamefont {Sun}}]{Shao2016}%
  \BibitemOpen
  \bibfield  {author} {\bibinfo {author} {\bibfnamefont {D.~F.}\ \bibnamefont
  {Shao}}, \bibinfo {author} {\bibfnamefont {R.~C.}\ \bibnamefont {Xiao}},
  \bibinfo {author} {\bibfnamefont {W.~J.}\ \bibnamefont {Lu}}, \bibinfo
  {author} {\bibfnamefont {H.~Y.}\ \bibnamefont {Lv}}, \bibinfo {author}
  {\bibfnamefont {J.~Y.}\ \bibnamefont {Li}}, \bibinfo {author} {\bibfnamefont
  {X.~B.}\ \bibnamefont {Zhu}},\ and\ \bibinfo {author} {\bibfnamefont {Y.~P.}\
  \bibnamefont {Sun}},\ }\bibfield  {title} {\bibinfo {title} {\emph
  {Manipulating charge density waves in {1T}-{TaS\s2} by charge-carrier doping:
  A first-principles investigation}},\ }\href
  {https://doi.org/10.1103/PhysRevB.94.125126} {\bibfield  {journal} {\bibinfo
  {journal} {Phys. Rev. B}\ }\textbf {\bibinfo {volume} {94}},\ \bibinfo
  {pages} {125126} (\bibinfo {year} {2016})},\ \Eprint
  {https://arxiv.org/abs/1512.06553} {arXiv:1512.06553}\BibitemShut {NoStop}%
\bibitem [{\citenamefont {Sanders}\ \emph {et~al.}(2016)\citenamefont
  {Sanders}, \citenamefont {Dendzik}, \citenamefont {Ngankeu}, \citenamefont
  {Eich}, \citenamefont {Bruix}, \citenamefont {Bianchi}, \citenamefont {Miwa},
  \citenamefont {Hammer}, \citenamefont {Khajetoorians},\ and\ \citenamefont
  {Hofmann}}]{Sanders2016}%
  \BibitemOpen
  \bibfield  {author} {\bibinfo {author} {\bibfnamefont {C.~E.}\ \bibnamefont
  {Sanders}}, \bibinfo {author} {\bibfnamefont {M.}~\bibnamefont {Dendzik}},
  \bibinfo {author} {\bibfnamefont {A.~S.}\ \bibnamefont {Ngankeu}}, \bibinfo
  {author} {\bibfnamefont {A.}~\bibnamefont {Eich}}, \bibinfo {author}
  {\bibfnamefont {A.}~\bibnamefont {Bruix}}, \bibinfo {author} {\bibfnamefont
  {M.}~\bibnamefont {Bianchi}}, \bibinfo {author} {\bibfnamefont {J.~A.}\
  \bibnamefont {Miwa}}, \bibinfo {author} {\bibfnamefont {B.}~\bibnamefont
  {Hammer}}, \bibinfo {author} {\bibfnamefont {A.~A.}\ \bibnamefont
  {Khajetoorians}},\ and\ \bibinfo {author} {\bibfnamefont {P.}~\bibnamefont
  {Hofmann}},\ }\bibfield  {title} {\bibinfo {title} {\emph {Crystalline and
  electronic structure of single-layer {TaS\s2}}},\ }\href
  {https://doi.org/10.1103/PhysRevB.94.081404} {\bibfield  {journal} {\bibinfo
  {journal} {Phys. Rev. B}\ }\textbf {\bibinfo {volume} {94}},\ \bibinfo
  {pages} {081404(R)} (\bibinfo {year} {2016})},\ \Eprint
  {https://arxiv.org/abs/1606.05856} {arXiv:1606.05856}\BibitemShut {NoStop}%
\bibitem [{\citenamefont {Shao}\ \emph {et~al.}(2019)\citenamefont {Shao},
  \citenamefont {Eich}, \citenamefont {Sanders}, \citenamefont {Ngankeu},
  \citenamefont {Bianchi}, \citenamefont {Hofmann}, \citenamefont
  {Khajetoorians},\ and\ \citenamefont {Wehling}}]{Shao2019}%
  \BibitemOpen
  \bibfield  {author} {\bibinfo {author} {\bibfnamefont {B.}~\bibnamefont
  {Shao}}, \bibinfo {author} {\bibfnamefont {A.}~\bibnamefont {Eich}}, \bibinfo
  {author} {\bibfnamefont {C.}~\bibnamefont {Sanders}}, \bibinfo {author}
  {\bibfnamefont {A.~S.}\ \bibnamefont {Ngankeu}}, \bibinfo {author}
  {\bibfnamefont {M.}~\bibnamefont {Bianchi}}, \bibinfo {author} {\bibfnamefont
  {P.}~\bibnamefont {Hofmann}}, \bibinfo {author} {\bibfnamefont {A.~A.}\
  \bibnamefont {Khajetoorians}},\ and\ \bibinfo {author} {\bibfnamefont
  {T.~O.}\ \bibnamefont {Wehling}},\ }\bibfield  {title} {\bibinfo {title}
  {\emph {Pseudodoping of a metallic two-dimensional material by the supporting
  substrate}},\ }\href {https://doi.org/10.1038/s41467-018-08088-8} {\bibfield
  {journal} {\bibinfo  {journal} {Nat. Commun.}\ }\textbf {\bibinfo {volume}
  {10}},\ \bibinfo {pages} {1} (\bibinfo {year} {2019})},\ \Eprint
  {https://arxiv.org/abs/1807.00756} {arXiv:1807.00756}\BibitemShut {NoStop}%
\bibitem [{\citenamefont {Doran}(1978)}]{Doran1978}%
  \BibitemOpen
  \bibfield  {author} {\bibinfo {author} {\bibfnamefont {N.~J.}\ \bibnamefont
  {Doran}},\ }\bibfield  {title} {\bibinfo {title} {\emph {A calculation of the
  electronic response function in {2H}-{NbSe\s2} including electron-phonon
  matrix element effects}},\ }\href
  {https://doi.org/10.1088/0022-3719/11/24/005} {\bibfield  {journal} {\bibinfo
   {journal} {J. Phys. C: Solid State Phys.}\ }\textbf {\bibinfo {volume}
  {11}},\ \bibinfo {pages} {L959} (\bibinfo {year} {1978})}\BibitemShut
  {NoStop}%
\bibitem [{\citenamefont {Castro~Neto}(2001)}]{CastroNeto2001}%
  \BibitemOpen
  \bibfield  {author} {\bibinfo {author} {\bibfnamefont {A.~H.}\ \bibnamefont
  {Castro~Neto}},\ }\bibfield  {title} {\bibinfo {title} {\emph {Charge density
  wave, superconductivity, and anomalous metallic behavior in {2D} transition
  metal dichalcogenides}},\ }\href
  {https://doi.org/10.1103/PhysRevLett.86.4382} {\bibfield  {journal} {\bibinfo
   {journal} {Phys. Rev. Lett.}\ }\textbf {\bibinfo {volume} {86}},\ \bibinfo
  {pages} {4382} (\bibinfo {year} {2001})},\ \Eprint
  {https://arxiv.org/abs/cond-mat/0012147} {arXiv:cond-mat/0012147}\BibitemShut
  {NoStop}%
\bibitem [{\citenamefont {Johannes}\ and\ \citenamefont
  {Mazin}(2008)}]{Johannes2008}%
  \BibitemOpen
  \bibfield  {author} {\bibinfo {author} {\bibfnamefont {M.~D.}\ \bibnamefont
  {Johannes}}\ and\ \bibinfo {author} {\bibfnamefont {I.~I.}\ \bibnamefont
  {Mazin}},\ }\bibfield  {title} {\bibinfo {title} {\emph {Fermi surface
  nesting and the origin of charge density waves in metals}},\ }\href
  {https://doi.org/10.1103/PhysRevB.77.165135} {\bibfield  {journal} {\bibinfo
  {journal} {Phys. Rev. B}\ }\textbf {\bibinfo {volume} {77}},\ \bibinfo
  {pages} {165135} (\bibinfo {year} {2008})},\ \Eprint
  {https://arxiv.org/abs/0708.1744} {arXiv:0708.1744}\BibitemShut {NoStop}%
\bibitem [{\citenamefont {Ge}\ and\ \citenamefont {Liu}(2012)}]{Ge2012}%
  \BibitemOpen
  \bibfield  {author} {\bibinfo {author} {\bibfnamefont {Y.}~\bibnamefont
  {Ge}}\ and\ \bibinfo {author} {\bibfnamefont {A.~Y.}\ \bibnamefont {Liu}},\
  }\bibfield  {title} {\bibinfo {title} {\emph {Effect of dimensionality and
  spin-orbit coupling on charge-density-wave transition in {2H}-{TaSe\s2}}},\
  }\href {https://doi.org/10.1103/PhysRevB.86.104101} {\bibfield  {journal}
  {\bibinfo  {journal} {Phys. Rev. B}\ }\textbf {\bibinfo {volume} {86}},\
  \bibinfo {pages} {104101} (\bibinfo {year} {2012})}\BibitemShut {NoStop}%
\bibitem [{\citenamefont {Xiao}\ \emph {et~al.}(2012)\citenamefont {Xiao},
  \citenamefont {Liu}, \citenamefont {Feng}, \citenamefont {Xu},\ and\
  \citenamefont {Yao}}]{Xiao2012}%
  \BibitemOpen
  \bibfield  {author} {\bibinfo {author} {\bibfnamefont {D.}~\bibnamefont
  {Xiao}}, \bibinfo {author} {\bibfnamefont {G.-B.}\ \bibnamefont {Liu}},
  \bibinfo {author} {\bibfnamefont {W.}~\bibnamefont {Feng}}, \bibinfo {author}
  {\bibfnamefont {X.}~\bibnamefont {Xu}},\ and\ \bibinfo {author}
  {\bibfnamefont {W.}~\bibnamefont {Yao}},\ }\bibfield  {title} {\bibinfo
  {title} {\emph {Coupled spin and valley physics in monolayers of {MoS\s2} and
  other group-{VI} dichalcogenides}},\ }\href
  {https://doi.org/10.1103/PhysRevLett.108.196802} {\bibfield  {journal}
  {\bibinfo  {journal} {Phys. Rev. Lett.}\ }\textbf {\bibinfo {volume} {108}},\
  \bibinfo {pages} {196802} (\bibinfo {year} {2012})},\ \Eprint
  {https://arxiv.org/abs/1112.3144} {arXiv:1112.3144}\BibitemShut {NoStop}%
\bibitem [{Note4()}]{Note4}%
  \BibitemOpen
  \bibinfo {note} {It is instructive to rewrite the valley Hamiltonians and
  velocity operators in polar coordinates: \begin {align*} H_{p \phi } &= v_0 p
  \protect \tmspace +\thinmuskip {.1667em} \protect \bm {e}_p \cdot \protect
  \bm {\sigma }_\tau + \protect \frac \Delta 2 \sigma _z, \\ \protect \bm
  {v}_{p \phi } &= v_0 \left [ (\protect \bm {e}_p \cdot \protect \bm {\sigma
  }_\tau ) \protect \bm {e}_p + (\protect \bm {e}_\phi \cdot \protect \bm
  {\sigma }_\tau ) \protect \bm {e}_\phi \right ], \end {align*} where
  $\protect \bm {\sigma }_\tau = (\tau \sigma _x, \sigma _y)$, $p$ ($\phi $) is
  the radial (angular) momentum coordinate, and $\protect \bm {e}_p$ ($\protect
  \bm {e}_\phi $) are the corresponding unit vectors in the radial (angular)
  direction. The equal-energy contours are circles with constant $p$. We see
  that $\protect \bm {v}_{p \phi }$ has contributions perpendicular ($\protect
  \bm {e}_p$ direction) and parallel ($\protect \bm {e}_\phi $ direction) to
  the equal-energy contours.}\BibitemShut {Stop}%
\bibitem [{\citenamefont {Wilson}(1978)}]{Wilson1978}%
  \BibitemOpen
  \bibfield  {author} {\bibinfo {author} {\bibfnamefont {J.~A.}\ \bibnamefont
  {Wilson}},\ }\bibfield  {title} {\bibinfo {title} {\emph {Modelling the
  contrasting semimetallic characters of {TiS\s2} and {TiSe\s2}}},\ }\href
  {https://doi.org/10.1002/pssb.2220860102} {\bibfield  {journal} {\bibinfo
  {journal} {Phys. Status Solidi B}\ }\textbf {\bibinfo {volume} {86}},\
  \bibinfo {pages} {11} (\bibinfo {year} {1978})}\BibitemShut {NoStop}%
\bibitem [{\citenamefont {Rossnagel}\ \emph {et~al.}(2002)\citenamefont
  {Rossnagel}, \citenamefont {Kipp},\ and\ \citenamefont
  {Skibowski}}]{Rossnagel2002}%
  \BibitemOpen
  \bibfield  {author} {\bibinfo {author} {\bibfnamefont {K.}~\bibnamefont
  {Rossnagel}}, \bibinfo {author} {\bibfnamefont {L.}~\bibnamefont {Kipp}},\
  and\ \bibinfo {author} {\bibfnamefont {M.}~\bibnamefont {Skibowski}},\
  }\bibfield  {title} {\bibinfo {title} {\emph {Charge-density-wave phase
  transition in {1T}-{TiSe\s2}: Excitonic insulator versus band-type
  {Jahn}-{Teller} mechanism}},\ }\href
  {https://doi.org/10.1103/PhysRevB.65.235101} {\bibfield  {journal} {\bibinfo
  {journal} {Phys. Rev. B}\ }\textbf {\bibinfo {volume} {65}},\ \bibinfo
  {pages} {235101} (\bibinfo {year} {2002})}\BibitemShut {NoStop}%
\bibitem [{\citenamefont {Cercellier}\ \emph {et~al.}(2007)\citenamefont
  {Cercellier}, \citenamefont {Monney}, \citenamefont {Clerc}, \citenamefont
  {Battaglia}, \citenamefont {Despont}, \citenamefont {Garnier}, \citenamefont
  {Beck}, \citenamefont {Aebi}, \citenamefont {Patthey}, \citenamefont
  {Berger},\ and\ \citenamefont {Forr\'o}}]{Cercellier2007}%
  \BibitemOpen
  \bibfield  {author} {\bibinfo {author} {\bibfnamefont {H.}~\bibnamefont
  {Cercellier}}, \bibinfo {author} {\bibfnamefont {C.}~\bibnamefont {Monney}},
  \bibinfo {author} {\bibfnamefont {F.}~\bibnamefont {Clerc}}, \bibinfo
  {author} {\bibfnamefont {C.}~\bibnamefont {Battaglia}}, \bibinfo {author}
  {\bibfnamefont {L.}~\bibnamefont {Despont}}, \bibinfo {author} {\bibfnamefont
  {M.~G.}\ \bibnamefont {Garnier}}, \bibinfo {author} {\bibfnamefont
  {H.}~\bibnamefont {Beck}}, \bibinfo {author} {\bibfnamefont {P.}~\bibnamefont
  {Aebi}}, \bibinfo {author} {\bibfnamefont {L.}~\bibnamefont {Patthey}},
  \bibinfo {author} {\bibfnamefont {H.}~\bibnamefont {Berger}},\ and\ \bibinfo
  {author} {\bibfnamefont {L.}~\bibnamefont {Forr\'o}},\ }\bibfield  {title}
  {\bibinfo {title} {\emph {Evidence for an excitonic insulator phase in
  {1T}-{TiSe\s2}}},\ }\href {https://doi.org/10.1103/PhysRevLett.99.146403}
  {\bibfield  {journal} {\bibinfo  {journal} {Phys. Rev. Lett.}\ }\textbf
  {\bibinfo {volume} {99}},\ \bibinfo {pages} {146403} (\bibinfo {year}
  {2007})},\ \Eprint {https://arxiv.org/abs/0704.0159}
  {arXiv:0704.0159}\BibitemShut {NoStop}%
\bibitem [{\citenamefont {van Wezel}\ \emph {et~al.}(2010)\citenamefont {van
  Wezel}, \citenamefont {Nahai-Williamson},\ and\ \citenamefont
  {Saxena}}]{vanWezel2010}%
  \BibitemOpen
  \bibfield  {author} {\bibinfo {author} {\bibfnamefont {J.}~\bibnamefont {van
  Wezel}}, \bibinfo {author} {\bibfnamefont {P.}~\bibnamefont
  {Nahai-Williamson}},\ and\ \bibinfo {author} {\bibfnamefont {S.~S.}\
  \bibnamefont {Saxena}},\ }\bibfield  {title} {\bibinfo {title} {\emph
  {Exciton-phonon-driven charge density wave in {TiSe\s2}}},\ }\href
  {https://doi.org/10.1103/PhysRevB.81.165109} {\bibfield  {journal} {\bibinfo
  {journal} {Phys. Rev. B}\ }\textbf {\bibinfo {volume} {81}},\ \bibinfo
  {pages} {165109} (\bibinfo {year} {2010})}\BibitemShut {NoStop}%
\bibitem [{\citenamefont {Monney}\ \emph {et~al.}(2011)\citenamefont {Monney},
  \citenamefont {Battaglia}, \citenamefont {Cercellier}, \citenamefont {Aebi},\
  and\ \citenamefont {Beck}}]{Monney2011}%
  \BibitemOpen
  \bibfield  {author} {\bibinfo {author} {\bibfnamefont {C.}~\bibnamefont
  {Monney}}, \bibinfo {author} {\bibfnamefont {C.}~\bibnamefont {Battaglia}},
  \bibinfo {author} {\bibfnamefont {H.}~\bibnamefont {Cercellier}}, \bibinfo
  {author} {\bibfnamefont {P.}~\bibnamefont {Aebi}},\ and\ \bibinfo {author}
  {\bibfnamefont {H.}~\bibnamefont {Beck}},\ }\bibfield  {title} {\bibinfo
  {title} {\emph {Exciton condensation driving the periodic lattice distortion
  of {1T}-{TiSe\s2}}},\ }\href {https://doi.org/10.1103/PhysRevLett.106.106404}
  {\bibfield  {journal} {\bibinfo  {journal} {Phys. Rev. Lett.}\ }\textbf
  {\bibinfo {volume} {106}},\ \bibinfo {pages} {106404} (\bibinfo {year}
  {2011})},\ \Eprint {https://arxiv.org/abs/1012.1119}
  {arXiv:1012.1119}\BibitemShut {NoStop}%
\bibitem [{\citenamefont {Chen}\ \emph {et~al.}(2018)\citenamefont {Chen},
  \citenamefont {Singh}, \citenamefont {Lin},\ and\ \citenamefont
  {Pereira}}]{Chen2018}%
  \BibitemOpen
  \bibfield  {author} {\bibinfo {author} {\bibfnamefont {C.}~\bibnamefont
  {Chen}}, \bibinfo {author} {\bibfnamefont {B.}~\bibnamefont {Singh}},
  \bibinfo {author} {\bibfnamefont {H.}~\bibnamefont {Lin}},\ and\ \bibinfo
  {author} {\bibfnamefont {V.~M.}\ \bibnamefont {Pereira}},\ }\bibfield
  {title} {\bibinfo {title} {\emph {Reproduction of the charge density wave
  phase diagram in {1T}-{TiSe\s2} exposes its excitonic character}},\ }\href
  {https://doi.org/10.1103/PhysRevLett.121.226602} {\bibfield  {journal}
  {\bibinfo  {journal} {Phys. Rev. Lett.}\ }\textbf {\bibinfo {volume} {121}},\
  \bibinfo {pages} {226602} (\bibinfo {year} {2018})},\ \Eprint
  {https://arxiv.org/abs/1712.04967} {arXiv:1712.04967}\BibitemShut {NoStop}%
\bibitem [{\citenamefont {Kaneko}\ \emph {et~al.}(2018)\citenamefont {Kaneko},
  \citenamefont {Ohta},\ and\ \citenamefont {Yunoki}}]{Kaneko2018}%
  \BibitemOpen
  \bibfield  {author} {\bibinfo {author} {\bibfnamefont {T.}~\bibnamefont
  {Kaneko}}, \bibinfo {author} {\bibfnamefont {Y.}~\bibnamefont {Ohta}},\ and\
  \bibinfo {author} {\bibfnamefont {S.}~\bibnamefont {Yunoki}},\ }\bibfield
  {title} {\bibinfo {title} {\emph {Exciton-phonon cooperative mechanism of the
  triple-$q$ charge-density-wave and antiferroelectric electron polarization in
  {TiSe\s2}}},\ }\href {https://doi.org/10.1103/PhysRevB.97.155131} {\bibfield
  {journal} {\bibinfo  {journal} {Phys. Rev. B}\ }\textbf {\bibinfo {volume}
  {97}},\ \bibinfo {pages} {155131} (\bibinfo {year} {2018})},\ \Eprint
  {https://arxiv.org/abs/1711.08547} {arXiv:1711.08547}\BibitemShut {NoStop}%
\bibitem [{\citenamefont {Pasquier}\ and\ \citenamefont
  {Yazyev}(2018)}]{Pasquier2018}%
  \BibitemOpen
  \bibfield  {author} {\bibinfo {author} {\bibfnamefont {D.}~\bibnamefont
  {Pasquier}}\ and\ \bibinfo {author} {\bibfnamefont {O.~V.}\ \bibnamefont
  {Yazyev}},\ }\bibfield  {title} {\bibinfo {title} {\emph {Excitonic effects
  in two-dimensional {TiSe\s2} from hybrid density functional theory}},\ }\href
  {https://doi.org/10.1103/PhysRevB.98.235106} {\bibfield  {journal} {\bibinfo
  {journal} {Phys. Rev. B}\ }\textbf {\bibinfo {volume} {98}},\ \bibinfo
  {pages} {235106} (\bibinfo {year} {2018})},\ \Eprint
  {https://arxiv.org/abs/1805.11560} {arXiv:1805.11560}\BibitemShut {NoStop}%
\bibitem [{\citenamefont {Harrison}(2004)}]{Harrison2004}%
  \BibitemOpen
  \bibfield  {author} {\bibinfo {author} {\bibfnamefont {W.~A.}\ \bibnamefont
  {Harrison}},\ }\href {https://doi.org/10.1142/5432} {\emph {\bibinfo {title}
  {Elementary Electronic Structure}}},\ \bibinfo {edition} {rev.}\ ed.\
  (\bibinfo  {publisher} {World Scientific},\ \bibinfo {address} {Singapore},\
  \bibinfo {year} {2004})\BibitemShut {NoStop}%
\bibitem [{\citenamefont {Giannozzi}\ \emph {et~al.}(2009)\citenamefont
  {Giannozzi}, \citenamefont {Baroni}, \citenamefont {Bonini}, \citenamefont
  {Calandra}, \citenamefont {Car}, \citenamefont {Cavazzoni}, \citenamefont
  {Ceresoli}, \citenamefont {Chiarotti}, \citenamefont {Cococcioni},
  \citenamefont {Dabo}, \citenamefont {Corso}, \citenamefont {Gironcoli},
  \citenamefont {Fabris}, \citenamefont {Fratesi}, \citenamefont {Gebauer},
  \citenamefont {Gerstmann}, \citenamefont {Gougoussis}, \citenamefont
  {Kokalj}, \citenamefont {Lazzeri}, \citenamefont {Martin-Samos},
  \citenamefont {Marzari}, \citenamefont {Mauri}, \citenamefont {Mazzarello},
  \citenamefont {Paolini}, \citenamefont {Pasquarello}, \citenamefont
  {Paulatto}, \citenamefont {Sbraccia}, \citenamefont {Scandolo}, \citenamefont
  {Sclauzero}, \citenamefont {Seitsonen}, \citenamefont {Smogunov},
  \citenamefont {Umari},\ and\ \citenamefont {Wentzcovitch}}]{Giannozzi2009}%
  \BibitemOpen
  \bibfield  {author} {\bibinfo {author} {\bibfnamefont {P.}~\bibnamefont
  {Giannozzi}}, \bibinfo {author} {\bibfnamefont {S.}~\bibnamefont {Baroni}},
  \bibinfo {author} {\bibfnamefont {N.}~\bibnamefont {Bonini}}, \bibinfo
  {author} {\bibfnamefont {M.}~\bibnamefont {Calandra}}, \bibinfo {author}
  {\bibfnamefont {R.}~\bibnamefont {Car}}, \bibinfo {author} {\bibfnamefont
  {C.}~\bibnamefont {Cavazzoni}}, \bibinfo {author} {\bibfnamefont
  {D.}~\bibnamefont {Ceresoli}}, \bibinfo {author} {\bibfnamefont {G.~L.}\
  \bibnamefont {Chiarotti}}, \bibinfo {author} {\bibfnamefont {M.}~\bibnamefont
  {Cococcioni}}, \bibinfo {author} {\bibfnamefont {I.}~\bibnamefont {Dabo}},
  \bibinfo {author} {\bibfnamefont {A.~D.}\ \bibnamefont {Corso}}, \bibinfo
  {author} {\bibfnamefont {S.~d.}\ \bibnamefont {Gironcoli}}, \bibinfo {author}
  {\bibfnamefont {S.}~\bibnamefont {Fabris}}, \bibinfo {author} {\bibfnamefont
  {G.}~\bibnamefont {Fratesi}}, \bibinfo {author} {\bibfnamefont
  {R.}~\bibnamefont {Gebauer}}, \bibinfo {author} {\bibfnamefont
  {U.}~\bibnamefont {Gerstmann}}, \bibinfo {author} {\bibfnamefont
  {C.}~\bibnamefont {Gougoussis}}, \bibinfo {author} {\bibfnamefont
  {A.}~\bibnamefont {Kokalj}}, \bibinfo {author} {\bibfnamefont
  {M.}~\bibnamefont {Lazzeri}}, \bibinfo {author} {\bibfnamefont
  {L.}~\bibnamefont {Martin-Samos}}, \bibinfo {author} {\bibfnamefont
  {N.}~\bibnamefont {Marzari}}, \bibinfo {author} {\bibfnamefont
  {F.}~\bibnamefont {Mauri}}, \bibinfo {author} {\bibfnamefont
  {R.}~\bibnamefont {Mazzarello}}, \bibinfo {author} {\bibfnamefont
  {S.}~\bibnamefont {Paolini}}, \bibinfo {author} {\bibfnamefont
  {A.}~\bibnamefont {Pasquarello}}, \bibinfo {author} {\bibfnamefont
  {L.}~\bibnamefont {Paulatto}}, \bibinfo {author} {\bibfnamefont
  {C.}~\bibnamefont {Sbraccia}}, \bibinfo {author} {\bibfnamefont
  {S.}~\bibnamefont {Scandolo}}, \bibinfo {author} {\bibfnamefont
  {G.}~\bibnamefont {Sclauzero}}, \bibinfo {author} {\bibfnamefont {A.~P.}\
  \bibnamefont {Seitsonen}}, \bibinfo {author} {\bibfnamefont {A.}~\bibnamefont
  {Smogunov}}, \bibinfo {author} {\bibfnamefont {P.}~\bibnamefont {Umari}},\
  and\ \bibinfo {author} {\bibfnamefont {R.~M.}\ \bibnamefont {Wentzcovitch}},\
  }\bibfield  {title} {\bibinfo {title} {\emph {\textsc{Quantum ESPRESSO}: A
  modular and open-source software project for quantum simulations of
  materials}},\ }\href {https://doi.org/10.1088/0953-8984/21/39/395502}
  {\bibfield  {journal} {\bibinfo  {journal} {J. Phys. Condens. Matter}\
  }\textbf {\bibinfo {volume} {21}},\ \bibinfo {pages} {395502} (\bibinfo
  {year} {2009})},\ \Eprint {https://arxiv.org/abs/0906.2569}
  {arXiv:0906.2569}\BibitemShut {NoStop}%
\bibitem [{\citenamefont {Giannozzi}\ \emph {et~al.}(2017)\citenamefont
  {Giannozzi}, \citenamefont {Andreussi}, \citenamefont {Brumme}, \citenamefont
  {Bunau}, \citenamefont {Nardelli}, \citenamefont {Calandra}, \citenamefont
  {Car}, \citenamefont {Cavazzoni}, \citenamefont {Ceresoli}, \citenamefont
  {Cococcioni}, \citenamefont {Colonna}, \citenamefont {Carnimeo},
  \citenamefont {Corso}, \citenamefont {Gironcoli}, \citenamefont {Delugas},
  \citenamefont {DiStasio}, \citenamefont {Ferretti}, \citenamefont {Floris},
  \citenamefont {Fratesi}, \citenamefont {Fugallo}, \citenamefont {Gebauer},
  \citenamefont {Gerstmann}, \citenamefont {Giustino}, \citenamefont {Gorni},
  \citenamefont {Jia}, \citenamefont {Kawamura}, \citenamefont {Ko},
  \citenamefont {Kokalj}, \citenamefont {K\"u\c{c}\"ukbenli}, \citenamefont
  {Lazzeri}, \citenamefont {Marsili}, \citenamefont {Marzari}, \citenamefont
  {Mauri}, \citenamefont {Nguyen}, \citenamefont {Nguyen}, \citenamefont
  {{Otero-de-la-Roza}}, \citenamefont {Paulatto}, \citenamefont {Ponc\'e},
  \citenamefont {Rocca}, \citenamefont {Sabatini}, \citenamefont {Santra},
  \citenamefont {Schlipf}, \citenamefont {Seitsonen}, \citenamefont {Smogunov},
  \citenamefont {Timrov}, \citenamefont {Thonhauser}, \citenamefont {Umari},
  \citenamefont {Vast}, \citenamefont {Wu},\ and\ \citenamefont
  {Baroni}}]{Giannozzi2017}%
  \BibitemOpen
  \bibfield  {author} {\bibinfo {author} {\bibfnamefont {P.}~\bibnamefont
  {Giannozzi}}, \bibinfo {author} {\bibfnamefont {O.}~\bibnamefont
  {Andreussi}}, \bibinfo {author} {\bibfnamefont {T.}~\bibnamefont {Brumme}},
  \bibinfo {author} {\bibfnamefont {O.}~\bibnamefont {Bunau}}, \bibinfo
  {author} {\bibfnamefont {M.~B.}\ \bibnamefont {Nardelli}}, \bibinfo {author}
  {\bibfnamefont {M.}~\bibnamefont {Calandra}}, \bibinfo {author}
  {\bibfnamefont {R.}~\bibnamefont {Car}}, \bibinfo {author} {\bibfnamefont
  {C.}~\bibnamefont {Cavazzoni}}, \bibinfo {author} {\bibfnamefont
  {D.}~\bibnamefont {Ceresoli}}, \bibinfo {author} {\bibfnamefont
  {M.}~\bibnamefont {Cococcioni}}, \bibinfo {author} {\bibfnamefont
  {N.}~\bibnamefont {Colonna}}, \bibinfo {author} {\bibfnamefont
  {I.}~\bibnamefont {Carnimeo}}, \bibinfo {author} {\bibfnamefont {A.~D.}\
  \bibnamefont {Corso}}, \bibinfo {author} {\bibfnamefont {S.~d.}\ \bibnamefont
  {Gironcoli}}, \bibinfo {author} {\bibfnamefont {P.}~\bibnamefont {Delugas}},
  \bibinfo {author} {\bibfnamefont {R.~A.}\ \bibnamefont {DiStasio}}, \bibinfo
  {author} {\bibfnamefont {A.}~\bibnamefont {Ferretti}}, \bibinfo {author}
  {\bibfnamefont {A.}~\bibnamefont {Floris}}, \bibinfo {author} {\bibfnamefont
  {G.}~\bibnamefont {Fratesi}}, \bibinfo {author} {\bibfnamefont
  {G.}~\bibnamefont {Fugallo}}, \bibinfo {author} {\bibfnamefont
  {R.}~\bibnamefont {Gebauer}}, \bibinfo {author} {\bibfnamefont
  {U.}~\bibnamefont {Gerstmann}}, \bibinfo {author} {\bibfnamefont
  {F.}~\bibnamefont {Giustino}}, \bibinfo {author} {\bibfnamefont
  {T.}~\bibnamefont {Gorni}}, \bibinfo {author} {\bibfnamefont
  {J.}~\bibnamefont {Jia}}, \bibinfo {author} {\bibfnamefont {M.}~\bibnamefont
  {Kawamura}}, \bibinfo {author} {\bibfnamefont {H.-Y.}\ \bibnamefont {Ko}},
  \bibinfo {author} {\bibfnamefont {A.}~\bibnamefont {Kokalj}}, \bibinfo
  {author} {\bibfnamefont {E.}~\bibnamefont {K\"u\c{c}\"ukbenli}}, \bibinfo
  {author} {\bibfnamefont {M.}~\bibnamefont {Lazzeri}}, \bibinfo {author}
  {\bibfnamefont {M.}~\bibnamefont {Marsili}}, \bibinfo {author} {\bibfnamefont
  {N.}~\bibnamefont {Marzari}}, \bibinfo {author} {\bibfnamefont
  {F.}~\bibnamefont {Mauri}}, \bibinfo {author} {\bibfnamefont {N.~L.}\
  \bibnamefont {Nguyen}}, \bibinfo {author} {\bibfnamefont {H.-V.}\
  \bibnamefont {Nguyen}}, \bibinfo {author} {\bibfnamefont {A.}~\bibnamefont
  {{Otero-de-la-Roza}}}, \bibinfo {author} {\bibfnamefont {L.}~\bibnamefont
  {Paulatto}}, \bibinfo {author} {\bibfnamefont {S.}~\bibnamefont {Ponc\'e}},
  \bibinfo {author} {\bibfnamefont {D.}~\bibnamefont {Rocca}}, \bibinfo
  {author} {\bibfnamefont {R.}~\bibnamefont {Sabatini}}, \bibinfo {author}
  {\bibfnamefont {B.}~\bibnamefont {Santra}}, \bibinfo {author} {\bibfnamefont
  {M.}~\bibnamefont {Schlipf}}, \bibinfo {author} {\bibfnamefont {A.~P.}\
  \bibnamefont {Seitsonen}}, \bibinfo {author} {\bibfnamefont {A.}~\bibnamefont
  {Smogunov}}, \bibinfo {author} {\bibfnamefont {I.}~\bibnamefont {Timrov}},
  \bibinfo {author} {\bibfnamefont {T.}~\bibnamefont {Thonhauser}}, \bibinfo
  {author} {\bibfnamefont {P.}~\bibnamefont {Umari}}, \bibinfo {author}
  {\bibfnamefont {N.}~\bibnamefont {Vast}}, \bibinfo {author} {\bibfnamefont
  {X.}~\bibnamefont {Wu}},\ and\ \bibinfo {author} {\bibfnamefont
  {S.}~\bibnamefont {Baroni}},\ }\bibfield  {title} {\bibinfo {title} {\emph
  {Advanced capabilities for materials modelling with \textsc{Quantum
  ESPRESSO}}},\ }\href {https://doi.org/10.1088/1361-648X/aa8f79} {\bibfield
  {journal} {\bibinfo  {journal} {J. Phys. Condens. Matter}\ }\textbf {\bibinfo
  {volume} {29}},\ \bibinfo {pages} {465901} (\bibinfo {year} {2017})},\
  \Eprint {https://arxiv.org/abs/1709.10010} {arXiv:1709.10010}\BibitemShut
  {NoStop}%
\bibitem [{\citenamefont {Mostofi}\ \emph {et~al.}(2014)\citenamefont
  {Mostofi}, \citenamefont {Yates}, \citenamefont {Pizzi}, \citenamefont {Lee},
  \citenamefont {Souza}, \citenamefont {Vanderbilt},\ and\ \citenamefont
  {Marzari}}]{Mostofi2014}%
  \BibitemOpen
  \bibfield  {author} {\bibinfo {author} {\bibfnamefont {A.~A.}\ \bibnamefont
  {Mostofi}}, \bibinfo {author} {\bibfnamefont {J.~R.}\ \bibnamefont {Yates}},
  \bibinfo {author} {\bibfnamefont {G.}~\bibnamefont {Pizzi}}, \bibinfo
  {author} {\bibfnamefont {Y.-S.}\ \bibnamefont {Lee}}, \bibinfo {author}
  {\bibfnamefont {I.}~\bibnamefont {Souza}}, \bibinfo {author} {\bibfnamefont
  {D.}~\bibnamefont {Vanderbilt}},\ and\ \bibinfo {author} {\bibfnamefont
  {N.}~\bibnamefont {Marzari}},\ }\bibfield  {title} {\bibinfo {title} {\emph
  {An updated version of wannier90: A tool for obtaining maximally-localised
  {Wannier} functions}},\ }\href {https://doi.org/10.1016/j.cpc.2014.05.003}
  {\bibfield  {journal} {\bibinfo  {journal} {Comput. Phys. Commun.}\ }\textbf
  {\bibinfo {volume} {185}},\ \bibinfo {pages} {2309} (\bibinfo {year}
  {2014})}\BibitemShut {NoStop}%
\bibitem [{\citenamefont {Ponc\'e}\ \emph {et~al.}(2016)\citenamefont
  {Ponc\'e}, \citenamefont {Margine}, \citenamefont {Verdi},\ and\
  \citenamefont {Giustino}}]{Ponce2016}%
  \BibitemOpen
  \bibfield  {author} {\bibinfo {author} {\bibfnamefont {S.}~\bibnamefont
  {Ponc\'e}}, \bibinfo {author} {\bibfnamefont {E.}~\bibnamefont {Margine}},
  \bibinfo {author} {\bibfnamefont {C.}~\bibnamefont {Verdi}},\ and\ \bibinfo
  {author} {\bibfnamefont {F.}~\bibnamefont {Giustino}},\ }\bibfield  {title}
  {\bibinfo {title} {\emph {{EPW}: Electron--phonon coupling, transport and
  superconducting properties using maximally localized {Wannier} functions}},\
  }\href {https://doi.org/10.1016/j.cpc.2016.07.028} {\bibfield  {journal}
  {\bibinfo  {journal} {Comput. Phys. Commun.}\ }\textbf {\bibinfo {volume}
  {209}},\ \bibinfo {pages} {116} (\bibinfo {year} {2016})},\ \Eprint
  {https://arxiv.org/abs/1604.03525} {arXiv:1604.03525}\BibitemShut {NoStop}%
\bibitem [{\citenamefont {Perdew}\ \emph {et~al.}(1996)\citenamefont {Perdew},
  \citenamefont {Burke},\ and\ \citenamefont {Ernzerhof}}]{Perdew1996}%
  \BibitemOpen
  \bibfield  {author} {\bibinfo {author} {\bibfnamefont {J.~P.}\ \bibnamefont
  {Perdew}}, \bibinfo {author} {\bibfnamefont {K.}~\bibnamefont {Burke}},\ and\
  \bibinfo {author} {\bibfnamefont {M.}~\bibnamefont {Ernzerhof}},\ }\bibfield
  {title} {\bibinfo {title} {\emph {Generalized gradient approximation made
  simple}},\ }\href {https://doi.org/10.1103/PhysRevLett.77.3865} {\bibfield
  {journal} {\bibinfo  {journal} {Phys. Rev. Lett.}\ }\textbf {\bibinfo
  {volume} {77}},\ \bibinfo {pages} {3865} (\bibinfo {year}
  {1996})}\BibitemShut {NoStop}%
\bibitem [{\citenamefont {Perdew}\ \emph {et~al.}(1997)\citenamefont {Perdew},
  \citenamefont {Burke},\ and\ \citenamefont {Ernzerhof}}]{Perdew1997}%
  \BibitemOpen
  \bibfield  {author} {\bibinfo {author} {\bibfnamefont {J.~P.}\ \bibnamefont
  {Perdew}}, \bibinfo {author} {\bibfnamefont {K.}~\bibnamefont {Burke}},\ and\
  \bibinfo {author} {\bibfnamefont {M.}~\bibnamefont {Ernzerhof}},\ }\bibfield
  {title} {\bibinfo {title} {\emph {Generalized gradient approximation made
  simple [{Phys. Rev. Lett.} 77, 3865 (1996)]}},\ }\href
  {https://doi.org/10.1103/PhysRevLett.78.1396} {\bibfield  {journal} {\bibinfo
   {journal} {Phys. Rev. Lett.}\ }\textbf {\bibinfo {volume} {78}},\ \bibinfo
  {pages} {1396(E)} (\bibinfo {year} {1997})}\BibitemShut {NoStop}%
\bibitem [{\citenamefont {Goedecker}\ \emph {et~al.}(1996)\citenamefont
  {Goedecker}, \citenamefont {Teter},\ and\ \citenamefont
  {Hutter}}]{Goedecker1996}%
  \BibitemOpen
  \bibfield  {author} {\bibinfo {author} {\bibfnamefont {S.}~\bibnamefont
  {Goedecker}}, \bibinfo {author} {\bibfnamefont {M.}~\bibnamefont {Teter}},\
  and\ \bibinfo {author} {\bibfnamefont {J.}~\bibnamefont {Hutter}},\
  }\bibfield  {title} {\bibinfo {title} {\emph {Separable dual-space {Gaussian}
  pseudopotentials}},\ }\href {https://doi.org/10.1103/PhysRevB.54.1703}
  {\bibfield  {journal} {\bibinfo  {journal} {Phys. Rev. B}\ }\textbf {\bibinfo
  {volume} {54}},\ \bibinfo {pages} {1703} (\bibinfo {year} {1996})},\ \Eprint
  {https://arxiv.org/abs/mtrl-th/9512004} {arXiv:mtrl-th/9512004}\BibitemShut
  {NoStop}%
\bibitem [{\citenamefont {Hartwigsen}\ \emph {et~al.}(1998)\citenamefont
  {Hartwigsen}, \citenamefont {Goedecker},\ and\ \citenamefont
  {Hutter}}]{Hartwigsen1998}%
  \BibitemOpen
  \bibfield  {author} {\bibinfo {author} {\bibfnamefont {C.}~\bibnamefont
  {Hartwigsen}}, \bibinfo {author} {\bibfnamefont {S.}~\bibnamefont
  {Goedecker}},\ and\ \bibinfo {author} {\bibfnamefont {J.}~\bibnamefont
  {Hutter}},\ }\bibfield  {title} {\bibinfo {title} {\emph {Relativistic
  separable dual-space {Gaussian} pseudopotentials from {H} to {Rn}}},\ }\href
  {https://doi.org/10.1103/PhysRevB.58.3641} {\bibfield  {journal} {\bibinfo
  {journal} {Phys. Rev. B}\ }\textbf {\bibinfo {volume} {58}},\ \bibinfo
  {pages} {3641} (\bibinfo {year} {1998})},\ \Eprint
  {https://arxiv.org/abs/cond-mat/9803286} {arXiv:cond-mat/9803286}\BibitemShut
  {NoStop}%
\bibitem [{\citenamefont {Liu}\ \emph {et~al.}(2013)\citenamefont {Liu},
  \citenamefont {Shan}, \citenamefont {Yao}, \citenamefont {Yao},\ and\
  \citenamefont {Xiao}}]{Liu2013}%
  \BibitemOpen
  \bibfield  {author} {\bibinfo {author} {\bibfnamefont {G.-B.}\ \bibnamefont
  {Liu}}, \bibinfo {author} {\bibfnamefont {W.-Y.}\ \bibnamefont {Shan}},
  \bibinfo {author} {\bibfnamefont {Y.}~\bibnamefont {Yao}}, \bibinfo {author}
  {\bibfnamefont {W.}~\bibnamefont {Yao}},\ and\ \bibinfo {author}
  {\bibfnamefont {D.}~\bibnamefont {Xiao}},\ }\bibfield  {title} {\bibinfo
  {title} {\emph {Three-band tight-binding model for monolayers of group-{VIB}
  transition metal dichalcogenides}},\ }\href
  {https://doi.org/10.1103/PhysRevB.88.085433} {\bibfield  {journal} {\bibinfo
  {journal} {Phys. Rev. B}\ }\textbf {\bibinfo {volume} {88}},\ \bibinfo
  {pages} {085433} (\bibinfo {year} {2013})},\ \Eprint
  {https://arxiv.org/abs/1305.6089} {arXiv:1305.6089}\BibitemShut {NoStop}%
\end{thebibliography}%

\end{document}